\documentclass[twocolumn,tighten,floatfix]{aastex62}
\usepackage{natbib}
\usepackage{gensymb}
\usepackage{hyperref}
\usepackage{amsmath}
\usepackage{multirow}
\usepackage[maxfloats=1000]{morefloats}

\newcommand{\mus}{$\mu$s}
\newcommand{\tres}{t_\mathrm{res}}
\newcommand{\FAup}{\mathrm{FA_{up}}}

\newcommand{\nbrst}{2118}

\def\MB{Minbar catalogue}
\def\Gwt{\citet{Galloway2008}}

\submitjournal{ApJS}

\shorttitle{Search for thermonuclear burst oscillations in RXTE data}
\shortauthors{Bilous \& Watts}

\begin{document}

\title{A uniform search for thermonuclear burst oscillations in the RXTE legacy dataset}

\correspondingauthor{Anna Bilous}
\email{A.Bilous@uva.nl, hanna.bilous@gmail.com}

\author{Anna~V.~Bilous}
\author{Anna~L.~Watts}
\affiliation{Anton Pannekoek Institute for Astronomy, 
                 University of Amsterdam, Science Park 904, 
                 1098 XH Amsterdam}

\begin{abstract}
We describe a blind uniform search for thermonuclear burst oscillations (TBOs) in the majority of Type-I 
bursts observed by RXTE (\nbrst\ bursts from 57 neutron stars). We examined 2--2002\,Hz power spectra from 
the Fourier transform in sliding 0.5--2\,s windows, using fine-binned light curves in 2-60\,keV energy range. 
The significance of the oscillation  candidates was assessed by simulations which took into account light 
curve variations, dead time and sliding time windows. Some of our sources exhibited multi-frequency variability 
at $\lesssim15$\,Hz that cannot be readily removed with light-curve modeling and may have an astrophysical 
(non-TBO) nature.  Overall, we found that the number and strength of potential candidates depends strongly 
on the parameters of the search.  We found candidates from all previously known RXTE TBO sources, with 
pulsations that had been detected at similar frequencies in multiple independent time windows, and discovered 
TBOs from SAX J1810.8$-$2658. We could not confirm most previously-reported tentative TBO detections or identify 
any obvious candidates just below the detection threshold at similar frequencies in multiple bursts. We computed 
fractional amplitudes of all TBO candidates and placed upper limits on non-detections. Finally, for a few sources 
we noted small excess of candidates with powers comparable to fainter TBOs, but appearing in single independent 
time windows at random frequencies. At least some of these candidates may be noise spikes that appear interesting 
due to selection effects. The potential presence of such candidates calls for extra caution if claiming single-window 
TBO detections. 
\end{abstract}


\section{Introduction} \label{sec:intro}

Thermonuclear burst oscillations (TBOs) are fast (typically, with a frequency of a few hundreds of Hz), 
relatively faint (fractional amplitude of a few percent) oscillations of photon count rate, detected 
in about 20\% of known Type I X-ray bursts\footnote{\url{http://www.sron.nl/~jeanz/bursterlist.html}, 
see also \citet{Galloway2008}, \citet{Watts2012}, and references therein.}. The phenomenon of TBOs is 
attributed to the development of bright patches during thermonuclear explosions on the surface of 
accreting neutron stars. Several theories of patch formation have been proposed: flame spreading from the 
ignition point of the bursts \citep[e.g.][]{Strohmayer1997b}, cooling wakes \citep{Mahmoodifar2016}, 
convective patterns \citep{Garcia2018}, or  large-scale (magneto)hydrodynamical oscillations in the 
burning material, induced by the spreading flame \citep[e.g.][]{Heyl2004}. 
However, none of them can explain all of the observed TBO properties, motivating the development of
better physical models for the ignition and progression of thermonuclear reactions on the neutron star 
surface (see the review by \citealt{Watts2012}). 

From the observational side, it is important to establish as complete a picture of TBOs as possible. 
Finding TBOs and constraining their properties is not always straightforward: although oscillations 
are highly coherent, their frequencies can drift (or jump) by a few Hz during the typical few-second 
duration of the TBO, with oscillations sometimes disappearing and reappearing throughout the burst 
\citep{Muno2002a,Muno2002b}. The standard TBO search method relies on the Fourier transform (or 
calculation of $Z^2$-statistics) in a series of closely overlapping windows covering the burst duration 
\citep[e.g.][]{Strohmayer1998a}. Blind searches assume a constant frequency within a single time window, 
since searching for frequency derivatives adds an extra dimension to parameter space and is thus 
computationally expensive.

Estimates of signal significance are traditionally done analytically, based on simple photon counting 
statistics \citep{Groth1975,vanderKlis1989}. At the same time, it has been recognized that the
distribution of noise powers in real spectra is more complicated, being influenced by the burst 
envelope and dead time of the detector \citep{vanderKlis1989,Zhang1995}. Using overlapping time 
windows and custom data filters complicates calculations of the number of independent trials, and 
thus estimates of TBO candidate significance. Some previous studies addressed these issues by 
discarding low frequencies affected by variation of the photon count rate due to the burst envelope
\citep[e.g.][]{Ootes2017}, directly measuring the dead time-affected average noise power 
\citep[e.g.][]{Thompson2005}, using a conservative number of trials \citep[e.g.][]{Bhattacharyya2007b}, 
or estimating candidate significance with the simulation of data for a small number of bursts 
\citep[e.g.][]{Kaaret2007}.

Searching for TBOs requires a sensitive instrument, operating in hard (1--30\,keV) X-rays and 
capable of providing $\mu$s time resolution. So far, the majority of TBO studies have been 
performed using the large set of observations from the \textit{Rossi X-ray Timing Explorer} 
\citep[RXTE,][]{Jahoda2006}, although other telescopes such as \textit{Swift} \citep{Strohmayer2008} 
and \textit{AstroSat} \citep{VerdhanChauhan2017} have  also been used to search for TBOs. The 
relatively quiet period that followed the termination of \textit{RXTE}'s mission in Dec 2011 
ended with the recent launch of NICER \citep{Arzoumanian2014} in 2017; and ongoing studies 
for the next generation of instruments, such as eXTP \citep{Zhang2019} and STROBE-X \citep{Ray2018}, 
are also providing new impetus for TBO studies. 

Previous TBO searches explored relatively small subsets
of data and were conducted using different methods or search parameters.
In order to prepare for searches with new satellites and to provide a uniform picture for 
theoretical modeling of TBOs, we have undertaken the first comprehensive blind search for 
TBOs across almost the entire archival RXTE burst data set, using the 2015 pre-release version 
of the Multi-INstrument Burst ARchive (MINBAR\footnote{\url{https://burst.sci.monash.edu/minbar/}}, 
Galloway et al., in prep). 
We use the standard approach of constant frequency and overlapping 
windows, but estimate the significance of candidates through data simulation that takes into account 
lightcurve (LC) variations, dead time and number of trials. 
This more realistic noise model allows us to search for TBOs in the regions of parameter
space that were omitted or poorly examined before, such as low TBO frequencies ($\nu\lesssim 200$\,Hz) or
TBOs at very high count rates. We also search for clusters of candidates just below our detection threshold.
We pay special attention to bursts with reported TBO candidates, which
were afterwards deemed as tentative or controversial, and 
 prepare our own list of potentially interesting candidates  for subsequent follow-up with 
 upcoming missions. Finally, the obtained frequencies and fractional amplitudes of TBOs
or the upper limits on FAs of non-detections will provide important input for the TBO mechanisms (see e.g. \citet{Heyl2004}
and \citealt{Piro2005}).



\section{Significance estimate of PS for idealized light curves}
\label{sec:stsearch}

\begin{figure}
 \centering
 \includegraphics[scale=1.0]{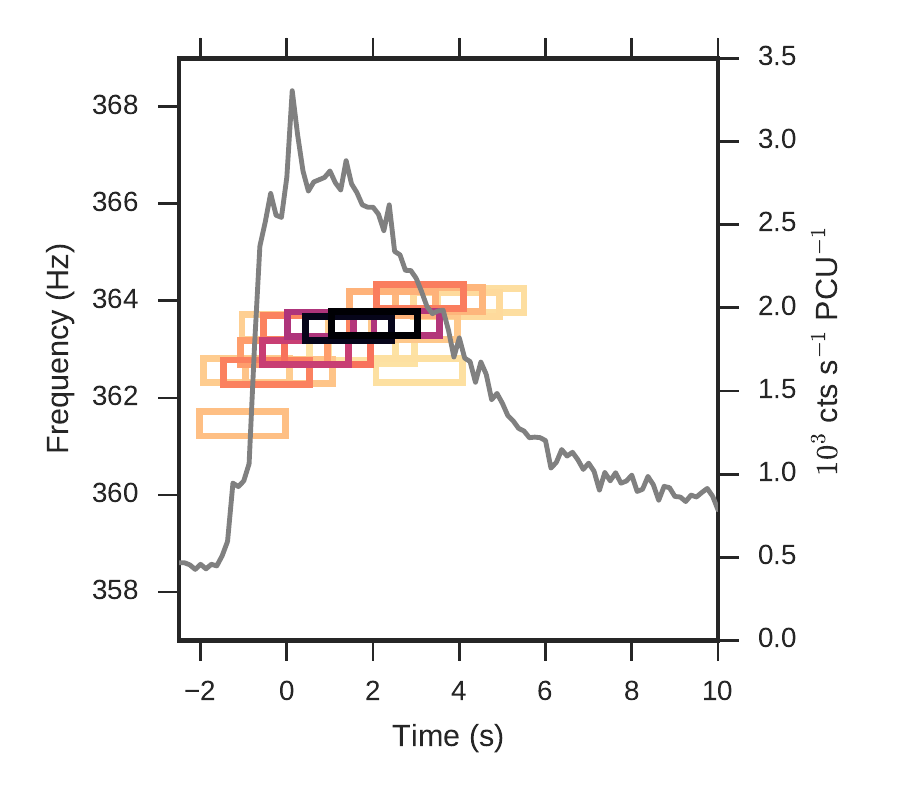}
   \caption{An example of TBOs from the burst observed on MJD 50711.4 from 4U~1728$-$34. The grey 
   line shows photon count rate binned into 0.125-s bins. Color boxes show the TBO candidates 
   from the Fourier power spectrum in a series of 2-s sliding windows, with the start time of 
   each window shifted every time by 0.25\,s. The size of the boxes matches the length of the 
   time window and the frequency resolution. A small random jitter was added to the central 
   frequency of the boxes for better visual clarity. Box colors reflect Leahy-normalized
   power, running from 15 (lightest) to 190 (darkest). }
 \label{fig:FFT_example}
\end{figure}

Burst oscillations are very coherent pulsations which typically last for several seconds. 
During this time the oscillation frequency may jump or drift by several Hz. The search for 
TBOs is therefore usually conducted in separate, but often heavily overlapping time windows of about
0.25--5\,s. Within each window, the photon arrival times are binned into sub-ms bins, then the 
Fast Fourier Transform (FFT) is taken from the number of photons versus time and the obtained 
power spectrum (PS) is examined for outliers. An example of such spectrum is shown in 
Fig.~\ref{fig:FFT_example}. An alternative to a FFT PS is to use $Z^2$ statistics \citep{Buccheri1983}, 
which do not require binning of photon arrival times, and can be computed on a finer grid of 
frequencies (including variable frequency). $Z^2$ statistics are more computationally expensive 
than FFTs, so they are usually used to search for TBOs from sources with known oscillations 
in a narrow range of frequencies \citep[e.g.][]{Watts2005}.

\begin{figure*}
 \centering
 \includegraphics[scale=1.0]{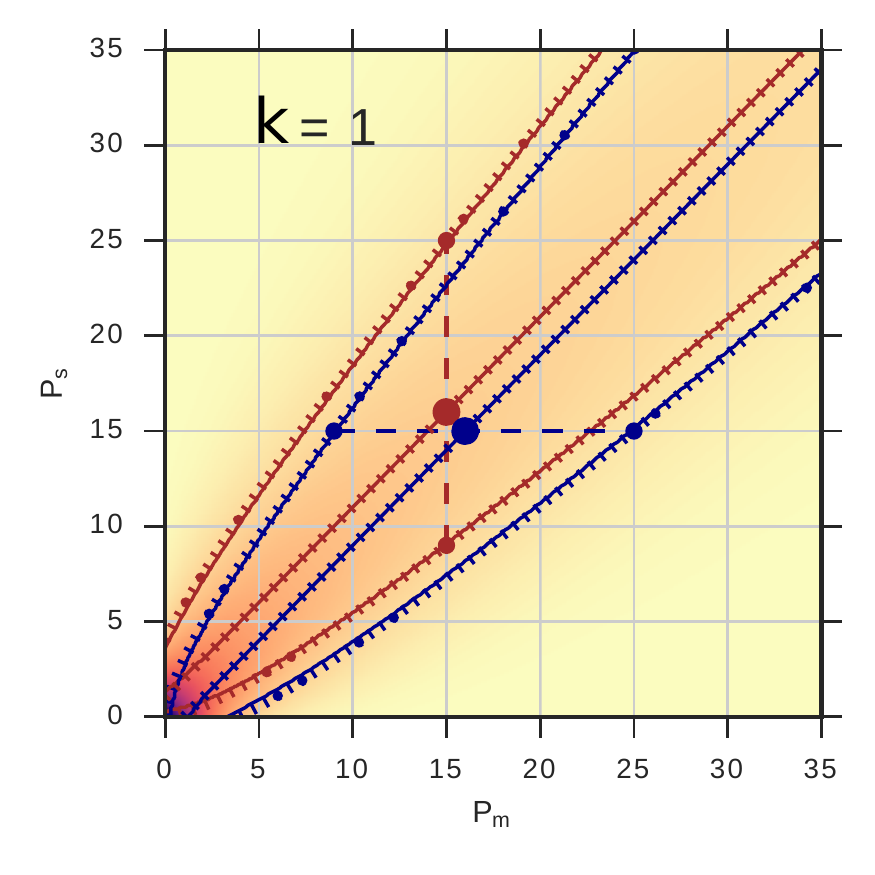}\includegraphics[scale=1.0]{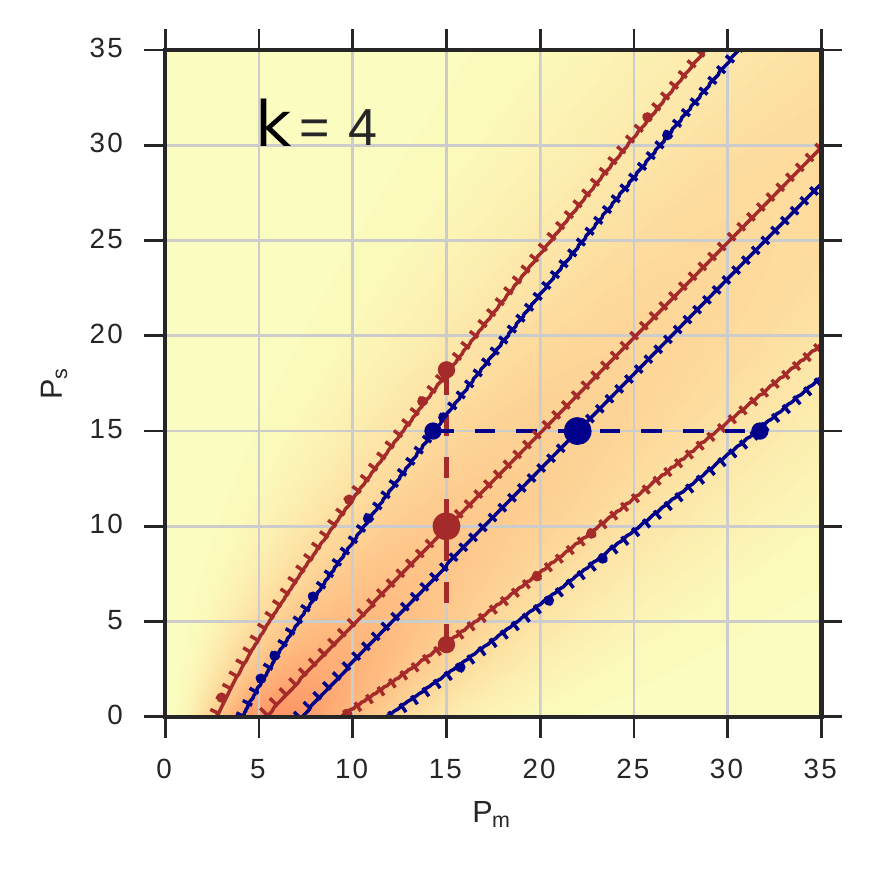}
 \caption{Color: probability density $p_k(P_\mathrm{m}, P_\mathrm{s})$ from \citet{Groth1975} for two 
 different values of number of averaged harmonics, $k$. The color scale is the same for both plots, with 
 the lightest and darkest colors corresponding to $p_k\approx 0$ and $p_k=0.5$, respectively. Solid lines 
 show the numerically calculated median and [0.159, 0.841] percentiles for $p_k(P_\mathrm{s}| P_\mathrm{m})$ 
 (light brown)  and $p_k(P_\mathrm{m}| P_\mathrm{s})$ (dark blue). Fine-dashed lines show the median 
 and mean $\pm$ standard deviation  from Table~\ref{table:grothprob}. The vertical and horizontal 
 lines mark the median and 68\% confidence interval of $P_\mathrm{s}$ given $P_\mathrm{m}=15$ 
 and $P_\mathrm{m}$ given $P_\mathrm{s}=15$, respectively.  For $k>1$ the probability density of 
 $P_\mathrm{s}$ at $P_\mathrm{m} \lesssim 2(k-\sqrt{k})$ has a very sharp peak close to 0, so the 
 median and percentiles of $p(P_\mathrm{s}| P_\mathrm{m})$ become progressively locked at $\approx 0$ 
 with decreasing $P_\mathrm{m}$.}
 \label{fig:probGroth}
\end{figure*}

\subsection{Probability of false detection and the distribution of intrinsic signal power}

Once the PS has been computed, potential oscillation candidates are identified as harmonics 
exceeding a certain threshold. Two questions immediately emerge: (a) for a given candidate, 
what is the probability of obtaining this power purely due to noise fluctuations, and (b) 
given the recorded power, what is the distribution of true signal power? The answers to both 
questions were given in the work of \citet{Groth1975}. Below, we repeat the author's derivations 
in a somewhat modified form,  using a different (nowadays, standard) normalization for the 
power spectrum.

In \citet{Groth1975} the data time series is assumed to be composed of signal and noise:
\begin{equation}
N_\mathrm{m}(t) = N_\mathrm{s}(t) + N_\mathrm{n}(t),
\end{equation}
where $N$ is the number of photons in a given time bin and the  subscripts correspond to 
measurement (m), signal (s) and noise (n). The coefficients of the Fourier transform of $N(t)$, 
$R$ for real and $I$ for imaginary, are the sum of the corresponding coefficients for signal 
and noise:
\begin{equation}
\begin{split}
R_\mathrm{m}(\nu) & = R_\mathrm{s}(\nu) + R_\mathrm{n}(\nu),\\
I_\mathrm{m}(\nu) & = I_\mathrm{s}(\nu) + I_\mathrm{n}(\nu).
\end{split}
\end{equation}
If $N_\mathrm{n}$ has a Poisson distribution, then both $R_\mathrm{n}$ and $I_\mathrm{n}$ have 
normal distributions. For the so-called Leahy-normalized $P_\mathrm{n}$ \citep{Leahy1983}:
\begin{equation}
 P_\mathrm{n} = \frac{2}{\sum N_\mathrm{n}}(R_\mathrm{n}^2 + I_\mathrm{n}^2),
\end{equation}
normal distributions are standard, with mean of 0 and variance of 1. In this case $P_\mathrm{n}$ 
will have a $\chi^2$ distribution with 2 degrees of freedom.

Assuming that the signal is deterministic, \citeauthor{Groth1975} derived an analytical 
expression for the joint probability distribution of the measured power $P_\mathrm{m}$ and the 
signal of power $P_s$:
\begin{equation}
\label{eq:grothprob}
\begin{split}
 p_k(P_\mathrm{m}, P_\mathrm{s}) = \frac{1}{2}\left(\frac{P_\mathrm{m}}{P_\mathrm{s}}\right)^{(k-1)/2} & \exp\left[-\frac{P_\mathrm{m}+P_\mathrm{s}}{2}\right] \times \\ 
 & \times I_{k-1}\left(\sqrt{P_\mathrm{m}P_\mathrm{s}}\right),
 \end{split}
 \end{equation}
where in this equation $I$ is a modified Bessel function of the first kind and $k$ is the 
number of PS samples summed. Here, both $P_\mathrm{m}$ and $P_\mathrm{s}$ are Leahy-normalized 
and the whole derivation is valid if the total number of noise photons in the time window, 
$\sum N_\mathrm{n}$, is larger than approximately ten.

Eq.~\ref{eq:grothprob} can be used to estimate the probability
distribution of $P_\mathrm{s}$ as a function of the measured power $P_\mathrm{m}$,
$ p_k(P_\mathrm{s}| P_\mathrm{m})$, or alternatively 
the probability distribution of measured power $P_\mathrm{m}$ as a function of the signal power $P_\mathrm{s}$,
$p_k(P_\mathrm{m}| P_\mathrm{s})$.
Fig.~\ref{fig:probGroth} shows an 
example of the 2D probability density $p_k(P_\mathrm{m}, P_\mathrm{s})$ for $k=1$ and $k=4$, 
together with the median and $[0.159,\,0.841]$ percentiles for $p_k(P_\mathrm{m}| P_\mathrm{s})$ 
and $p_k(P_\mathrm{s}| P_\mathrm{m})$.

Table~\ref{table:grothprob} gives expressions for the median, mean and standard deviation of 1-D distributions
$p_k(P_\mathrm{m}| P_\mathrm{s})$ and $p_k(P_\mathrm{s}| P_\mathrm{m})$. The mean and standard deviation of $p_k(P_\mathrm{m}| P_\mathrm{s})$ 
were given in \citet{Groth1975} and are exact. The rest of the moments are useful approximations, 
obtained using numerically computed values for $1\leq k \leq 20$ and both $P_\mathrm{m}$ and 
$P_\mathrm{s}$ smaller than 200. For $p_k(P_\mathrm{s}| P_\mathrm{m})$, the approximations 
are valid when  $P_\mathrm{m} \gtrsim 2(k+\sqrt{k})$. For these $P_\mathrm{m}$ the absolute 
value of discrepancies between the approximation and the numerically computed moments are
$\lesssim 0.2k$, $\lesssim 0.02k$ and $\lesssim 0.03k$, for the median, mean, and standard deviation, respectively. 
For  $p_k(P_\mathrm{m}| P_\mathrm{s})$, the median value deviates from 
$P_\mathrm{s}+2k-1$ by $\lesssim 0.1k$ for $P_\mathrm{s} > k$.  

\begin{table}
\begin{center} 
\caption{Moments of the 1-D distributions from Eq.~\ref{eq:grothprob}.  
The moments for  $p_k(P_\mathrm{s}| P_\mathrm{m})$ and median for  $p_k(P_\mathrm{m}| P_\mathrm{s})$  
are approximations that are not valid for $P_\mathrm{m} \lesssim 2k+3\sqrt{n}$.
 \label{table:grothprob}}
\begin{tabular}{lll} 
\hline\\ 
\parbox{2.2cm}{\centering  } &
\parbox{2.2cm}{\centering  $P_\mathrm{m}$ given $P_\mathrm{s}$} &
\parbox{2.2cm}{\centering  $P_\mathrm{s}$ given $P_\mathrm{m}$ } 
\\ [0.1cm]
\hline\\
Median &	$P_\mathrm{s} + 2k - 1$	& $P_\mathrm{m}-2k+3$ \\
Mean & $P_\mathrm{s} +2k$ & $P_\mathrm{m} -2k + 4$ \\
Standard deviation	& $2\sqrt{P_\mathrm{s}+k}$	& $2\sqrt{P_\mathrm{m}-k+2}$ \\
\hline 
\end{tabular} 
\end{center}
\end{table}

For $P_\mathrm{s}=0$, Eq.~\ref{eq:grothprob} expresses the probability of obtaining $P_\mathrm{m}$ 
without any signal, due to noise alone. This is used to estimate the significance of a potential signal detection. 
For $k=1$, Eq.~\ref{eq:grothprob} reduces to:
\begin{equation}
\label{eq:chi2}
 p_1(P_\mathrm{m}, 0) = \frac{1}{2}\exp\left[-P_\mathrm{m}/2\right],
\end{equation}
which is the probability density function (pdf) for the $\chi^2$ distribution with two degrees of freedom. 
It can be also shown that for $k>1$ the pdf is the $\chi^2$ distribution with $2k$ degrees of freedom.

\subsection{Fractional amplitudes}

Besides oscillation frequency, power spectra contain information about the amplitude of pulsations.
For example, for a Leahy-normalized PS, the rms fractional amplitude (FA) is defined as:
\begin{equation}
\label{eq:framp_simp}
 A = \left(\frac{P_\mathrm{s}}{\sum{ N_\mathrm{m} }}\right)^{1/2}. 
\end{equation}

Let us show that for the simplest case of a purely sinusoidal wave on a constant-rate background, described by a 
measured photon count of $N_\mathrm{m} = \mathrm{Poisson}(C) + B\sin(2\pi\nu t)$, the fractional amplitude calculated 
from Eq.~\ref{eq:framp_simp} is, on average, equal to $B/(C\sqrt{2})$. In this specific case the noise count 
rate is described by the Poisson process with a mean rate of $C$, and the signal is 
$N_\mathrm{s} = B\sin(2\pi\nu t)$. 
The Leahy-normalized 
power spectrum of the signal is, by definition:
\begin{equation}
 P_\mathrm{s} = \frac{2}{\sum{N_\mathrm{n}}}(R^2_\mathrm{s} + I^2_\mathrm{s}).
\end{equation}
Here, the average total number of noise photons is equal to its average rate times the number of time bins: 
$\sum{N_\mathrm{n}} = CN_\mathrm{bin}$. Since $\sum{N_\mathrm{s}} \approx 0 $, 
$\sum{N_\mathrm{n}} \approx \sum{N_\mathrm{m}} = C N_\mathrm{bin}$. For a sine wave with amplitude $B$, 
\texttt{scipy} DFT\footnote{\url{https://docs.scipy.org/doc/numpy-1.13.0/reference/routines.fft.html}} 
yields PS harmonic with an amplitude of $R^2_\mathrm{s} + I^2_\mathrm{s} = B^2N^2_\mathrm{bin}/4$. 
Thus, the Leahy-normalized $P_\mathrm{s} = B^2 N_\mathrm{bin}/(4C)$ and $A = B/(C\sqrt{2})$. 

In the limit of very small noise, $A$ approaches $A = 1/\sqrt{2}\approx 70\%$. However, formal 
calculation of fractional amplitudes may result in arbitrarily large $A$. If there is no signal 
present, $B=0$, $P_\mathrm{s}=0$, and $P_\mathrm{m}$ is distributed as $\chi^2$ with $2n$ degrees 
of freedom. The formally estimated $P_\mathrm{s}$ given observed $P_\mathrm{m}$, regardless of its 
actual probability, yields a median value of $P_\mathrm{s} = P_\mathrm{m}-2k+3$ (Table~\ref{table:grothprob}). 
For sufficiently large $P_\mathrm{m}$, $P_\mathrm{s}$ can be such that $A>1$. 

Rms fractional amplitude is not a uniformly accepted way of describing oscillation amplitude.
Some authors quote a full 
fractional amplitude ($2B/C$) or a half fractional amplitude ($B/C$). 

For the actual burst observations, the noise level has a contribution from background unrelated 
to the observed low-mass X-ray binary (both astrophysical and instrumental), and the persistent 
emission from the source itself. The fractional amplitudes are usually calculated for photons 
in the burst only:
\begin{equation}
\label{eq:framp}
 A = \left(\frac{P_\mathrm{s}}{ N_\mathrm{m} }\right)^{1/2}\frac{N_\mathrm{m}}{N_\mathrm{m}-N_\mathrm{bkg}}, 
\end{equation}
with $N_\mathrm{bkg}$ being an estimate of the number of background photons collected during the 
burst interval.

While analyzing fractional amplitudes one should bear in mind that there may be additional 
complications biasing the obtained values: the persistent emission may increase during the burst 
\citep{Worpel2013,Worpel2015} and there may be pulsed background, unrelated to TBOs, such as 
accretion-powered pulsations (APPs) or a pulsed reflection component from the accretion disk.

\subsection{Complications and caveats}

The standard approach described above provides a simple and relatively fast method for 
searching for TBOs. However, there exist some complications and caveats:

\textit{(a):} The burst count rate may vary significantly within the typical time window 
(e.g. during the burst rise). This results in excess power at low frequencies, and biases 
estimates of $P_\mathrm{s}$ and its significance. 

\textit{(b):} Because of dead time (time during which the detector is busy processing 
the current event and cannot record the next one), noise statistics deviate from $\chi^2$. 
The influence of dead time is larger at higher count rates.

\textit{(c):} Searching for TBOs in overlapping windows complicates the assessment of 
the number of independent trials, and thus the significance of $P_\mathrm{s}$.

\textit{(d):} Abrupt variations of the burst count rate cause covariance between harmonics 
in the PS spectra. This may create the illusion of rapid drift or splitting of the TBO 
frequency.

In the following sections we are going to address caveats \textit{(a)}-\textit{(c)}, 
complementing the traditional approach with more realistic noise modeling.  The influence 
of rapid count variation on the recorded TBO frequency will be explored in a subsequent work 
(Bilous \& Watts, in prep).

\section{RXTE data set}
\label{sec:rxte}

\begin{figure*}
 \centering
 \includegraphics[scale=0.9]{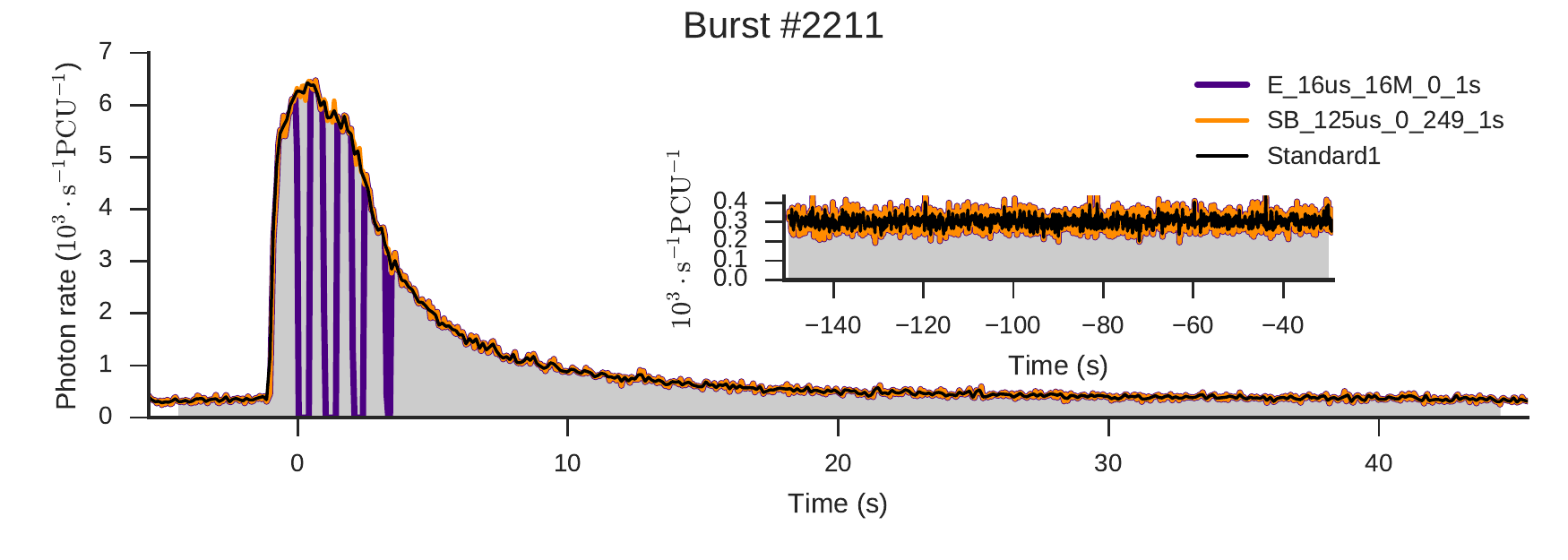}
 \includegraphics[scale=0.9]{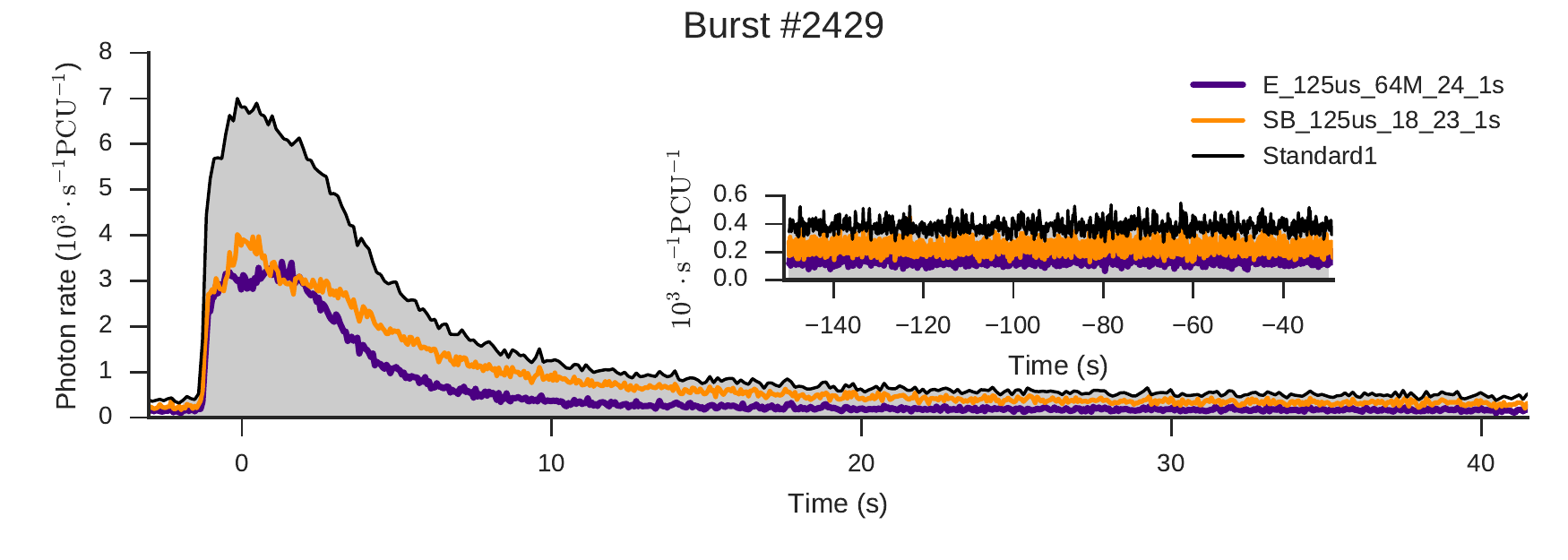}
 \caption{The light curves of two bursts from 4U~1728$-$34. 
 The black line shows total count rate from the Standard-1 data, with 
 the native 0.125-s resolution. The violet and orange lines show Science Event (data mode 
 \texttt{E\_125us\_64M\_24\_1s}) and Science Array (\texttt{SB\_125us\_18\_23\_1s}) data files 
 from the same observations, binned with 31.25-ms bins. Time is measured from the peak of the 
 Standard-1 light curve, binned with 0.5-s bins. The grey area marks the on-burst or baseline 
 (inset) windows. For the top subplot, the high-resolution data was recorded in whole energy 
 band available to RXTE, and Science Event data suffer from gaps due to the high photon count 
 rate. For the bottom subplot, only parts of the band were recorded with high time resolution, 
 resulting in different LC shapes. }
 \label{fig:LCexample}
\end{figure*}

The observations were performed in 1996--2011 with the proportional counter array \citep[PCA, ][]{Jahoda1996} 
on board the RXTE telescope. The PCA consists of five identical co-aligned proportional counter 
units (PCUs), each with a $r=1\degree$ circular field of view. The number of active PCUs varied 
between observing sessions and over the course of RXTE's mission two PCUs went out of order 
permanently\footnote{\url{https://heasarc.gsfc.nasa.gov/docs/xte/recipes/stdprod_guide.html}}.
The PCA is sensitive to photons in the energy range between 2 and 60\,keV. Photon counts are processed 
independently by up to six Event Analyzers (EAs). Two EAs record data in the standard modes, namely
Standard-1 ($\tres=0.125$\,s, one energy channel) and Standard-2 ($\tres=16$\,s, 128 energy channels). 
The rest of the EAs can be configured in a variety of modes, representing the trade-off between 
time and spectral resolution due to finite data transfer capacities while streaming the data 
from the satellite to Earth.

Incoming photons can be recorded in two data modes:  either with all photon arrival times recorded 
separately (``Science Event'' mode) or with arrival times binned in small time bins (``Science Array'' 
mode). Typically, Science Event files have good spectral resolution, but suffer from data losses 
at high count rates. Those Science Array files, which have the bin size suitable for oscillation 
analysis, usually have little information about photon energies, but are less prone to data losses. 
Often, the data were recorded in both Scientific Event and Scientific Array modes and sometimes 
different time resolutions and energy cuts are available for a single observation.

Burst selection was based on the information available in the 2015 pre-release version of the MINBAR 
database, which contains the times of arrival, source associations and other properties of Type I 
bursts observed with different satellites. We selected all bursts observed with RXTE, excluding
bursts which: (a) did not have high-$\tres$ ($\tres<1$\,ms) data during the burst and either 
immediately before or after the burst (b) did not pass the extended good time interval (GTI) 
criterion\footnote{\url{https://heasarc.gsfc.nasa.gov/docs/xte/abc/data_files.html}}
(c) were missing spacecraft housekeeping data, or (d) had variable bin size in Scientific Array 
mode. The final sample contained  \nbrst\ bursts from 57 sources\footnote{We omitted two known 
RXTE superbursts \citep{Zand2017}.}. In this work we use the MINBAR catalogue burst entry number 
as a unique burst identifier. 

For this paper we did not include Burst Catcher data, which can be also used for TBO searches
\citep{Zhang1998,Kaaret2002}. This omission was not crucial as there were no bursts that were 
missing high time-resolution data on the burst rise, or throughout the burst, that would have 
been covered by the relevant Burst Catcher mode (the one with time resolution of 500\,\mus\ or less). 
Nevertheless, several bursts with severe data gaps may benefit from the use of burst catcher data 
(e.g. burst \#2266 from Aql X-1, see also \citealt{Zhang1998}).

For each of the \nbrst\ bursts, we downloaded the data for the observations\footnote{\url{legacy.gsfc.nasa.gov}} 
covering the MINBAR burst arrival time. We made a reference light curve, using the Standard-1 
data re-binned to 0.5\,s and searched for a LC peak within  $\pm1$\,min of the MINBAR burst arrival 
time. The peak time $t_\mathrm{peak}$ served as the absolute reference point within each burst. 
LCs were visually inspected and the baseline window was selected manually for each burst. For 
most of the bursts the baseline window lay within ($t_\mathrm{peak}-150$, $t_\mathrm{peak}-30$)\,s, 
but often it was placed in the burst tail (if no pre-burst data were available) or was shorter 
because of observation duration constraints or the presence of other bursts nearby. The on-burst 
window, ($t_\mathrm{peak}-dt_\mathrm{rise}$,  
$t_\mathrm{peak}+dt_\mathrm{decay}$), was confined 
to the region where photon count exceeded the baseline mean plus two of its standard deviations.
The on-burst window was manually adjusted for bursts with peculiar shapes and faint bursts. 
Figure~\ref{fig:LCexample} shows an example of LCs, the baseline and on-burst windows for 
two bursts from 4U~1728$-$34, recorded in different data modes.

In the event that more than one high-$t_\mathrm{res}$ data mode was available for a single burst, 
we selected files with the largest number of photons per on-burst window, fewer gaps or with 
finer time resolution. We did not discard any photons based on their energy, and merged together 
data files which covered parts of the energy band (e.g. \texttt{SB\_250\_us\_0\_13\_2s} and \texttt{SB\_250\_us\_14\_35\_2s}). 
Sometimes files recorded in different data modes contained 
completely the same information; in this case the data mode was chosen arbitrarily. For uniformity, 
we converted Scientific Array files to the pseudo-Scientific Event format by recording the counts 
in each time bin as individual photons with time of arrival equal to the bin start time.

Photon arrival times were converted from the Mission Elapsed Time (MET) seconds to the UTC time 
system with the TIMEZERO value. This leads to timing  accuracy\footnote{\url{https://heasarc.gsfc.nasa.gov/docs/xte/abc/time.html}} of 100\,\mus, 
which is sufficient for searching for burst oscillations in small windows (up to 4\,s) if one 
is not trying to phase-connect oscillations between different bursts. Since the noise modeling 
relies on the housekeeping data that provides information at regularly sampled time intervals 
in the MET, in this work we chose not to barycenter the data. 

Appendix~A provides more detailed information about the burst sample. Two overview figures 
show burst times of arrival, source-by-source (Fig.~\ref{fig:arrivals}) and Standard-1 LCs
(Fig.~\ref{fig:allbursts}). Table~\ref{table:data} available in its entirety in machine-readable form lists
entry \# and burst arrival time from MINBAR, $t_\mathrm{peak}$ in MET, rise, half-peak and decay 
times, S/N of the burst peak, data mode and notes for each burst. 
Notes indicate manual windows for faint bursts and bursts with peculiar shapes, presence of 
data gaps, partial GTI coverage or incomplete burst coverage.

\section{Noise simulation and data analysis}
\label{sec:noisesim}

In order to estimate the significance of oscillation candidates, for each burst we created a 
number of artificial oscillation-free bursts, which followed the properties of real data as closely 
as possible. The same search analysis was then conducted on real and simulated data.

Originally, we simulated the bursts by scrambling the intervals between photon arrival times in 
$\sim0.1$\,s windows. The size of the window was chosen to be much larger than the presumed TBO 
period, but smaller than the timescales of most large-scale LC variations. A similar technique 
was used by \citet{Fox2001}, however the authors scrambled the LC bins, not the time intervals 
between individual photons. 

Such scrambling preserves deviations from the $\chi^2$ noise distribution, but destroys any 
oscillation signal. However, it appeared that this method failed to produce enough statistically 
independent realizations of noise for some of the count rates. Statistical 
independence was assessed in the following way: we generated a sequence of 100 photons with 
constant rate and random arrival times, then reshuffled the time differences between them 1000 times, 
and computed the power spectra. For each harmonic, the Kolmogorov-Smirnov test was performed to 
test whether the 1000 realizations were consistent with being drawn from a $\chi^2$ distribution. 
For about 45\% of cases the $p$-value was smaller than $0.05$, indicating large deviations 
from $\chi^2$. Acceptable $p$-values were obtained only for a number of photons larger than about 10000.

Thus, the scrambling method was abandoned. Instead, we performed random generation of photon time 
of arrivals (TOAs) using the approximated LC, with subsequent pruning according to known dead time.

 \begin{figure}
 \centering
 \includegraphics[scale=0.8]{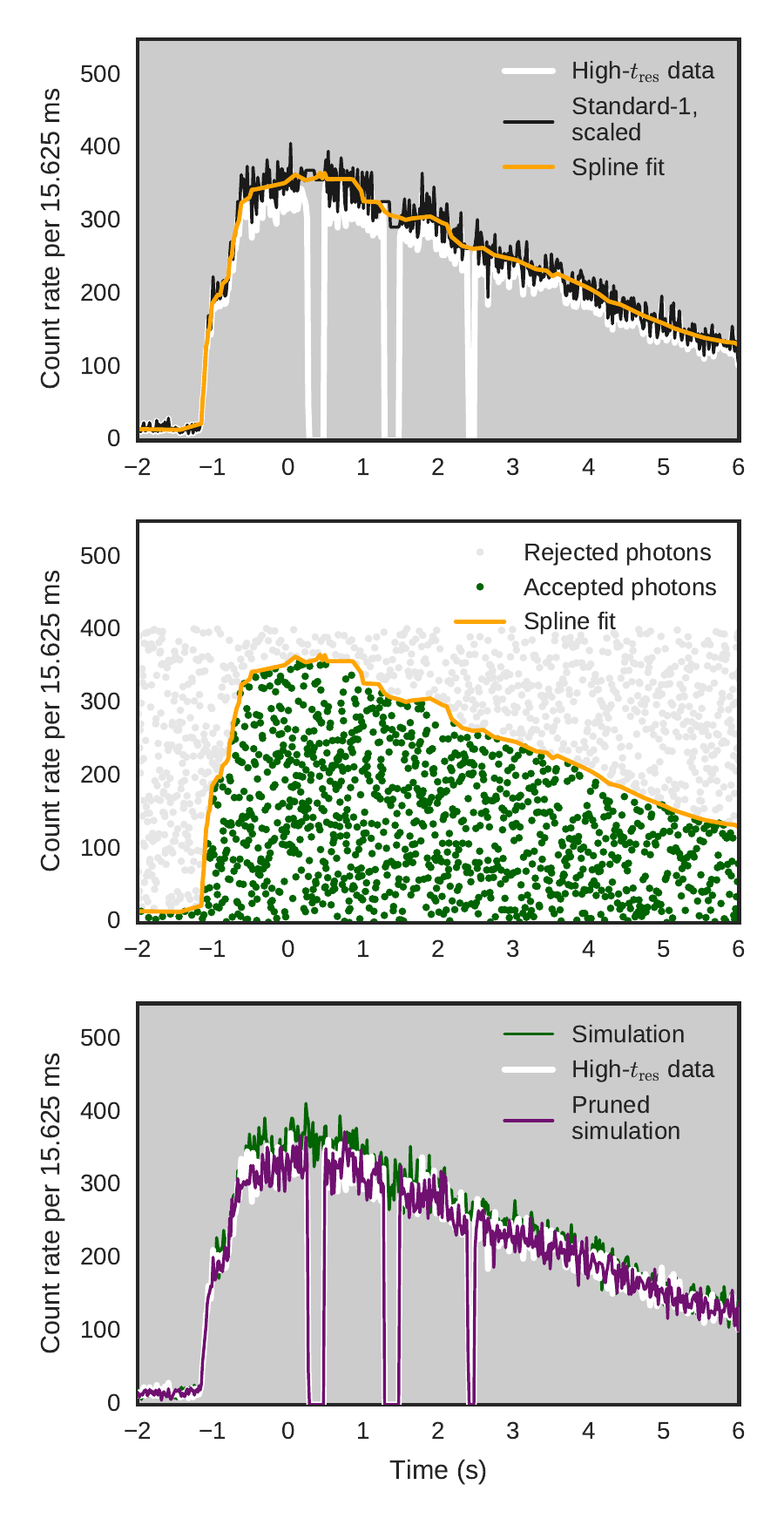}
 \caption{\textit{Top:} the LCs of the high-$t_\mathrm{res}$ data, binned with 15.625\,ms time 
 bins (white), together with Standard-1 LC, scaled according to the fraction of dead time, and re-normalized to mimic the time resolution of the 
 high-$t_\mathrm{res}$ LC outside the data 
 gaps. The orange line shows a spline fit to the scaled and renormalized Standard-1 LC, 
 smoothed with a 0.5-s median filter. \textit{Middle:} Simulation of photon arrival time with 
 the ``acceptance-rejection'' method. Only 1\% of all simulated photons are plotted. 
 \textit{Bottom:} LC from the simulated photons before and after pruning by dead time and 
 data gaps (green and purple, respectively). }
 \label{fig:photon_gen}
\end{figure}

\subsection{Light curve modeling and simulation of photon TOA}

RXTE records four types of events: ``Good Xenon Events'' (events which pass all of the 
discriminators and anti-coincidence vetoes, the desired astrophysical signal) and three 
types of events considered to be mostly due to instrumental background: ``Coincidence Events'', 
``Very Large Events'', or ``Propane Events'' \citep{Jahoda2006}. ``Good Xenon'' events 
are recorded per PCU; for the rest the sum over all active PCUs is saved. All four types 
of events were simulated, since all of them cause dead time, influencing the number of 
recorded Good Xenon events. 

In order to simulate the arrival times we used the information from the Standard-1 files, 
which contain events from the whole energy band, binned in 0.125\,s bins. Sometimes, this 
binning is not sufficient for characterizing the burst rise properly. We therefore 
re-normalized the Standard-1 LC\footnote{Namely, Good Xenon and Propane $+$ Coincidence 
LCs. The Very Large Event LC was not renormalized because its count rate is usually low 
and does not change much during the burst.} with the weights obtained from the 
higher-$t_\mathrm{res}$ data, binned to 1/8 of the Standard-1 resolution (15.625 ms), 
keeping the total number of Standard-1 events in 0.125\,s bins unchanged. If the 
higher-$t_\mathrm{res}$ data were unavailable due to data gaps, uniform weights were 
applied (Fig.~\ref{fig:photon_gen}, top). LC count rates were adjusted for the dead time 
(see more details in Sect.~\ref{subsec:dead_time}).

We followed two different procedures for simulating Good Xenon events and the instrumental 
background. For the Good Xenon events we used a method which was more expensive 
computationally, but which resulted in better reproduction of the stochastic variation 
of the count rate.

The method was as follows. First, the dead time-corrected LC was smoothed with a median 
filter with a typical length of 0.5\,s and a linear spline fit was performed to obtain the 
estimate of the count rate $\mathrm{LC}(t)$ in any arbitrary moment of time 
(Fig.~\ref{fig:photon_gen}, top). The tolerance of spline fit and the size of the median 
filter window were adjusted on a per-source basis, so that the fit was maximally smooth, 
yet preserved, by eye, the short-timescale variations in the LC shape. However, for frequencies 
below $~\sim\,5$\,Hz it becomes complicated to distinguish between potential TBOs and 
non-TBO LC variation, so in this region spurious candidates may be present or the 
significance of TBO candidates may be underestimated (see discussion in 
Sect.~\ref{sec:results}). Light curves for each individual PCU were obtained by multiplying 
the spline fit by the total number of Standard-1 photons recorded by a given PCU, 
divided by the sum from all PCUs.

Then, the arrival times of the Good Xenon events were simulated for each PCU separately 
using the acceptance\,-\,rejection method \citep{vonNeumann1951}. We used the standard 
Python random number generator based on the Mersene Twister algorithm \citep{Matsumoto1998} 
to generate $\mathrm{LC_{max}}\times N_\mathrm{bins}$ pairs of random variables $(L,\,T)$.
Here, $\mathrm{LC_{max}}$ is the maximum value of $\mathrm{LC}(t)$ and $N_\mathrm{bins}$ 
is the number of 0.125-s time bins in the on-burst window 
(Fig.~\ref{fig:photon_gen}, middle). Both $L$ and $T$ were uniformly distributed within
$[0,\,\mathrm{LC_{max}}]$ and the on-burst window, respectively.
Then, the pairs with $L<\mathrm{LC}(T)$ were discarded. This way, a Poisson distribution of photon 
TOAs was created, with the instantaneous rate closely matching $\mathrm{LC}(t)$, but 
devoid of any oscillations with periods smaller than the characteristic time scale of 
the features of the modeled burst envelopes.

For the Propane, Coincidence and Very Large events we used a simpler procedure. The light 
curve counts were divided by the number of PCUs, and for each time bin with local count 
rate $C$ we generated the following number of uniformly distributed TOAs:
\begin{equation}
N = \mathrm{floor}(C) + \mathrm{Binom}[N=1,\,p=\mathrm{frac}(C)],
\end{equation}
where floor denotes the floor function and Binom is a binomial random variable with 
number of trials $N$ and success probability $p$. This way the number of simulated photons
is very close to the real data value and thus the simulation does not emulate the Poisson 
noise which is present in the data.

\subsection{Dead time pruning}
\label{subsec:dead_time}

\begin{figure}
 \centering
 \includegraphics[scale=0.95]{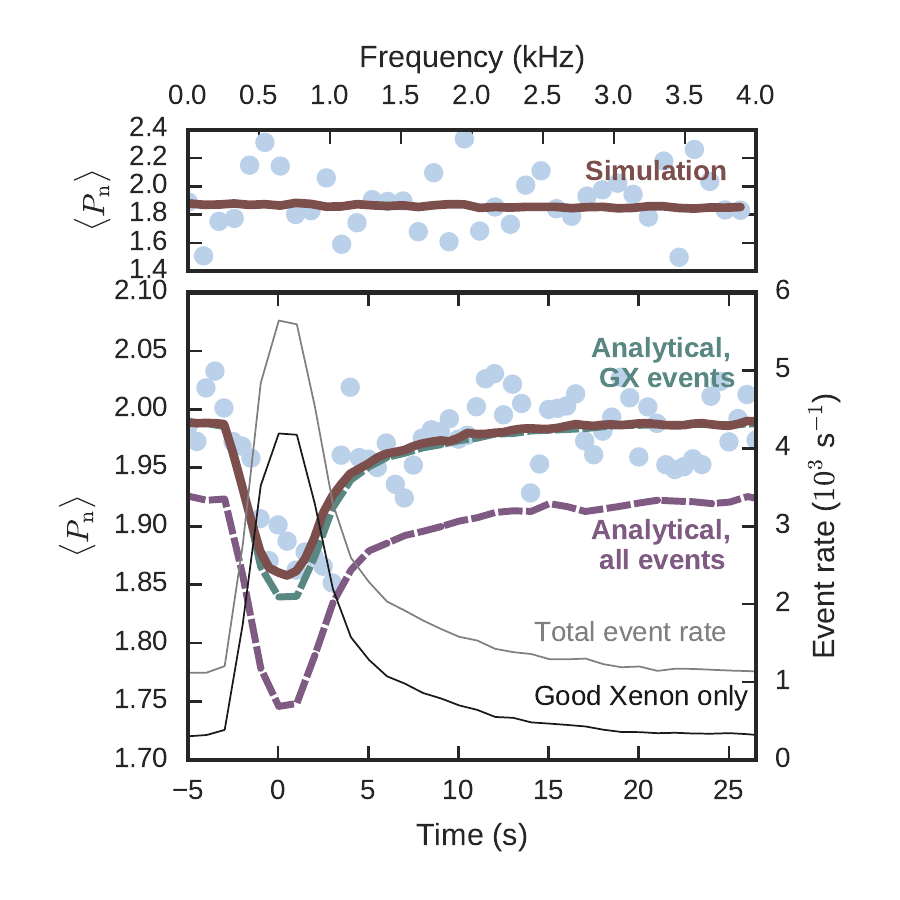}
 \caption{The average noise power $\langle P_\mathrm{n}\rangle$ for one of the PCUs from burst~\#2980 from Aql~X-1 for 
 $t_b = 122$\,$\mu$s and 1-s FFT windows, overlapping by 0.5\,s.  \textit{Top}: $\langle P_\mathrm{n}\rangle$ 
 for the window with the largest photon  count rate versus FFT frequency, averaged by 108 
 harmonics.  Light circles show the actual data, while the dark line shows the mean of 1000 
 simulations. \textit{Bottom:} frequency-averaged $\langle P_\mathrm{n}\rangle$ in each of the FFT windows 
 for the actual data (light circles) and the simulation (dark solid lines).  The dashed 
 curves show the analytically calculated $\langle P_\mathrm{n}\rangle$ from Eq.~\ref{eq:dead_time} for two 
 event rates:  GX only (upper curve) and total event rate (lower curve). Overlayed are the 
 GX and total event count rates. }
 \label{fig:dead_sim}
\end{figure}

Simulated events were subsequently pruned to account for the detector's dead time. Dead time 
calculation for RXTE is rather complex and thus deserves a detailed examination. According to 
\citet{Jahoda2006}, RXTE PCUs process events independently and the dead time is caused by all 
events recorded by the PCU. Any event recorded belongs to one and only to one of the four  
classes\footnote{\url{http://heasarc.nasa.gov/docs/xte/recipes/pca_deadtime.html}}:
Good Xenon, Coincidence, Very Large, or Propane Events.

In general, there exist two types of dead time:  paralyzable (cumulative), where events 
entering the detector during dead time themselves cause further dead time (even though they 
are not recorded); and non-paralyzable, where events entering detector during dead time are 
completely ignored. For RXTE, the actual dead time is a mixture of both types, depending also 
on event class and assigned  energy. However, for most of cases $t_d$ can be approximated
as 10\,$\mu$s non-paralyzable dead time (set by the analog-to-digital converter, ADC) for all 
classes except VL events. For VL events, the dead time can vary between 70\,$\mu$s and 
500\,$\mu$s and depends on the instrumental setting, most of the time being approximately 
170\,$\mu$s \citep{Jahoda2006}. 

\citet{Zhang1995} developed an analytic formula for the Leahy-normalized noise PS in the 
presence of dead time. The mean value $\langle P_\mathrm{n}\rangle$ is always less than 2 by 
some amount which depends on event rate, the type of dead time and its value, the LC bin size, 
$t_b$, and the FFT frequency. The analytic formula for paralyzable dead time has a much 
simpler form than that for the non-paralyzable dead time.

\citet{Jahoda2006} gives an example showing how to calculate the dead-time modifications to 
pure noise for a count rate below $10^4$\,cts~PCU$^{-1}$~s$^{-1}$, applying the correction for 
the paralyzable dead time. Disregarding larger $t_d$ for the VL events:
\begin{equation}
\label{eq:dead_time}
\begin{split}
 \langle P_\mathrm{n}\rangle = 2 & - 4r_0 t_d \left( 1-\frac{t_d}{2t_b}\right)  \\ 
 & - 2r_0 t_d \frac{N-1}{N}\left(\frac{t_d}{t_b}\right)\cos\left(\frac{\pi \nu}{\nu_\mathrm{Nyq}}\right). 
\end{split} 
\end{equation}
Here $\mathrm{n}$ indicates noise (as in Sect.~\ref{sec:stsearch}), $r_0$ is the output rate of 
all events (all four types combined), $t_d$ is the dead time, $t_b$ is the bin size, $\nu$ is 
the FFT frequency, $N$ is the number of frequencies in the PS, $\nu_\mathrm{Nyq}$ is the Nyquist 
frequency. The authors also give an ad hoc correction for the larger $t_\mathrm{d}$ of VL events, 
which is much smaller than the one introduced by Eq.~\ref{eq:dead_time} for our $t_d$ and $t_b$ 
and VL event rates.

Although technically RXTE dead time is a complex mixture of both paralyzable and non-paralyzable 
dead times, with the non-paralyzable dead time of ADC contributing the most at  energies below 
$\approx 20$\,keV, for the typical RXTE rates and $t_b$ size used in this work, 
the formulas for paralyzable and non-paralyzable dead times
yield essentially the same corrections. Thus, the more simple Eq.~\ref{eq:dead_time} can be 
used to estimate the dead time influence on the average noise power.

For our simulations, we treated dead time as purely
non-paralyzable. Note that we do not model the absence of dead time caused by events which 
triggered only $\mathrm{V_X}$ or alpha chains. According to \citet{Jahoda2006}, not doing this 
leads to a small overestimation of dead time fraction by $\delta t_\mathrm{d}/t_\mathrm{b} \approx 0.0014$, 
an amount nearly constant throughout the mission. 

Fig.~\ref{fig:dead_sim} shows the average signal power in Burst~\#2980 from Aql~X-1 (in which 
no TBOs were detected). Photons from only one PCU were selected, and a bin size of
$1/8192\,\mathrm{s}\approx 122\,\mu$s  was used, together with an FFT window of 1\,s. For this 
choice of binning and the rates of Good Xenon (GX) and other events the noise power has only a 
minuscule dependence on FFT frequency. We have summed all harmonics above 10\,Hz (below 10\,Hz 
the PS may be affected by red noise) and compared the noise power to the mean 
obtained from 1000 simulations, which appeared to match the data well (Fig.~\ref{fig:dead_sim}).

\begin{figure}
 \centering
 \includegraphics[scale=0.95]{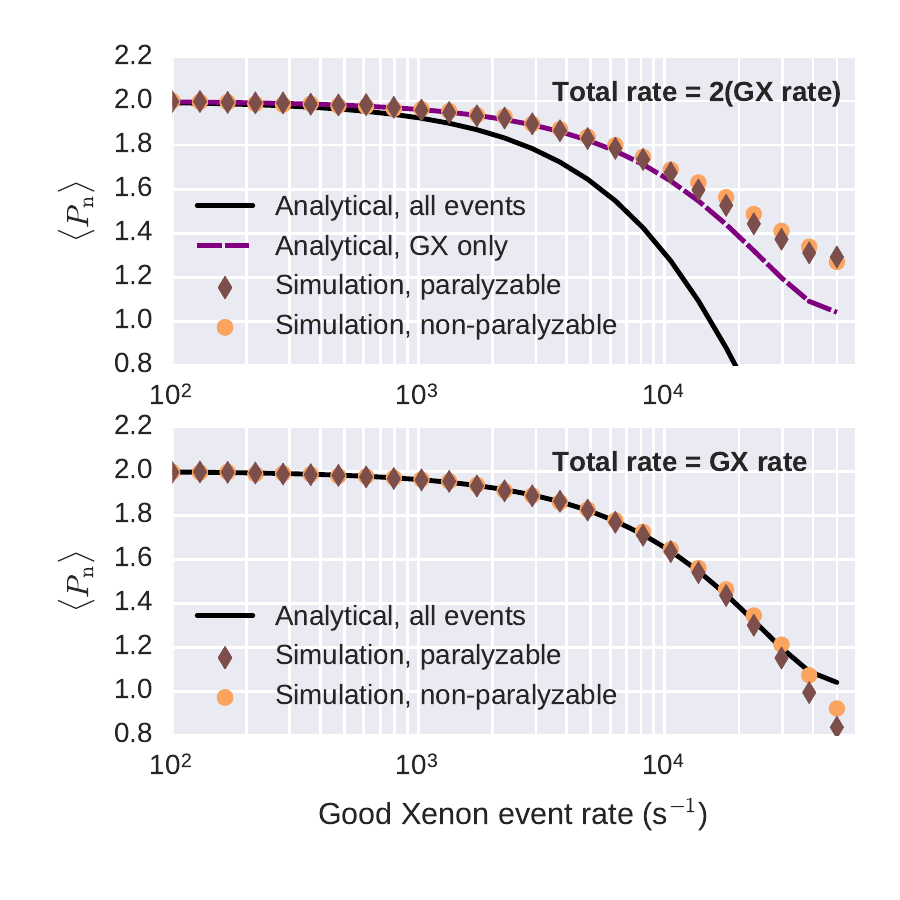}
 \caption{Simulation of Leahy-normalized $P_\mathrm{n}$ for the 1-s Poisson event sequence 
 with $t_d=10\,\mu$s, $t_b= 122$\,$\mu$s, and various event rates.  $P_\mathrm{n}$ is 
 averaged for 50 simulations and 4094 FFT harmonics (omitting two lowest harmonics). 
 Diamonds and circles show paralyzable and non-paralyzable dead time, respectively. The 
 analytic formula is Eq.~\ref{eq:dead_time} for the total event rate (solid line) and the 
 GX event rate (dashed line). \textit{Top:} only half of the simulated events were 
 considered to be GX events, the rest were deleted before computing the PS. \textit{Bottom}: 
 all simulated events were considered to be GX events.} 
 \label{fig:dead_comp}
\end{figure}

Attempting to apply Eq.~\ref{eq:dead_time} yielded interesting results: if the rate was taken 
to be the rate of all events (since all of them cause dead time), as stated in \citet{Jahoda2006}, 
the dead time correction was considerably and consistently larger than that required to match 
both data and simulations. However the correction did match observations well if only the GX 
events were taken into account. It appears that for the given range of event rates and the 
bin/dead time windows, the operations of dead time pruning and selection of GX events are 
commutative, meaning that GX photons have the same noise power distribution as if they were 
not affected by the dead time from all other event types. As simulations have shown, this does 
not hold true for larger count rates (Fig.~\ref{fig:dead_comp}), however, such large
count rates do not occur in our burst sample. 

To summarize, dead time was accounted for in the following way: initially LC curves for 
all four event types were re-normalized using the fraction of dead time calculated with 
the observed rates\footnote{The RXTE Cook Book states that "the VLE rate is not affected 
by dead time", this was not reproduced in the simulations.}. Then the TOAs of the four types 
of events were simulated separately and combined to form a ``mixed-bag'' event sequence 
which mimicked the real data. Then events which arrived within 10\,$\mu$s after previous 
non-VL event and variable $t_d$ after VL event were removed. Those events were assumed not 
to cause dead time themselves, so the dead time was by definition non-paralyzable. This 
procedure was performed for each PCU separately, and then the resulting photon lists were 
merged together.

\begin{figure*}
 \centering
 \includegraphics[scale=0.9]{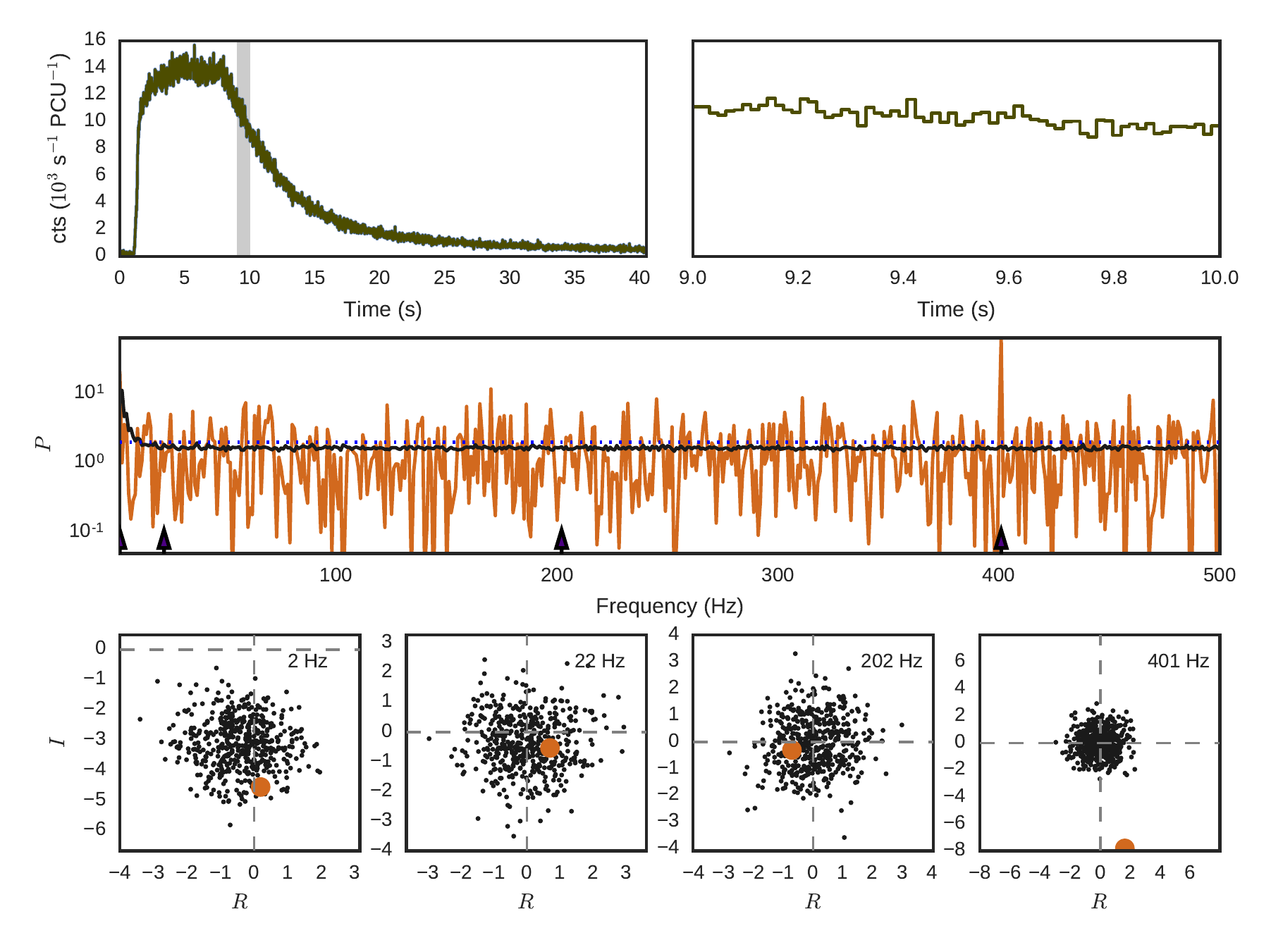}
 \caption{\textit{Upper row:} LC of burst \#3038 from SAX J1808.4$-$3658 detected on 2002-10-17 
 07:19:24 UT. On the left, grey shaded area marks the 1-s FFT window.  The zoomed-in LC in 
 the same window is shown on the right. \textit{Middle row:} PS from the real data (orange) 
 and the mean of  500 simulations (black). The arrows show four sample FFT frequencies (rightmost with TBOs)
 and  the blue dotted line shows $P=2$. \textit{Bottom row:} the distribution of Fourier coefficients 
 for the four sample FFT frequencies (black dots -- simulated data, orange circle -- real data).
 $I=0$ and $R=0$ are marked with dashed lines. }
 \label{fig:noisesim}
\end{figure*}

\subsection{Accounting for data gaps and occasional limited energy range}

Data gaps are losses of data due to saturated telemetry occurring for bursts with high count 
rate. They typically last from a fraction of a second to several seconds and are not reflected 
in the Good Time Interval table of the data files. To search for data gaps, we selected all 
time intervals with gaps between two successive photons being larger than 0.02\,s. In order 
to exclude ``natural'' gaps in bursts with intrinsically low count rate, we calculated the 
estimated number of photons inside the potential gap and selected only those gaps where 
this number was larger than two. The number of photons inside the gap was estimated from the 
mean photon count rate of Standard-1 data and adjusted for the overall difference in the 
energy band (multiplied by a sum of all Standard-1 counts divided by the sum of all 
high-$\tres$ counts for the time bins where the latter was larger than 0). All simulated 
photons inside the data gaps were deleted. 

Finally, the photon sequences were adjusted for the difference in energy band between 
high-$\tres$ data and the Standard-1 data (e.g. Fig.~\ref{fig:LCexample}). For each 0.125-s bin, 
the ratio between Standard-1 LC and the high-$\tres$ data LC was estimated and the simulated 
data were pruned by deleting the appropriate fraction of photons at random 
(Fig.~\ref{fig:photon_gen}, bottom).

Both real and simulated photon sequences were binned with sub-ms time bins. For observations 
in Good Xenon data mode, the bin size was $t_b=2^{-14}\,\mathrm{s}\approx 61\,\mu\mathrm{s}$, 
yielding about 8 phase bins in pulse profile at the highest frequency searched. For all other 
modes, the bin size was two times larger, corresponding to the typical $\tres=122\,\mu$s. 
Although 24 bursts had larger $\tres$, they were still binned with the smaller bin size. 
Among those bursts, five had $\tres=2^{-11}\approx 488\,\mu\mathrm{s}$, all of them 
from GRS~1747$-$312. For these bursts any candidates at frequencies above the Nyquist 
frequency of 1024\,Hz were discarded.

\subsection{Fourier transform}
\label{subsec:FFT}

Fourier transforms were taken for the binned sequences in series of 0.5, 1, 2 and 4-s sliding 
windows, each window starting 0.5\,s later than the previous one. The FFT coefficients $R$ and 
$I$ were recorded for harmonics between 2 and 2002\,Hz. The lower limit was set by the smallest 
non-zero harmonic for the PS in 0.5\,s windows. The upper limit reflects the largest possible 
NS spin frequency allowed by current reasonable models of the neutron star equation of state 
\citep{Haensel2009}.

In what follows, we will operate with FFT coefficients in a $(\nu, t, T_\mathrm{win})$ cell, 
with $\nu$ being the FFT frequency, $t$ referring to the center of the time window and 
$T_\mathrm{win}$ being the given window size. The 500 simulation runs were used to make 
distributions of $R_\mathrm{n}$ and $I _\mathrm{n}$ in each cell. The $I_\mathrm{n}$ and 
$R_\mathrm{n}$ had, most of the time, Gaussian distributions with the mean influenced by 
the baseline variation and the standard deviation influenced by the dead time 
(Fig.~\ref{fig:noisesim}). We used the mean and unbiased estimate of standard deviation 
of the first $N_\mathrm{smp}=400$ simulation runs to normalize $R_\mathrm{m}$ and 
$I_\mathrm{m}$ of the real data. Power spectra from the remaining 100 runs, normalized 
as the same way as the real data, were used to estimate detection significance.

The mean and standard deviation $R_\mathrm{n}$ and $I _\mathrm{n}$ used for re-normalizing 
are inevitably influenced by the limited number of simulation runs. For pure Poisson 
noise, the means are random variables with normal distribution, having $\mu=0$ and 
$\sigma^2=1/N_\mathrm{smp}=0.05$. Since $N_\mathrm{smp}\gg 1$, the standard deviation 
is also distributed normally, with $\mu=1$ and $\sigma^2=2/(N_\mathrm{smp}-1)$. Since the 
dead time influence has negligible dependence on Fourier frequency, for the normalization 
we averaged the standard deviation by $800\times T_\mathrm{win}$ harmonics in order to 
reduce the stochastic error caused by the limited number of simulations. Simulation of 
$10^9$ pairs of standard normal random variables showed that re-normalizing them by the
mean and standard deviation drawn from the appropriate Gaussian distributions causes 
about 10\% of detections above the threshold set by $p_{\chi^2}=2\times10^{-7}$ (adopted as the 
detection criterion, see Sect.~\ref{subsec:filt}) to be false positives. Another 10\% of 
un-normalized candidates had power below the threshold after normalization. It is hard 
to estimate the rate of false negatives or false positives for the real-data candidates, 
since it depends on the intrinsic distribution of TBO powers. However, we checked the 
normalization values for all candidates that were deemed interesting, for example 
occurring in an unusual place in the burst or being detected from a burst without 
previously reported TBOs. 

In rare cases of a gap occupying most of the FFT window, $I_\mathrm{n}$ and 
$R_\mathrm{n}$ become covariant at the lowest Fourier frequencies. In such cases the 
subsequent analysis is not applicable, so we discarded candidates from those cells. The 
covariance threshold was estimated as follows: we simulated 500 independent normal random 
variables and the distribution of covariance was calculated. The threshold was set as 5 
times the standard deviation of the covariances.

We also checked for the covariance along the frequency axis. Such covariance stems 
from abrupt changes in photon count within the window, caused by data gaps or even 
on the burst rise if the latter is sharp. For each time window and simulation we 
calculated autocorrelation function (ACF) from the renormalized simulated 
$P_\mathrm{n}(\nu)$. ACFs from all 500 simulations were added together and the 50\% 
half-width of the peak was measured, with the baseline levels subtracted from the 
peak. We discarded PS in the given time window (regardless of frequency) if the 50\% 
half-width of the ACF peak was larger than 5 harmonics. Such bursts were marked with ``freq cov''
comment in the notes column of Table~\ref{table:data}.

Finally, we removed the cells covering regions where the simulated LC deviated 
substantially from the real data due to narrow gaps or spikes. Substantial deviation 
was defined being larger than 10 standard deviations of the simulated photon count in 
given 0.125\,s or 15.625\,ms time bins. Such bursts were marked with ``bad LC''
comment in the notes column of Table~\ref{table:data}.

\subsection{Filtering potential oscillation candidates and computing fractional amplitudes}
\label{subsec:filt}

In order to filter potential TBO candidates, we selected all cells with renormalized
$P_\mathrm{m}>P_\mathrm{up}$, where $P_\mathrm{up}$ corresponded to a $\chi^2$
probability of getting $2\times10^{-4}$ chance candidates per single spectrum:
\begin{equation}
p_{\chi^2}(P_\mathrm{up})= \frac{2\times10^{-4}}{2000\times T_\mathrm{win}}.
\label{eq:pthr}
\end{equation}
The choice of $p_{\chi^2}(P_\mathrm{up})$ was arbitrary and motivated by the requirement 
to have a manageable number of candidates for the given data sample. For 0.5, 1, 2 and 
4-s FFT windows $P_\mathrm{up}$ was 30.85, 32.24, 33.62, and 35.01, respectively. 
$P_\mathrm{up}$ was adopted as the upper limit in the event that no candidate detections 
were found. 

Since the power values in adjacent cells are covariant both in time and (to a smaller 
extent) in frequency, the number of trials is not equal to the number of cells, 
$N_\mathrm{cell}$, and the simple significance formula 
$p_{\chi^2}(P_\mathrm{m})\times N_\mathrm{cell}$ is not readily applicable. 
To assess the significance of detections, we performed the same candidate search for 
the simulated data and compared the number of oscillation candidates 
from the real data with the distribution of the same values from the simulated data. 

For each potential candidate we computed fractional amplitude (FA) using Eq.~\ref{eq:framp}
and the median value of $P_\mathrm{s}|P_\mathrm{m}$ from Table~\ref{table:grothprob}, 
$P_\mathrm{s}=P_\mathrm{m}+1$. The uncertainties in fractional amplitudes were calculated 
by linear error propagation of the independent parameters in Eq.~\ref{eq:framp}
\citep{Ootes2017}. For the uncertainty on $P_\mathrm{s}$, we took [0.159, 0.841] percentiles 
of the $P_\mathrm{s}|P_\mathrm{m}$ distribution. The uncertainty on the number of photons 
in the FFT window was taken to be Poissonian and the uncertainty in the background level
was taken to be the standard deviation of count rates in the baseline window, computed 
in the overlapping windows of the length equal to the current FFT window.

A few potential uncertainties are not included in the given FA errors. Firstly, we do not 
include the variation of background within the burst from \citet{Worpel2015}, since it is 
not available for all bursts in our sample. We also do not correct for the possibility 
of the TBO frequency falling between FFT harmonics. Simulation showed that with our choice of 
FFT windows and oversampling in time, fractional amplitude can be underestimated by 
as much as a factor of 0.68. However only in 9\% of cases (assuming no prior knowledge of 
TBO frequency) suppression of FAs is stronger than 0.85.

Finally, we did not account for bias caused by the limited number of trials. Simulated 
distribution of $P_\mathrm{s}|P_\mathrm{m}$ for the normalized data had a mean and median 
consistent with the ones from Table~\ref{table:grothprob}. The standard deviation of 
$P_\mathrm{s}|P_\mathrm{m}$  for the normalized data was larger by a small value of 
$\lesssim 0.4\%$. 

\subsection{Organizing the results}
\label{subsec:org}

Table~\ref{table:data} gives a general picture of the maximum power recorded 
for each individual burst, as well as the smallest FA of potential detection
(defined by the threshold detection probability, Eq.~\ref{eq:pthr}). 
Besides renormalized power $P_\mathrm{m}$, it lists its frequency
and $T_\mathrm{win}$, as well as smallest $\FAup$ for all four $T_\mathrm{win}$.

Table~\ref{table:cand} contains basic properties for each source (number of bursts, total 
duration, median S/N of the burst peak, minimum and median upper limits on FAs in 1-s
window at the burst peak), providing an overview of the amount and quality of observational 
material for each source as well as the extent of FA range that can be probed by our analysis. 
The properties of oscillation candidates are given per frequency group, with a group being 
defined as the candidates with $|\delta \nu|\leq2$\,Hz (matching the lowest frequency 
resolution).  For each frequency group we list the number of bursts with TBO candidates 
in this frequency range and the number of bursts with candidates in one of three non-overlapping 
regions: ``R'', defined as the region between rise and peak of the burst; ``B'' starting 
at the peak and spanning three times half-peak width (more or less corresponding to the 
traditional on-burst windows); and ``T'' covering the rest of the burst tail. For each 
frequency group we list also the average $P_\mathrm{m}$ of the candidates in each of the 
four Fourier windows. The remaining three columns give a handle on the number of spurious 
candidates in both real and simulated data, listing the total number of real-data 
non-TBO candidates (counting the ones from overlapping windows as independent and omitting 
low-frequency ones), the average number of candidates in the simulated data 
(averaged over 100 simulation runs) and the percentile of the real-data number with respect 
to a 100-run simulation sample, $p$. 

Table~\ref{table:det} gives more detailed information about each group of candidates (except 
for the low-frequency ones) for each individual burst, listing MINBAR burst entry, MINBAR TOA, 
frequency range, location of candidates within the burst (R, B or T), number of independent 
time windows and the maximum FA for each size of Fourier window.

\begin{figure*}
 \centering
\includegraphics[scale=0.9]{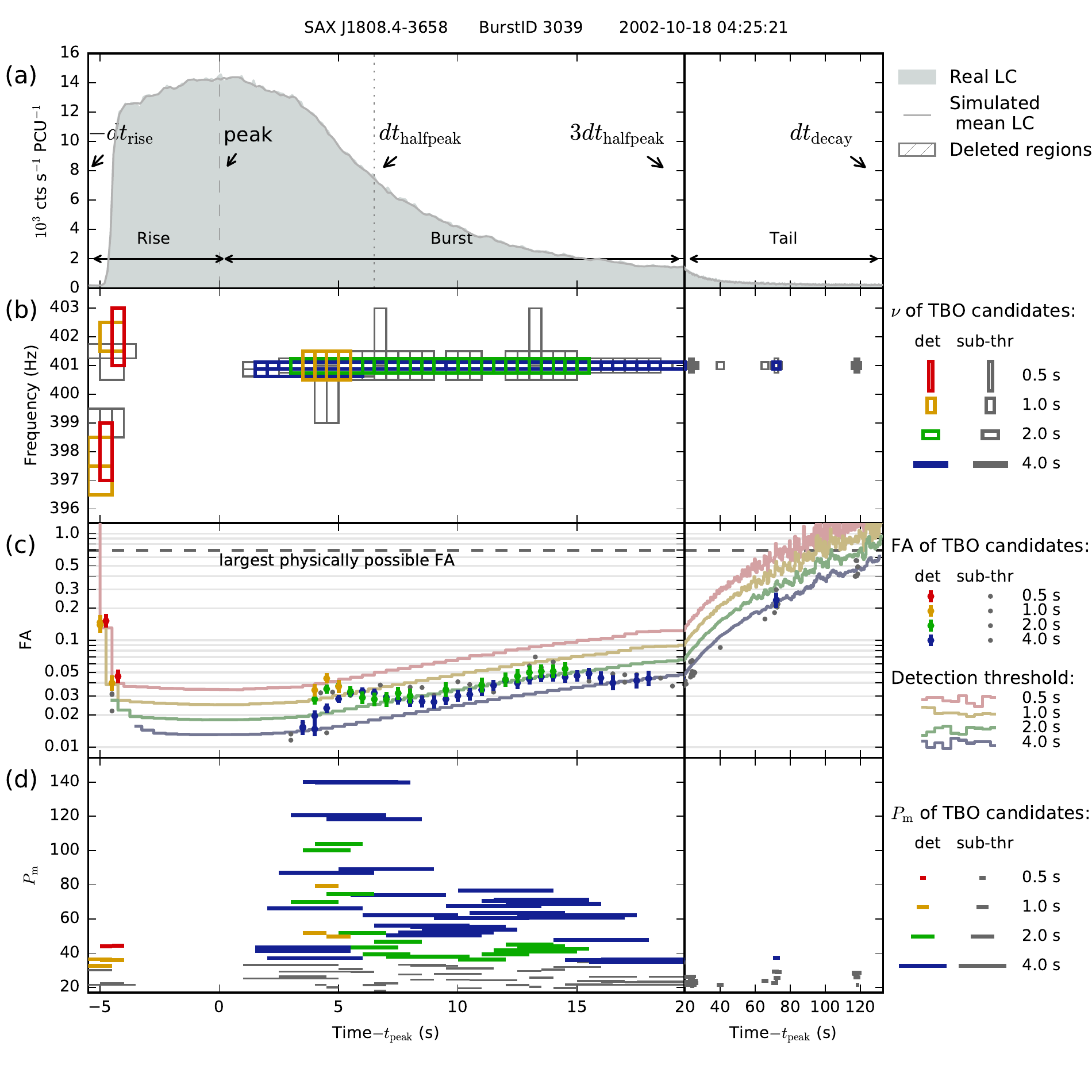}
\caption{An example of a diagnostic plot for a cluster of TBO candidates belonging to 
the same frequency group. For each plot pair, the left panel corresponds to RB (Rise+Burst) regions, 
the right to T (Tail, note the different time scale). 
\textit{(a)} LC, binned in 0.125-s time bins for the real (shaded) and mean of the simulated 
(line) data sets. Vertical lines mark $t_\mathrm{peak}$ and $dt_\mathrm{halfpeak}$. The regions excluded 
from the analysis (outside GTI, with large frequency covariance in the power spectra, or 
bad LC modeling) are shown as dashed (absent in this particular plot). 
\textit{(b)} Time and frequency of detected (det) TBO candidates, represented by boxcars with width 
equal to the width of the time window and the height to the Fourier spectral resolution. 
The color encodes the length of the window (red, yellow, green and blue for 0.5--4-s windows,
respectively). Sub-threshold (sub-thr) candidates, with $p_{\chi^2}<10^{-1} /(2000\times T_\mathrm{win})$ 
are plotted as grey. \textit{(c)}  FAs of TBO candidates vs time (color error bars, 
see Sect.~\ref{subsec:filt} for details on the uncertainty calculation) and sub-threshold 
candidates (grey dots) The FAs are given at the center of each sliding time window and are 
not corrected for the lack of signal during data gaps. Adopted upper limits on FA are set by the power 
corresponding to the threshold probability (lines). The dashed horizontal line marks FA 
of 0.7, the maximum rms FA which is allowed physically. \textit{(d)} Normalized 
$P_\mathrm{m}$ of candidates vs. time, with the length and the color of the mark representing 
the length of the FFT window.
}
 \label{fig:diag}
\end{figure*}

Finally,  for each group of candidates we provide reference plots, aggregating information 
about frequency, time of arrival, power and FA of candidates, as well as upper limits on FAs.
Fig.~\ref{fig:diag} gives an example of such a plot, with separate panels for the burst LC and
the frequency, FA, and power of the candidates. To conserve space, the rest of the reference plots are made
more condensed, without legends and with individual panels merged together.

\section{Results}
\label{sec:results}

 \begin{figure}
 \centering
 \includegraphics[scale=0.95]{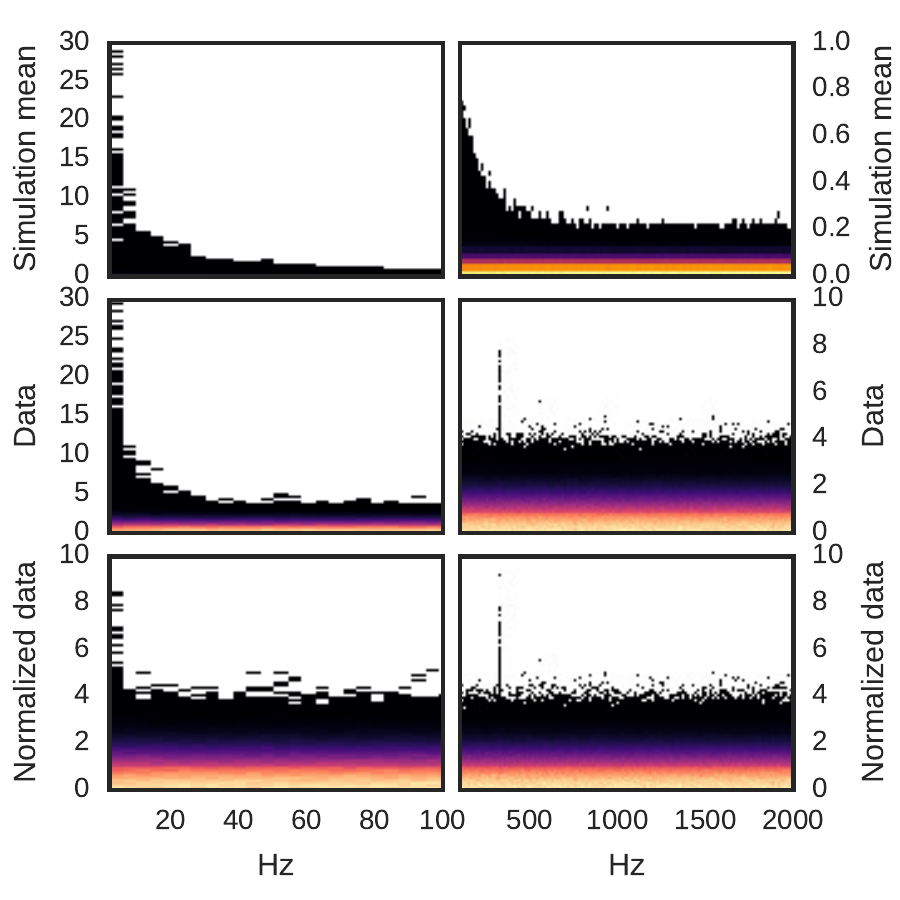}
\caption{An example of the frequency-dependent distribution of both Fourier coefficients 
$|I|$ and $|R|$. The distribution was computed in 1\,s windows using the data from 4U 
1702$-$429. For plotting purposes, the harmonics from 2 to 2000\,Hz were grouped in 4\,Hz 
bins. Color marks histogram values in a given 2D (frequency/coefficient value) bin. White 
corresponds to zero counts in a given bin, black to one or a few. The color becomes 
progressively lighter as the number of counts increases. The left and right panels 
highlight different parts of the distribution, below and above 100\,Hz, respectively. For 
both panels, the top subplot shows the average coefficients from the simulated photon 
sequences. Middle subplot shows the data and the lower subplot shows the normalized data, 
with the mean value of the simulated coefficients subtracted. Simulations removed most of 
the low-frequency noise, but some of it is still present at the very low frequencies.}
 \label{fig:InRn_vs_F}
\end{figure}

 \begin{figure*}
 \centering
\includegraphics[scale=0.71]{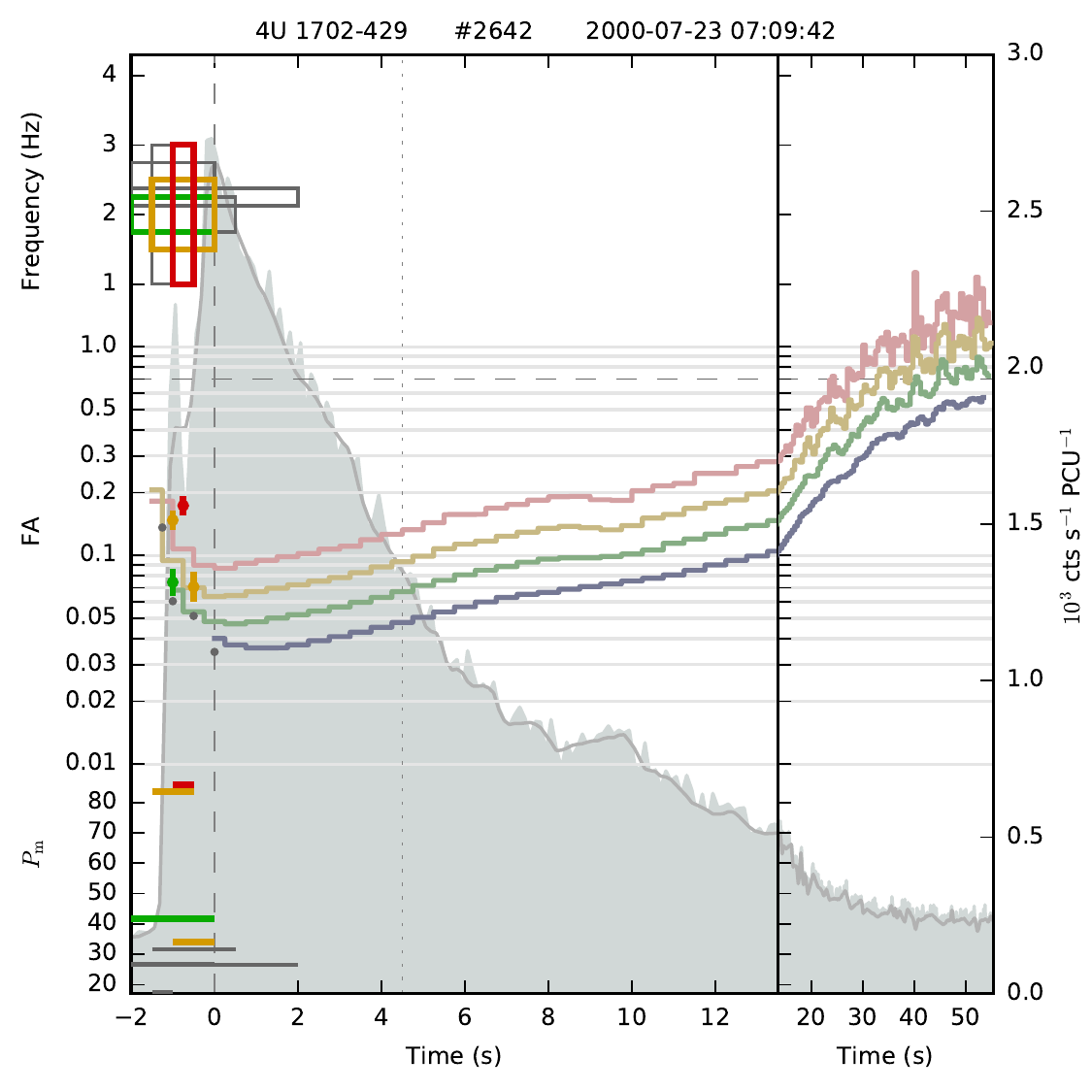}\includegraphics[scale=0.71]{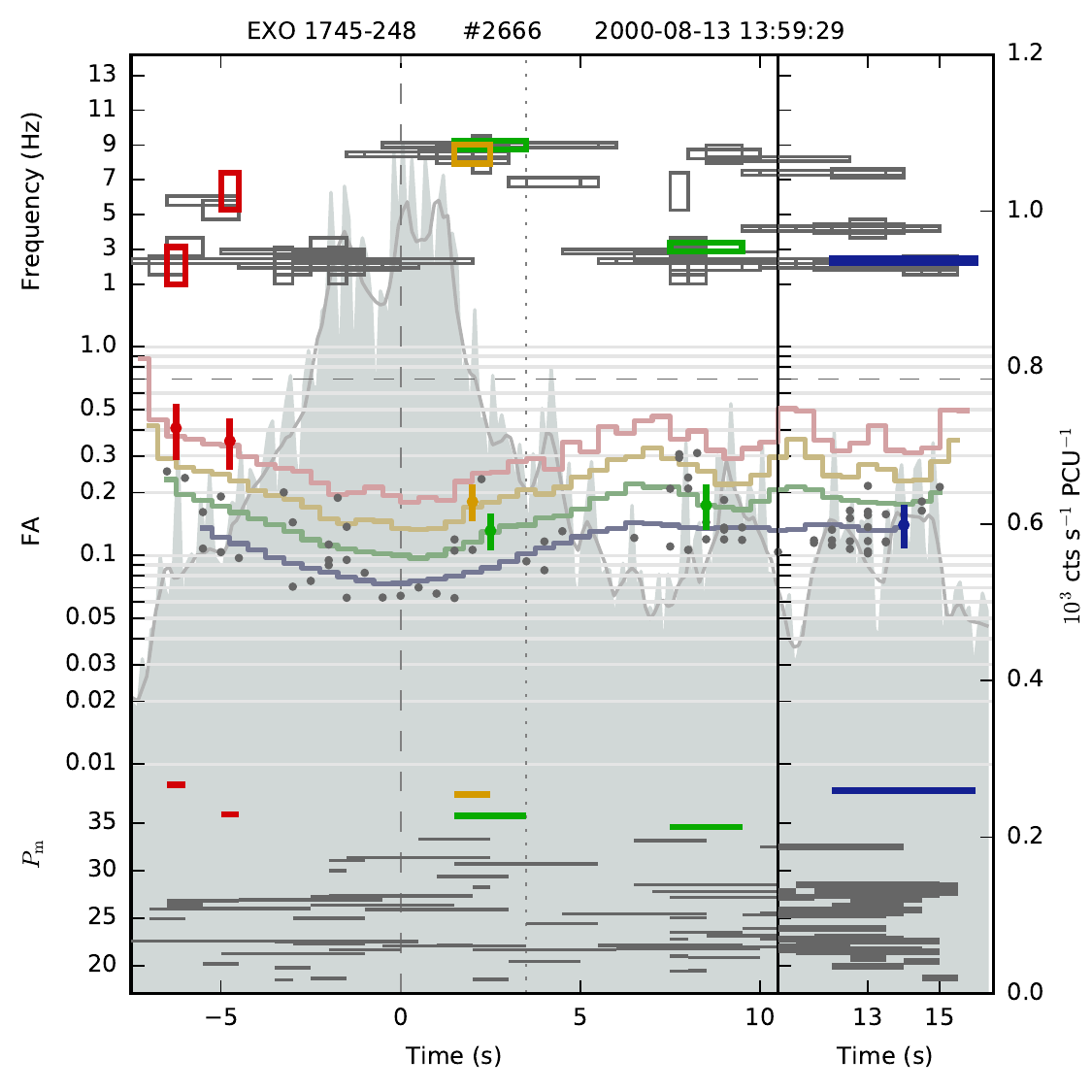}
\caption{\textit{Left:} a group of low-frequency oscillation candidates appearing due to 
imperfect modeling of burst LC, which did not reproduce a short spike around $t=-1$\,s 
(type I low-frequency candidates). \textit{Right:} clusters of low-frequency candidates 
not immediately connected to flaws in LC modeling (type II).}
 \label{fig:rednoise}
\end{figure*}

\subsection{Low-frequency noise}
\label{subsec:RnIn}

The mean values of simulated Fourier coefficients, $\langle R_\mathrm{n}\rangle$ 
and $\langle I_\mathrm{n}\rangle$, give us a handle on how much the power spectrum 
is affected by the change of photon count rate during Fourier window. 
Fig.~\ref{fig:InRn_vs_F} shows an example of the frequency-dependent distribution 
of the absolute values of Fourier coefficients for all 
$(\nu,\,t,\, T_\mathrm{win}=1\,\mathrm{s})$ cells in the bursts from 4U 1702$-$429,
omitting the cells with large frequency covariance or large discrepancy between 
the simulated and real LCs (see Sect.~\ref{subsec:FFT}). At higher Fourier frequencies 
the spread of   $\langle R_\mathrm{n}\rangle$ and $\langle I_\mathrm{n}\rangle$ 
is mostly determined by the finite number of simulation runs, whereas at the lower 
frequencies we record an excess of large coefficient values. For the majority of 
the sources this excess continues to at least 100\,Hz. In rare cases the frequencies 
as high as 1000\,Hz can be still affected.

Re-normalization of Fourier coefficients allowed us to remove most of the described
power excess. However, we still record a relatively large number of strong candidates 
at frequencies  below $\sim 15$\,Hz. Some of these candidates can be associated with 
obvious flaws in LC modeling, where our spline failed to reproduce short peaks or 
drops in the LC (Fig.~\ref{fig:rednoise}, left, hereafter ``type I'' low-frequency candidates). Other 
low-frequency candidates cannot be immediately connected to the imperfect LC modeling 
(hereafter ``type II'' low-frequency candidates). Several sources (e.g. Cyg X-2, EXO 0748$-$76, EXO 1745$-$248) 
exhibited such unexplained bursts of low-frequency candidates spanning multiple harmonics
and sometimes showing at distinctly separate frequencies  (Fig.~\ref{fig:rednoise}, 
right). The origin of this type II low frequency noise remains unclear: it could well 
be astrophysical, associated with either the bursting surface or the accretion disk 
\citep[see for example][]{vanderKlis2006}.

 \begin{figure}
 \centering
 \includegraphics[scale=0.605]{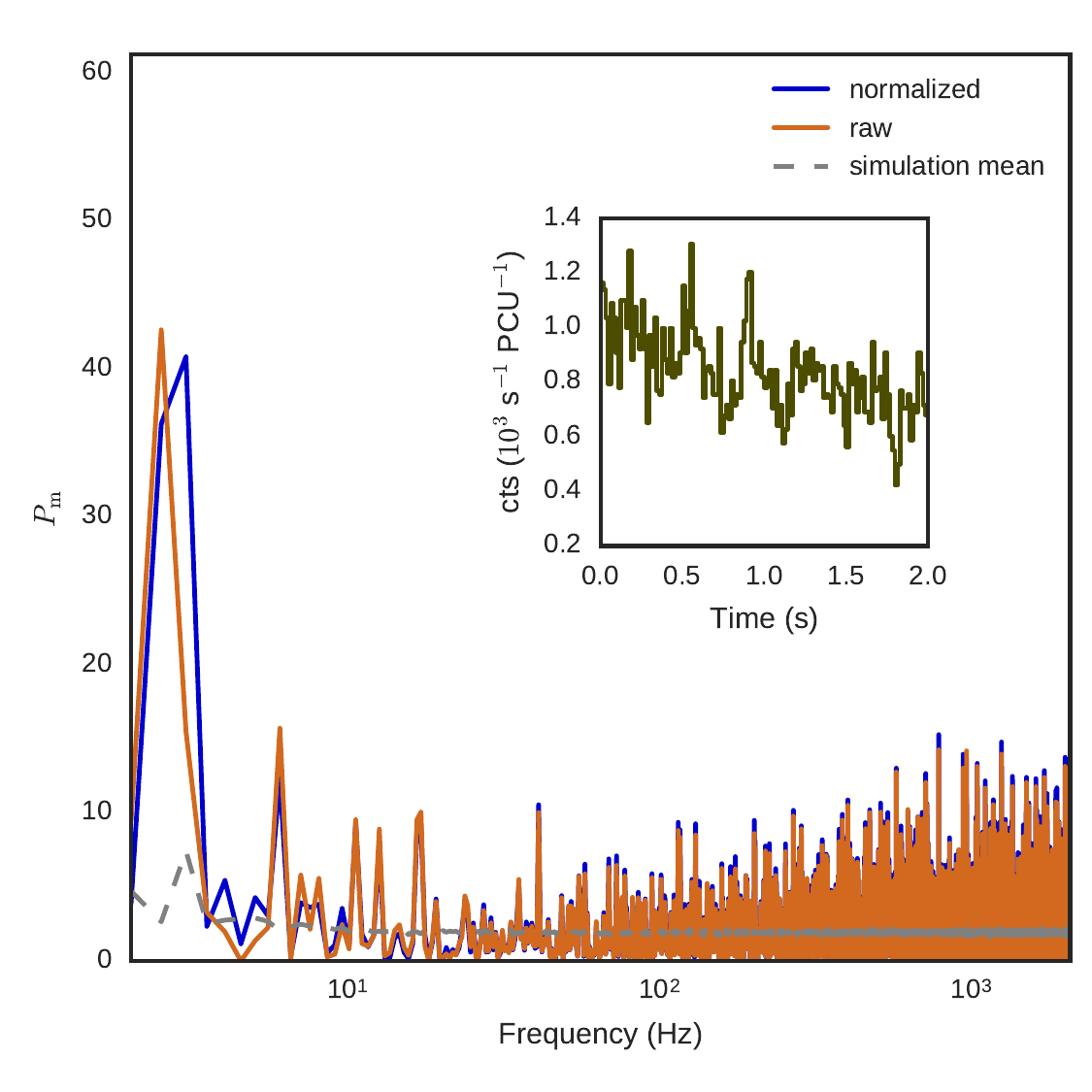}
\caption{An example of 1D power spectrum in 2-s window right after the peak of a faint 
burst (\#3306) from Cyg X-2. Blue/brown line shows normalized and un-normalized spectra, 
respectively, with the mean of the simulated spectra shown with a dashed line. The inset 
features the LC in the same window, with $\sim3$-Hz oscillations readily visible.  }
 \label{fig:CygX2_rednoise}
\end{figure}

\begin{figure*}
 \centering
 \includegraphics[scale=0.9]{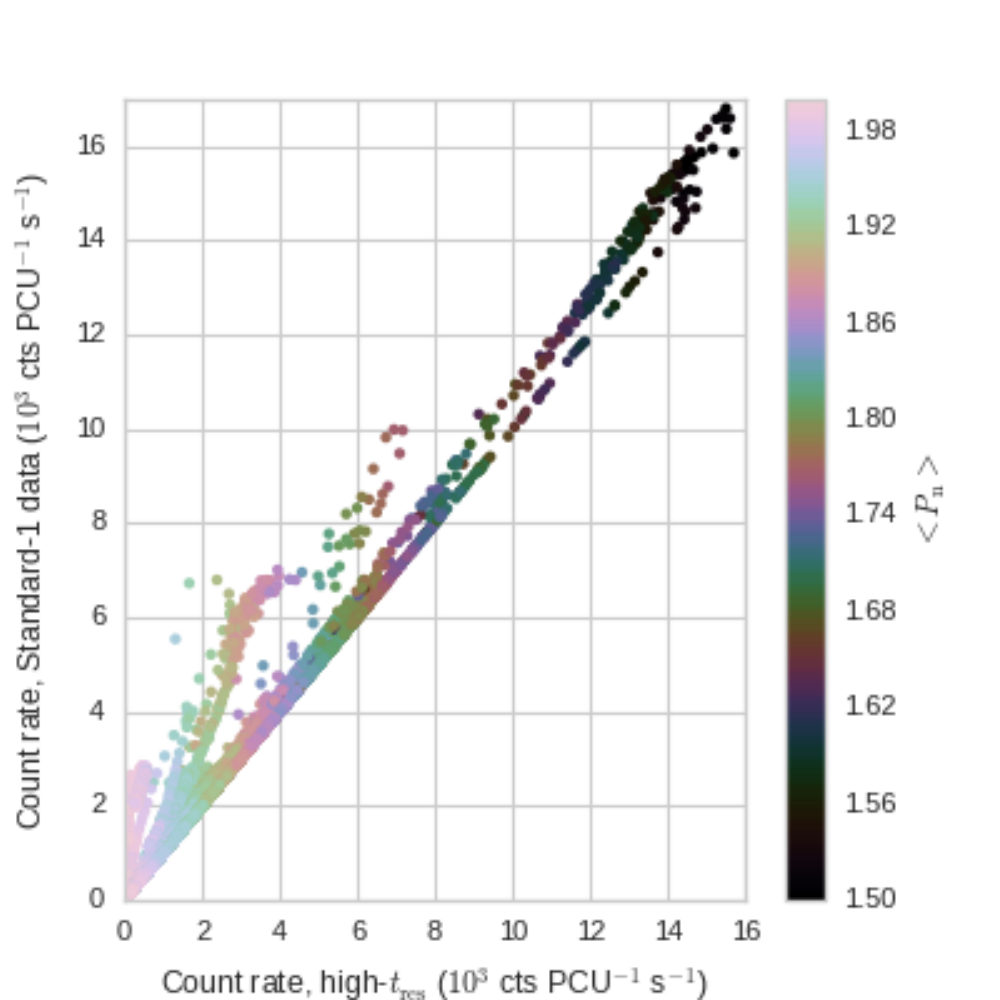}\includegraphics[scale=1.0]{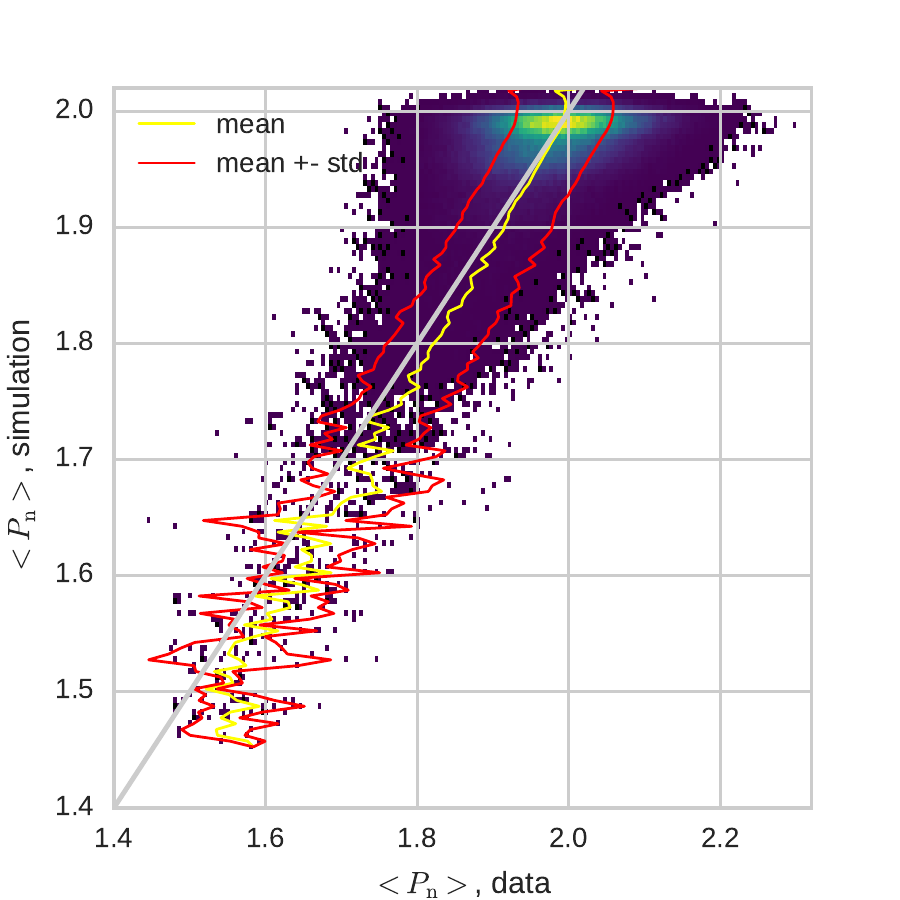}
   \caption{\textit{Left:} simulated noise power vs. count rate in Standard-1 and high-$t_\mathrm{res}$
   files for 1-s Fourier windows for all sources. \textit{Right:} 2D histogram of the mean simulated vs.
   real data noise power averaged over frequencies above $1$\,Hz. Yellow/red lines show mean and 
   $\pm$std of real data noise power for a given value of simulated power. The color represents the 
   number of counts in each cell of the histogram, with the darkest one being the smallest (1 count). 
   On both plots the simulated power is averaged over 100 simulation runs and all harmonics above 1\,kHz. }
 \label{fig:Pn_countrate}
\end{figure*}

Although mostly recorded from fast-spinning neutron stars, TBOs can occur at frequencies as low 
as 11\,Hz \citep[IGR J17480$-$2446,][]{Cavecchi2011}. So far, only one such slow TBO source is 
known and finding another one (or the one spinning at even lower frequency) would be very 
interesting. In general, however, we found that power spectra at the lowest frequencies of 
few Hz are difficult to interpret, since it is hard to distinguish between LC variation and 
oscillations here. An example of such a problematic spectrum  is shown in Fig.~\ref{fig:CygX2_rednoise}. 
Oscillations with a frequency of about 3\,Hz are clearly visible in the lightcurve. 
With the given choice of smoothing and spline fitting parameters, LC modeling removes some 
of the count rate variation and changes the shape of the peak in the power spectrum.
More stringent LC models can reproduce the observed LC variations, removing the peak completely

We have inspected visually all diagnostic plots featuring type II low-frequency candidates looking for  
signals resembling the 11 Hz TBOs from IGR J17480$-$2446: with detections in multiple independent 
time windows and multiple bursts, at frequencies larger than the lowest recorded frequency 
of 2\,Hz and without candidates of comparable strength at the nearby, but distinctly separate 
frequencies. No such candidates were found.

\subsection{Dead time}

\begin{figure*}
 \centering
  \includegraphics[scale=1.0]{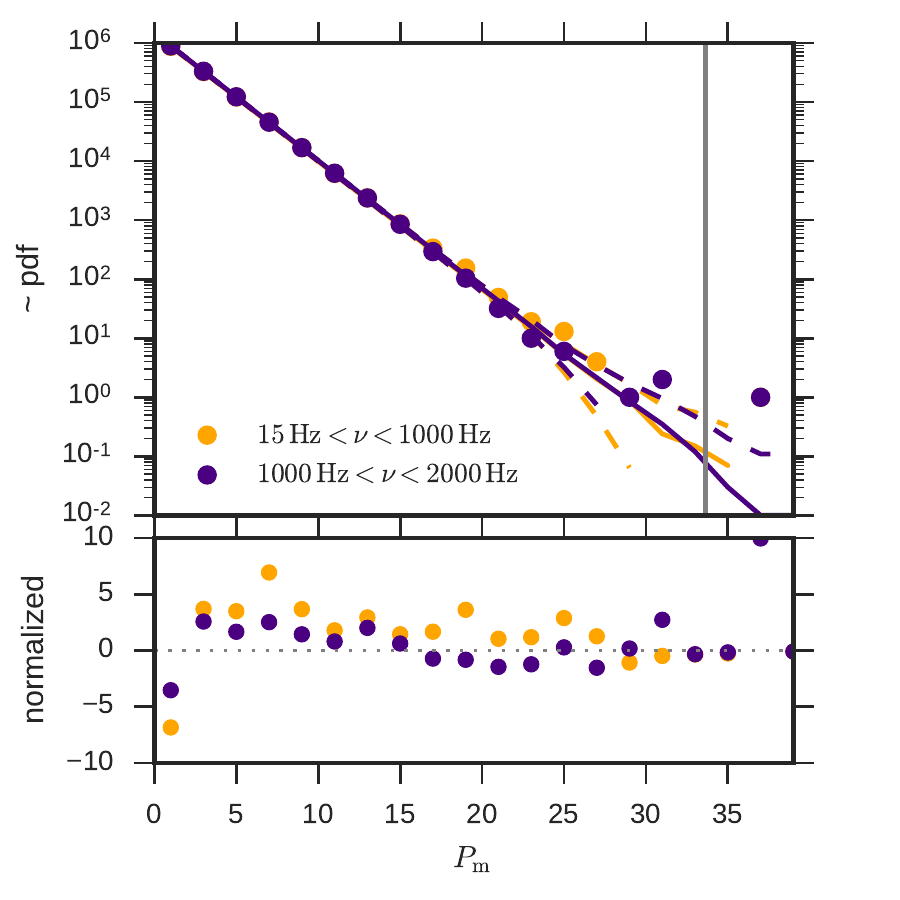}\includegraphics[scale=1.0]{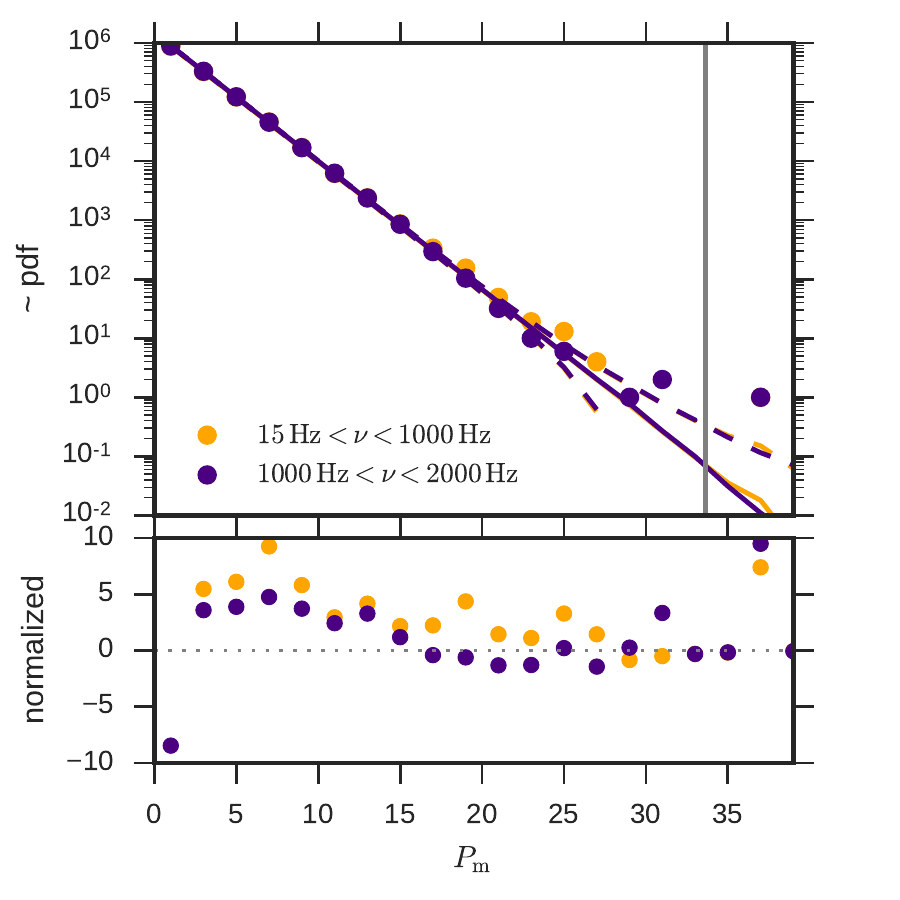}
\caption{Histogram of power values for 2-s time windows for all bursts from Cyg X-2 in two 
frequency groups (dots) compared to the mean (solid line)  of 100 simulations (left panels) 
and 4000 $\chi^2$-distributed random values (right panels, for conservative number of trials).  
Dashed lines show one standard deviation offset from the simulation mean. Vertical grey lines 
show the candidate selection threshold. Bottom panels show  histogram counts for the real data, 
normalized by the mean and standard deviation of the corresponding simulated or $\chi^2$ 
distributions.}
 \label{fig:pdf}
\end{figure*}

In Section~\ref{subsec:dead_time} we reviewed the methods of estimating the influence of the 
instrument's dead time on the observed noise statistic $P_\mathrm{n}$. In this section we will 
examine the $P_\mathrm{n}$ from the whole set of observations in our sample.

In order to investigate the dead time influence, we recorded the mean non-normalized simulated 
power at frequencies above 1\,kHz. At these frequencies the bias caused by LC variation is 
small for all of our sources (Sect.~\ref{subsec:RnIn}). We found that the simulated noise power 
did not have any discernible dependence on the number of PCUs that were on, and varied between 
1.5 and 2, depending on the total photon count rate recorded by the PCUs (obtained from 
Standard-1 data files). This count rate is always equal  to or larger than the count rate 
derived from the high-$t_{\mathrm{res}}$ data. 
 
Fig.~\ref{fig:Pn_countrate} (left) provides the reference for average simulated noise power 
versus count rate per PCU in high-$t_{\mathrm{res}}$  and Standard-1 files. For count rates 
larger than $8\times10^3$\,cts\,s$^{-1}$\,PCU$^{-1}$, $P_\mathrm{n}$ is smaller than about 1.7, differing 
dramatically from the value of 2 prescribed by the ideal $\chi^2$ noise model. Thus, neglecting 
dead time influence for bright bursts can lead to an underestimation of the potential signal 
significance by orders of magnitude. The average power is considerably smaller than 1.7 at the 
peaks of at least one burst from 4U 0614$+$09, 4U 1608$-$552, Aql X-1, HETE J1900.1$-$2455, and 
SAX J1808.4$-$3658.

Comparison of noise powers between real and simulated data is not straightforward because of 
the large intrinsic noise variance: for $P_\mathrm{n}$ obeying $\chi^2$ distribution with 
two degrees of freedom, the standard deviation of noise powers is 2, the same as the mean 
value. Averaging all harmonics above 1\,kHz reduces the standard deviation to $\approx 0.045$ 
for $T_\mathrm{win} = 1$\,s and allows us to pinpoint the influence of dead time. For simulated 
data, additional averaging by 100 simulation runs further reduces the standard deviation by a 
factor of 10. Fig.~\ref{fig:Pn_countrate} (right) shows the 2D distribution of such average 
noise powers in real and simulated data. For most values of simulated power, the noise powers 
for the real data appeared to be statistically slightly larger than the corresponding 
simulated power, most probably stemming from the simplifications we made during dead time pruning.
The discrepancy can reach as much as 0.15 for 
$\langle P_\mathrm{n}\rangle\lesssim 1.6$, but is not larger than 0.02 for the more common 
$\langle P_\mathrm{n}\rangle\gtrsim1.8$. This leads to an overestimation of the real data 
candidate significance by a factor that can be as large as few (for the largest count rates), 
but more commonly of about a few percent. 

It is worth mentioning that dead time also biases the measured fractional amplitudes of TBO 
candidates, since the fraction of dead time is different during the crests and troughs of 
the oscillation trains. Using non-normalized $P_\mathrm{m}$ in Eq.~\ref{eq:framp} may bias 
FAs by as much as a factor of $(1.5/2)^{0.5} \approx 0.87$.

\subsection{Overall simulation quality and glimmer candidates}

In order to assess whether our simulations are adequately reproducing the data, we compared 
the distributions of normalized powers $P_\mathrm{m}$ for the real and simulated data sets. 
For each source and $ T_\mathrm{win}$ we combined powers in two frequency regions: between 
15 and 1000\,Hz (thus, excluding any low-frequency noise), and between 1000 and 2000\,Hz 
(see Fig.~\ref{fig:pdf}, left, for an example). The distributions for real data and the mean 
distribution of 100 simulation runs match reasonably well. Moreover the distribution of 
normalized $P_\mathrm{m}$ is well described by $\chi^2$ statistics, assuming a conservative 
number of trials (i.e. treating all time windows as independent, Fig.~\ref{fig:pdf}, right). 
The same is true for the candidates from all four $ T_\mathrm{win}$ combined -- the 
estimates of the average number of candidates using Eq.~\ref{eq:chi2} and assuming all 
windows and harmonics to be independent are not dramatically different from the average 
number of candidates from the simulation runs (see also Table~\ref{table:cand}).

However, for some of the sources the match is not perfect. After normalizing the real-data distribution by the 
corresponding mean and standard deviation of 100 simulated-data distributions, one can 
see that there is a systematic discrepancy between the two for $P_\mathrm{m}\lesssim 20$.
The amount of discrepancy is usually larger for frequencies below 
1000\,Hz and varies considerably from source to source. It may stem from imperfect 
dead time or LC modeling, or any weak broadband astrophysical signal.

\begin{figure}
 \centering
  \includegraphics[scale=0.75]{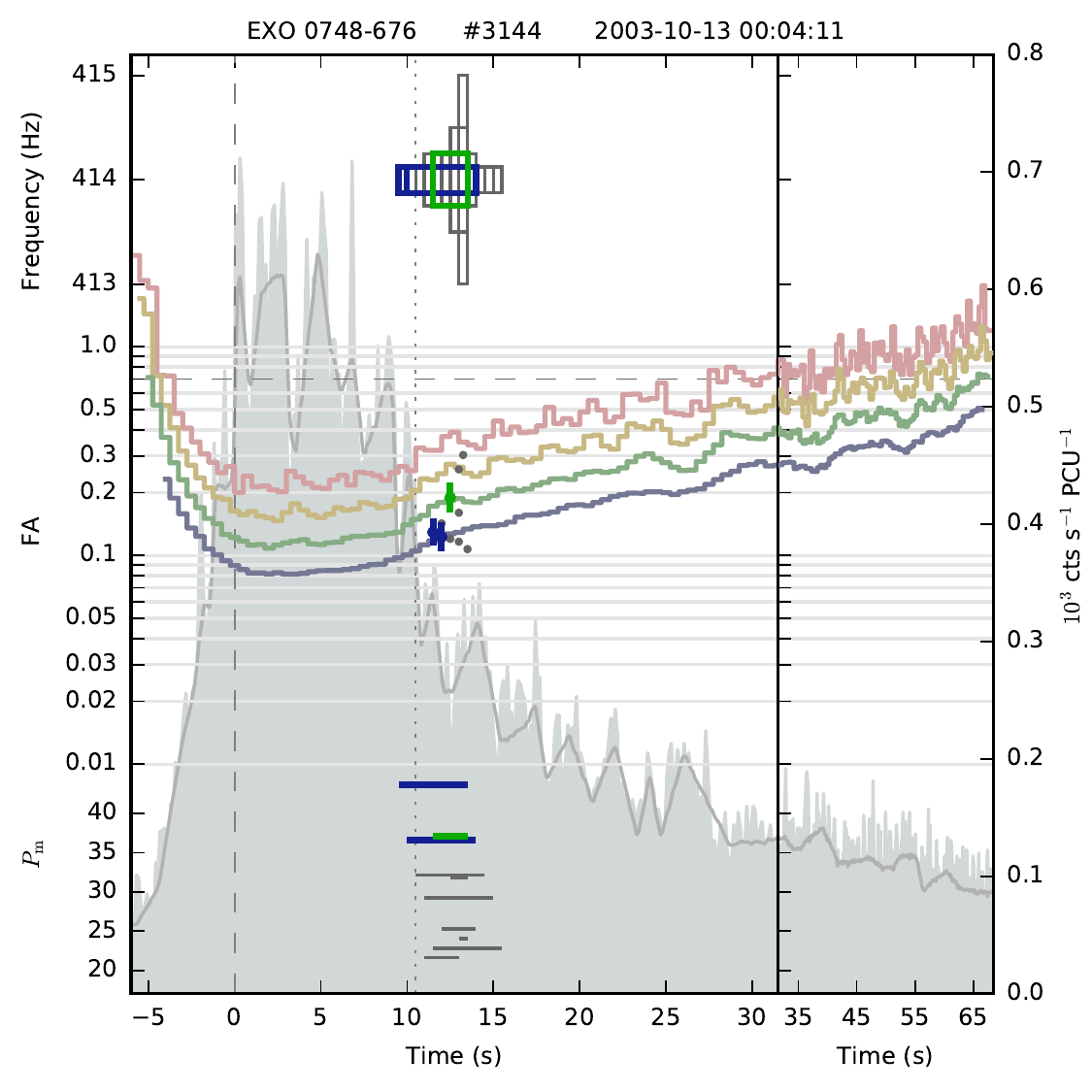}
    \caption{An example of a glimmer candidate, one of the marginally significant noise candidates 
    from EXO 0748$-$676. This source has TBOs
    at $\approx553$\,Hz (see Sect.~\ref{subsec:EXO0748}).}
 \label{fig:glimmer}
\end{figure}

For some of the sources, we found a small excess of higher-power candidates (e.g. with 
$P_\mathrm{m}>35$ on Fig.~\ref{fig:pdf}). This excess can be present in either of the 
two frequency groups and is equivalent to 5--10 standard deviations in a given power bin.
The examination of the cumulative versions of the normalized power distributions
for all $T_\mathrm{win}$ combined showed that 
for several sources, e.g. 4U 1608$-$522, EXO 0748$-$676, Cyg X-2, and others,
the number of candidates above the detection threshold on the real data (excluding 
frequency ranges of known TBO) is larger than the corresponding number in $\geq 99$\% 
of the simulation runs ($p\leq0.01$, see Table~\ref{table:cand}). On the other hand, the prolific TBO 
source 4U 1636$-$536 yielded fewer real-data noise candidates than any of 100 simulations. 
The origin of this discrepancy is unclear.

Such marginally significant noise candidates (dubbed ``glimmer candidates'', reflecting 
potential attractiveness) are detected in a single independent time window, at seemingly 
random, never repeating frequencies\footnote{Except for two 1108-Hz candidates from EXO 0748$-$676, see
Table~\ref{table:cand}.} and throughout all on-burst windows. 
Some of the 
candidates occur at lower frequencies in time windows with substantial count rate variation 
and their significance is very sensitive to LC modeling. Glimmer candidates can be present 
in the bursts with TBOs, sometimes even in the same time bins as TBOs. Folded glimmer
candidates produce sinusoidal profiles and some of them are stronger than weak TBOs 
(e.g. Fig.~\ref{fig:glimmer}). 

It must be noted that whether the source has glimmer candidates
depends on the detection threshold. For example SAX J1808.4$-$3658 has $p=0.22$ 
at standard detection threshold, however the power of some of the noise candidates is
much larger than any of the simulated powers. On the other hand, for MXB 1658$-$298 $p$ drops from 0.06
to 0.01 if the threshold probability is multiplied by 3.7 (Sect.~\ref{subsec:MXB1658}). 

One must be very careful in interpreting glimmer candidates. 
By definition, the source has glimmer candidates if the number of detections 
outside TBO frequencies and not immediately connected to low-frequency noise
is larger than the number of candidates in 99\% of simulation runs. 
This means that assuming a sufficiently large
number of bursts per source, 1\% of all sources will have glimmer candidates purely due 
to chance. At our detection threshold, five sources, or 8.8\% had $p\leq0.01$, 
which is larger than the expected 1\%,
suggesting that at least some of the glimmer candidates 
may have an astrophysical origin. Some may for example be connected to type II low-frequency 
candidates which happen to occur at somewhat higher frequency and  more than 2 Hz apart from 
other candidates, thus being placed in a separate frequency group by our grouping procedure.

\subsection{TBOs from known oscillation sources}

\begin{figure}
 \centering
 \includegraphics[scale=0.72]{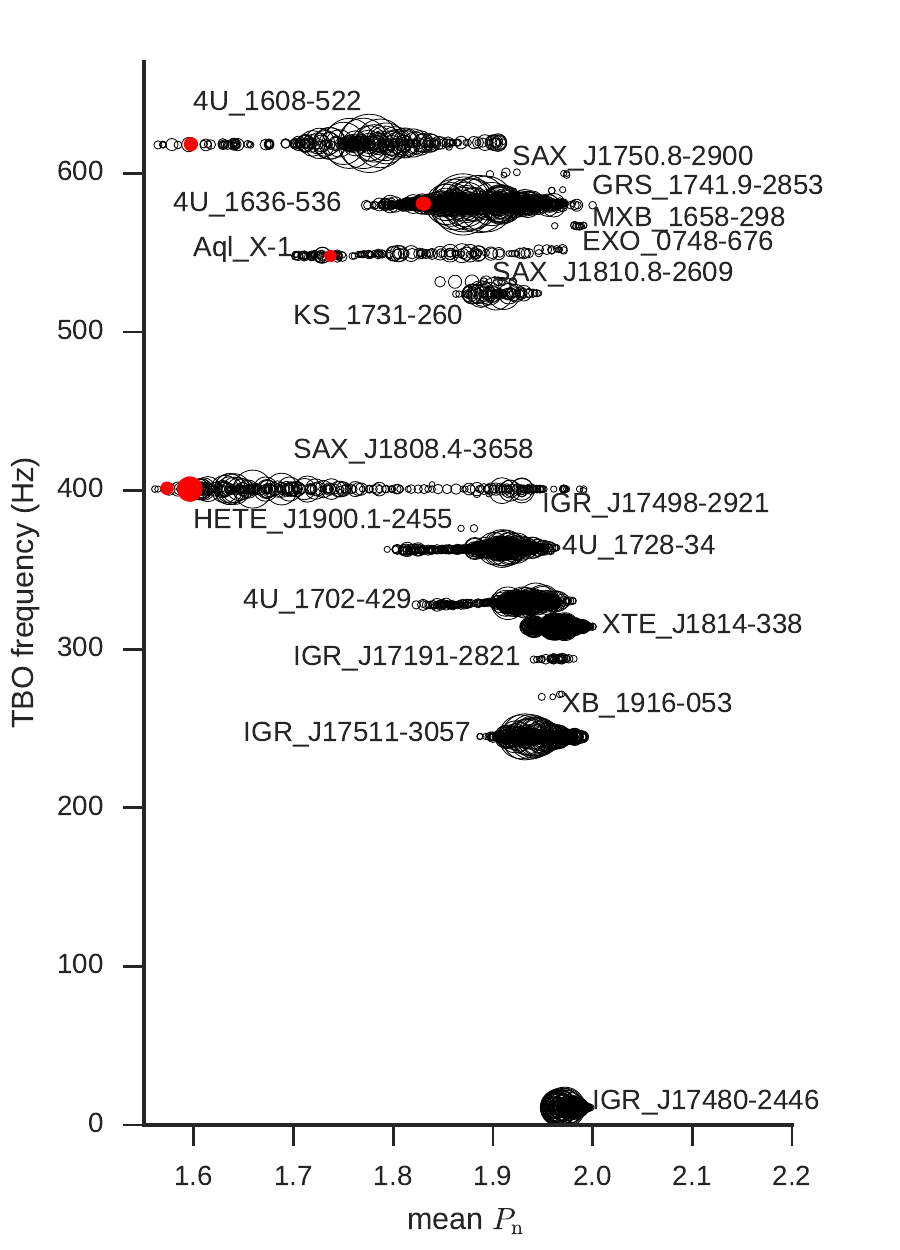}
   \caption{Frequency against mean simulated noise power for candidates in the groups covering 
   the frequencies of known TBO sources. The size of the marker is set by $P_\mathrm{m}$ of a 
   candidate.  Red markers show candidates, for which at least one of the absolute values
   of the mean simulated Fourier coefficients was likely to be influenced by the
   variation of count rate within the Fourier window. }
 \label{fig:Pn_TBO}
\end{figure}

Seventeen TBO sources were known prior to our analysis. These are the sources with TBOs 
detected at similar frequencies, in several independent time bins, several bursts or at 
frequencies close to the frequency of APPs. All of these sources yielded candidates at 
the frequencies close to those reported previously and all but one had more candidates
than any of the simulation runs ($p=0$) in a purely blind search. We note that the TBO candidates from 
the accreting MSP HETE J1900.1$-$2455 would not have been significant in our blind search 
which neglected closeness to the known APP frequency for this source, since the TBOs come 
from one independent time window and have moderate power. For the other sources, sometimes 
we did not have any candidates (including sub-threshold, see Sect.~\ref{subsec:subthr} 
for our definition of sub-threshold candidates) where they have been reported previously. 
This may be explained by differences in data processing. 

In general, we find that the measured power of TBO candidates depends strongly on the 
size of the Fourier window and its location. Because we searched in several windows 
of different length, we were able to give a more complete picture of the fractional 
amplitudes, which may be important for bursts with more than one oscillation train, 
i.e. bursts with photospheric radius expansion (PRE), with a short train of TBOs in 
the rise and a longer train after the burst peak). We have compiled an extensive dataset 
of fractional amplitudes (Table~\ref{table:det}) as well as upper limits for each of 
the four window lengths (Table~\ref{table:data}), to support future studies of TBO physics 
(see Sect. \ref{sec:concl}).

For many of the TBO sources a considerable fraction of TBO candidates came from the data 
regions with large influence from dead time (Fig.~\ref{fig:Pn_TBO}). Low-frequency noise 
does not have much of an influence -- only in a few cases the absolute mean values of 
the simulated Fourier coefficients were larger than $6\times 0.05$ (see Sect.~\ref{subsec:FFT}, 
\ref{subsec:RnIn} and Fig.~\ref{fig:InRn_vs_F}).  More information about each TBO source 
is given in Section~\ref{subsec:knownTBO}.

\subsection{Tentative TBO detections from the literature}

Eight sources in our sample had tentative TBO detections prior to our analysis
\citep[see Table 2 in][]{Watts2012}. Here we summarize the results of our analysis of 
these sources, more detailed information about each of them is given in Section~\ref{subsec:tentTBO}.

No TBO candidates were detected in the only burst from 4U 0614$+$09, observed by \textit{RXTE}.
Previously, the 415-Hz oscillations were reported from one burst observed with \textit{Swift}. 
The FA limits from the \textit{RXTE} burst are more stringent than the detection reported from 
the \textit{Swift}, but as we know from other sources, TBOs are not consistently detectable 
in all bursts from a given source.  

The previously reported 529-Hz TBO candidate from 1A 1744$-$361 was also detected in 
our analysis, however it was too faint to be significant given the number of trials. 
The presence of sub-threshold candidates tracing out a small frequency drift argues in 
favour of this candidate being a genuine TBO, but a new detection is needed to confirm this. 

TBO candidates from 4U 1254$-$69 (95\,Hz), XTE J1739$-$285 (1122\,Hz), 
and SAX J1748.9$-$2021 (410\,Hz), discovered in time windows with sizes similar to the range of 
$T_\mathrm{win}$ used in this work were in our analysis not significant enough to pass 
the detection threshold. The sources did not have any clusters of sub-threshold 
candidates close to the reported frequencies. TBO from another two sources, MXB 1730$-$335 
(306\,Hz) and GS 1826$-$24 (611\,Hz) were claimed based on stacked power spectra. None of our 
candidates for these sources were close to the frequencies reported in the previous papers. 

XB 1916$-$053, with its pair of TBO candidates at frequencies 2\,Hz apart remained 
controversial in our analysis.  In addition to these two candidates, we detect four other 
strong candidates, all of them potentially due to type II low-frequency noise.  None of the 
simulated data sets had as many candidates as the real data, whether or not one counted the 
270-Hz candidates as a TBO.  A more precise estimate of the significance of this signal should 
take into account frequency separation between the candidates; however this is outside the 
scope of this current work.

\begin{figure*}
 \centering
\includegraphics[scale=0.95]{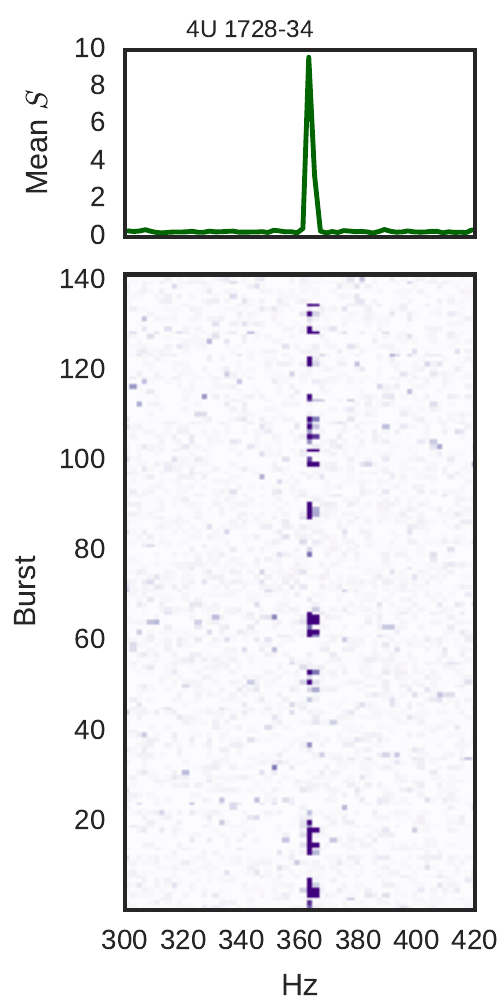}\includegraphics[scale=0.95]{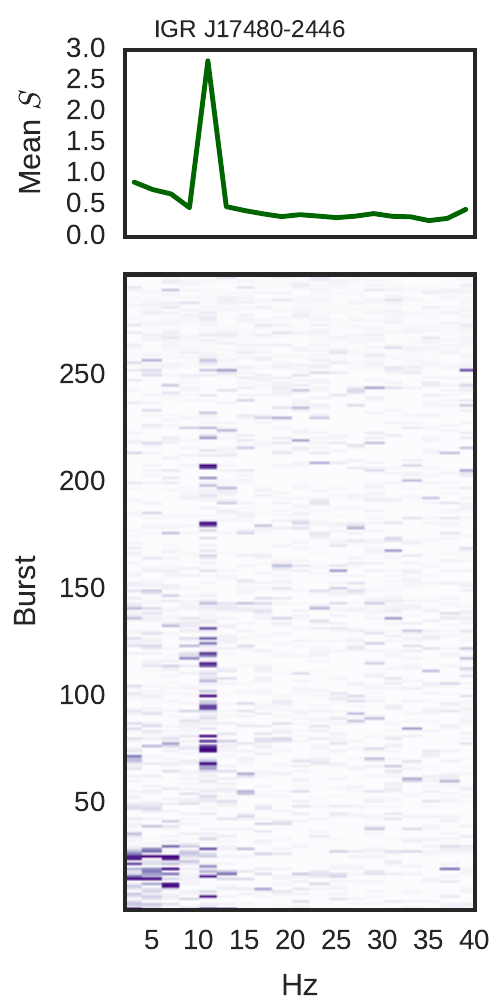}\includegraphics[scale=0.95]{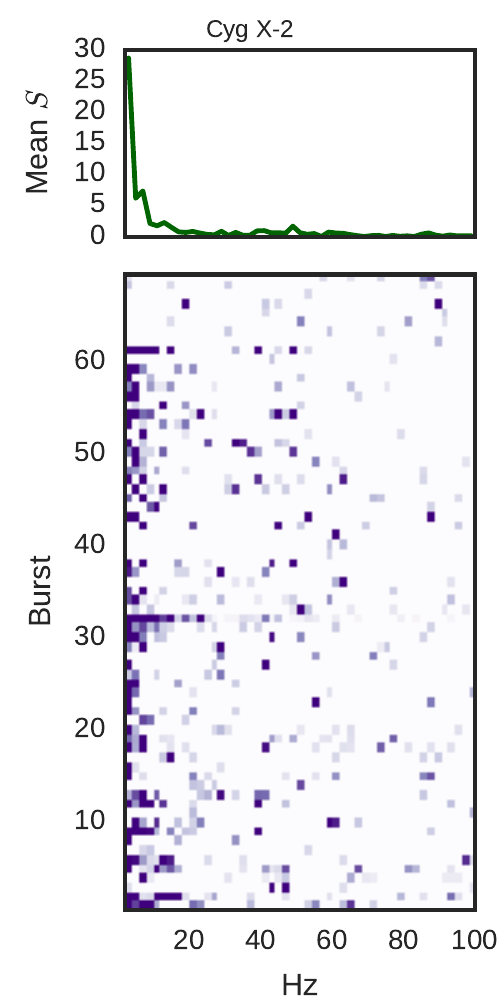}
\caption{Total power of candidates with $p_\mathrm{sb}(\chi^2)$ from Eq.~\ref{eq:psubthr},
normalized by the size of the on-burst window for three sources. For all three panels 
the color map has been saturated at the Leahy-normalized power of 15. \textit{Left:} 
Known TBO source 4U 1728$-$34. Bursts without regular candidates do not, in general, 
have sub-threshold candidates. \textit{Center}: IGR J17480$-$2446 has only two 11-Hz 
candidates above the standard detection threshold (Eq.~\ref{eq:pthr}), but many more 
sub-threshold candidates. \textit{Right:} Cyg X-2, exhibiting many sub-threshold 
candidates at random frequencies.}
 \label{fig:subthr}
\end{figure*}
 
 \subsection{New TBO discoveries}
 
One more TBO source has been discovered, SAX J1810.8$-$2609, yielding strong ($P_\mathrm{m}=79$) 
531-Hz oscillations in one independent time window (see Sect.~\ref{subsec:SAXJ1810}).
The candidate power so strong that it has small $p(\chi^2)$ probability assuming the most 
conservative number of trials (counting all harmonics and all time bins from the whole 
57-source sample as independent). Full details of this discovery are reported in \citet{Bilous2018}. 

Besides that, we recorded an interesting pair of $\sim 600$-Hz candidates from 
IGR J17473$-$2721 (see also Sect.~\ref{subsec:IGRJ17473}). These candidates were faint 
($p>0.5$), but came within 3\,Hz from each other and framed the burst peak during a burst 
with PRE, showing typical features of TBOs from established TBO sources (e.g. SAX J1750.8$-$2900).

\subsection{Other sources}

Out of the remaining thirty sources in our sample that had no published record of TBOs prior 
to our study  twenty-three were unremarkable, with the number of noise candidates reproduced 
well by simulations. Some of these sources also had low-frequency noise of type I or II. 
Seven more sources had a marginally significant number of candidates (albeit occurring 
at random frequencies), or stronger and more broadband low-frequency noise. More details 
can be found in Sect.~\ref{subsec:other}.

\subsection{Subthreshold candidates}
\label{subsec:subthr}

Even if an individual candidate has moderate power that is below our nominal threshold, 
a cluster of sub-threshold candidates in a relatively narrow frequency range may indicate 
the presence of TBOs. We performed a simple search for such a clustering of sub-threshold 
candidates by summing the powers of all candidates with:
\begin{equation}
\label{eq:psubthr}
p_\mathrm{sb}(\chi^2)<\frac{10^{-1}}{2000\times T_\mathrm{win}},     
\end{equation}
which corresponds to $P_\mathrm{m}$ of 17.03, 18.42, 19.81, and 21.19 for $T_\mathrm{win}$ 
of 0.5$-$2\,s. The sums, $S(\nu)$, were additionally added in 2-, 4- or 8-Hz frequency bins 
and normalized by burst duration. The stacks of $S(\nu)$ were then inspected visually for 
traces of power excess correlated in frequency.

Interestingly, for known TBO sources the bursts without TBO candidates did not necessarily
yield sub-threshold candidates, with $S(\nu)$ in the TBO frequency range being similar to 
$S(\nu)$ at other frequencies (e.g. Fig.~\ref{fig:subthr}, left).  Nevertheless, some of the 
known TBO sources did have sub-threshold TBO candidates, with the most prominent example 
being IGR J17480$-$2445. Only two bursts from this source have candidates at 10--11\,Hz in 
Table~\ref{table:cand}, but many more bursts have relatively large $S(\nu)$ 
(Fig.~\ref{fig:subthr}, middle). 

About half of sources in our sample exhibited low-frequency noise on $S(\nu)$ stacks, 
sometimes extending to $\sim 20$\,Hz.  For Cyg X-2, this frequency region was particularly 
noisy with multiple sub-threshold candidates at random frequencies (Fig.~\ref{fig:subthr}, right). 

None of the sources showed any obvious clustering of candidates at frequencies different 
from the frequencies of known TBOs. This was something of a surprise: we had anticipated 
that there would be sub-threshold candidates emerging from such a large data set.   
It also must be noted that, similarly to most TBO searches,
our analysis does not include the effects of any potential smearing due 
to intrinsic TBO frequency drift and Doppler shifts due to the motion of the Earth and the 
binary orbit, all of which would reduce detectability.

\section{Summary}
\label{sec:concl}

In this work, we conducted a large-scale blind search for thermonuclear burst oscillations 
for the majority of type-I X-ray bursts observed by RXTE. In comparison to previous work,
our analysis encompassed more sources, and probed potential signals on a range of time scales 
and further into burst tails, treating all sources in a uniform fashion. 

In order to estimate the significance of selected oscillation candidates, we developed a 
more realistic noise model by simulating photon sequences with variable count rate which 
mimicked the real light curves and was affected by dead time. Fourier spectra from simulated 
sequences were used to renormalize the corresponding Fourier spectra from the real data and 
thus to remove the low-frequency noise due to variable count rate, and to restore the 
dead-time-affected average power. 

Our noise model showed that abrupt LC variations, for example during the burst rise or 
data gaps, can bias the noise statistics in a frequency-dependent manner at frequencies up 
to approximately 100\,Hz, or, in several cases, even up to 1\,kHz (thus, not being confined
to low frequencies any more). LC modeling allowed us to remove most of this bias.  However,
in some cases we still detect strong candidates below 16\,Hz. These low-frequency candidates 
did not immediately resemble known low-frequency TBOs from IGR J17480$-$2446: with detections 
in multiple independent time windows and multiple bursts, at frequencies larger than the 
lowest recorded frequency of 2 Hz and without candidates of comparable strength at the 
nearby, but distinctly separate frequencies. Some of the detected low-frequency candidates 
are clearly generated by single, poorly modeled sharp peaks or dips in LCs, these we refer 
to as ``type I'' low-frequency candidates. 

Several more sources yielded candidates not immediately connected to flaws in LC modeling 
(e.g. Cyg X-2, 4U 1729$-$34, EXO 0748$-$676, EXO 1745$-$248, and others). Such candidates, 
dubbed ``type II'' low frequency candidates, frequently appeared to be grouped in time and/or frequency, 
sometimes appearing at distinct frequencies simultaneously. It is possible that these 
type II low frequency candidates may have an astrophysical origin:  perhaps a non-TBO 
process on the burning surface, or varying emission due to the effect of the burst on 
the accretion flow \citep[see for example][]{Worpel2013,Worpel2015}.  Generally, the 
signal at the lowest frequencies in our spectra (below approximately 5\,Hz) is quite 
hard to interpret, since its strength depends substantially on how closely the model 
LC follows the real one.  

The instrumental dead time had, somewhat surprisingly, a rather large influence on 
the power spectra statistics, with the average noise power dropping below 1.7 for the 
burst peaks of five sources, some of them with TBOs. Neglecting the influence of dead 
time can lead to underestimation of candidate TBO significance by as much as two orders 
of magnitude. 

Overall, our noise models provide an important insight into the statistics of RXTE power 
spectra, but they do not give a perfect description of the data, most probably because 
of the set of assumptions regarding the dead time influence and what constitutes a 
``real'' LC. Also, some bias is caused by the limited number of simulations 
run to derive the statistical properties of noise. From the computational point of view, 
it is much easier to estimate the average noise power using harmonics past 
$\gtrsim 1$\,kHz, renormalize the power spectra and use $\chi^2$ probability distribution 
with conservative number of trials (treating all time windows as independent, regardless 
of overlap) to estimate the candidate significance. However, this approach would not 
work at lower Fourier frequencies during the burst rise or during data gaps.

We have also found that abrupt changes in the LC rate (sharp rise or a data gap) can lead 
to covariance between adjacent Fourier harmonics and can manifest as a fast change of TBO 
frequency. A quantitative investigation of this phenomenon will be presented in subsequent 
work. Overall, data gaps obliterate part of the signal and bias the fractional amplitude 
evolution: using data with gaps should be avoided if at all possible.  Future X-ray 
telescopes aiming to study this phenomenon should aim for high throughput.  

For our study, we have selected all candidates with renormalized $\chi^2$ probabilities 
less than $2\times10^{-4}$ per spectrum. This resulted in the power thresholds varying with time window size.
Our choice of detection threshold was to some degree arbitrary, 
but was motivated by a wish to analyze a manageable number of candidates. The significance 
of these candidate detections was then estimated by comparing the number of candidates 
in the real data to a pool of an additional 100 of simulated spectra, renormalized in 
the same way as the real data. 

Our candidates included all previously known TBOs. For 
one of the sources, the accreting MSP HETE J1900.1$-$2455, the detection in a single 
time window was not significant because of the large number of trials in our analysis. 
The study that reported this finding originally searched a narrower frequency range around 
the known pulsar frequency \citep{Watts2009}. We find that the power of candidates depends 
dramatically on the specific window sizes and degrees of overlap used.  

Overall, we have compiled an extensive data set containing information on the frequency 
and fractional amplitudes of all selected candidates, as well as upper limits on fractional 
amplitudes derived from the threshold powers. We anticipate that this information will 
be a valuable resource for future studies of TBO properties, particularly when used in 
conjunction with the burst property database MINBAR. The conditions under which TBOs are 
excited and detectable are important factors in assessing the viability of physical 
models for the TBO mechanism \citep{Watts2012}. 

Eight sources in our dataset had prior claims of TBOs where the claimed detections
were either weak and came from one independent time window (or, in the case of XB 
1916$-$053, two close but separated frequencies) in a single burst or several stacked 
bursts. We were unable to confirm TBOs from any of those sources. Some of the previously 
claimed detections had smaller powers in our analysis (which can be sensitive to the 
choice of the time windows) and were not significant when compared to noise simulations.
For 4U 0614$+$09 we had different bursts than the ones with potential TBOs (which 
came from a different telescope); the burst in the RXTE sample showed no TBO candidates. 
Other claimed detections were based on analysis of stacked spectra and yielded no 
candidates in our time windows. 

One of the sources without previously reported TBOs, SAX J1810.8-269 yielded a strong, 
brief 531-Hz pulsation in one of the bursts.  The signal was detected in one independent 
time window, however its strength ($P_\mathrm{m}>70$) speaks in favour of it being a TBO 
\citep[for more in-depth significance analysis, see][]{Bilous2018}. The other sources did 
not provide any compelling TBO candidates, despite our removing most of the low-frequency 
noise and making better significance estimates for bright bursts. In addition, we found 
no groups of sub-threshold candidates, probing probabilities up to 100 higher than 
our adopted detection threshold.  This was somewhat surprising: we had anticipated finding 
at least some clusters of sub-threshold candidates in such a large burst sample.

An interesting (albeit not formally significant in our analysis) pair of $\sim 600$-Hz TBO 
candidates was recorded  from IGR J17473$-$2721. The candidates were rather faint, but 
came close in frequency (within 3\,Hz) and framed the burst peak during a burst with 
PRE. More than half of the simulation runs had as many or more candidates (at arbitrary 
frequencies) with at least the same significance. 

Another source with previously-reported 
potential TBOs with similar characteristics, XB 1916$-$053 had much more significant 
candidates, with as few as 2\% of the simulations runs having the candidates at least as 
strong as the strongest one on the real data. Overall, IGR J17473$-$2721 and XB 1916$-$053
would be interesting sources for subsequent follow-up.  

Our estimate of candidate significance treated all frequencies as independent and did not 
include important TBO features such as frequency drift coupled with signal disappearance 
during PRE. In the case of weaker signals, it is currently unclear how small a gap in 
frequency should be for the signals to be attributed to a single TBO. 

We have found that some of the sources exhibited a marginally significant number of noise 
candidates, meaning that 99\% or more simulations  runs had a smaller number 
of candidates. These candidates appeared at random frequencies both below and above 1\,kHz 
in single independent time windows and were often stronger than some of the TBO detections 
in individual bursts, reaching $P_\mathrm{m}\gtrsim40$. We dub them \textit{glimmer candidates}.  
It is possible that some of the glimmer candidates are of astrophysical origin (especially 
the ones at lower frequencies).

\section{Conclusions}

TBOs are transient phenomena with rapidly changing properties.
The measured power of potential TBO candidates depends greatly on the specific choices
regarding data selection, such as energy filters, time windows, degree of overlap,
summing harmonics or adjacent time windows and stacking spectra from different bursts. 
Thus, considering the researcher's natural desire to find TBOs, one must be very careful 
with estimating the number of trials resulting from tweaking the search parameters and exploring multiple sources.

While searching for high-power narrowband signals using Fourier transform in overlapping time windows,
it is generally reasonable to use a $\chi^2$ model of the distribution of the noise powers
with the conservative number of trials, after correcting for dead time influence, LC variation 
and making sure that the harmonics in Fourier spectra are not covariant. However, it is strongly advisable to 
verify that a $\chi^2$ distribution actually describes well the noise powers of a given dataset. 

Our search for TBOs resulted in several short (each detected in one independent time window)
candidates with powers comparable to those of the fainter TBOs ($P_\mathrm{m} \sim 30-40$). These candidates, dubbed ``glimmer''
are marginally significant, meaning that 99\% or more simulations  runs had a smaller number 
of candidates. They produce sinusoidal
oscillation profiles and are in all aspects resembling fainter TBOs. 
However, they occur at random frequencies within a single source and sometimes are coincident in time with real TBOs.
Partly, glimmer candidates may stem from selection bias, however an astrophysical origin is not excluded.
Regardless of their nature, the phenomenon of glimmer candidates may explain the large number of unconfirmed single-window 
detections, e.g \citet{Kaaret2002}, especially considering the tendency of underestimating
the number of trials. 

For the potential detections with the smaller power, the best corroboration of TBO nature is 
detecting the signal at the same frequency in independent time windows; however with the intrinsic 
frequency drift and Doppler modulations complicate it. It is therefore important to develop 
a procedure for estimating significance of signals with drifting or jumping signal. This would help
refining the significance of frequency jump in MXB 1658$-$298 \citep{Wijnands2001}, 
drifting candidate in 1A 1744$-$361 \citep{Bhattacharyya2006}, a pair of 270-Hz candidates XB 1916$-$053 \citep{Galloway2001} 
and potential pair of 600-Hz candidates from IGR 17473$-2721$ (this work).

Despite our efforts, we did not find TBOs below 200\,Hz and during high count rates.
Several measures can be undertaken in order to obtain a better, more complete picture of 
TBOs. Obtaining new data using better instrumentation with higher throughput 
leads to better sensitivity and the absence of data gaps allows a better characterization of the frequency evolution.
Continuing searching is also an option, since TBOs may appear from the sources without promising 
candidates, although having some theoretical guidance would be better, something the data 
could be used for.  
Further improvement of TBO searches can also be made by selecting only that part of energy 
spectrum where there  are most burst photons, in order to minimize the relative contribution 
of background. Having ephemerides would also help to correct for the Doppler change in 
frequency: this would be especially helpful for the ultra-compact binaries such as  
4U~1820$-$303 or the potential ultra-compact binary 2S~0918$-$549.

\acknowledgements

A.V.B. and A.L.W. acknowledge support from ERC Starting grant No. 639217 CSINEUTRON- STAR 
(PI: A.L. Watts). We would like to thank Duncan Galloway and the MINBAR team for sharing a 
pre-release version of the MINBAR database with us. A.V.B. thanks Hauke Worpel for sharing
the data on background variation during type I bursts.

\appendix


\section{Times of arrival, light curves, and tables}

\begin{figure*}
 \centering
 \includegraphics[scale=0.77]{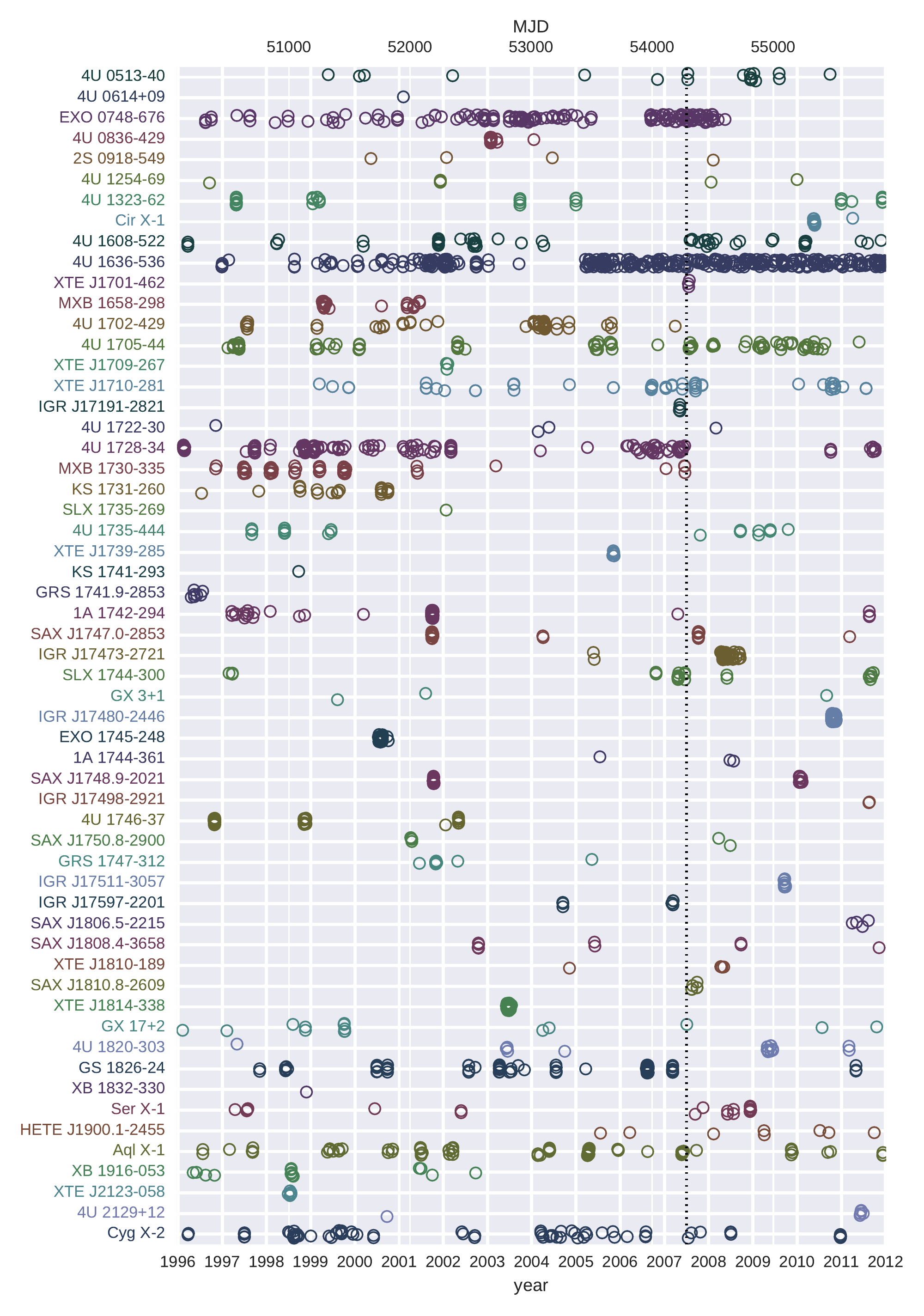}
 \caption{Times of arrival of \nbrst\ bursts from MINBAR catalogue, for which high $\tres$
data were available. The sources are ordered by their coordinates (right ascension first). 
Dotted vertical line marks the end of the time cut for the sample of \Gwt, however not all 
of the sources before that date have been examined in \Gwt. Some of the bursts are 
so close to each other that the markers overlap completely. }
 \label{fig:arrivals}
\end{figure*}

\begin{figure*}
 \centering
 \includegraphics[scale=0.9]{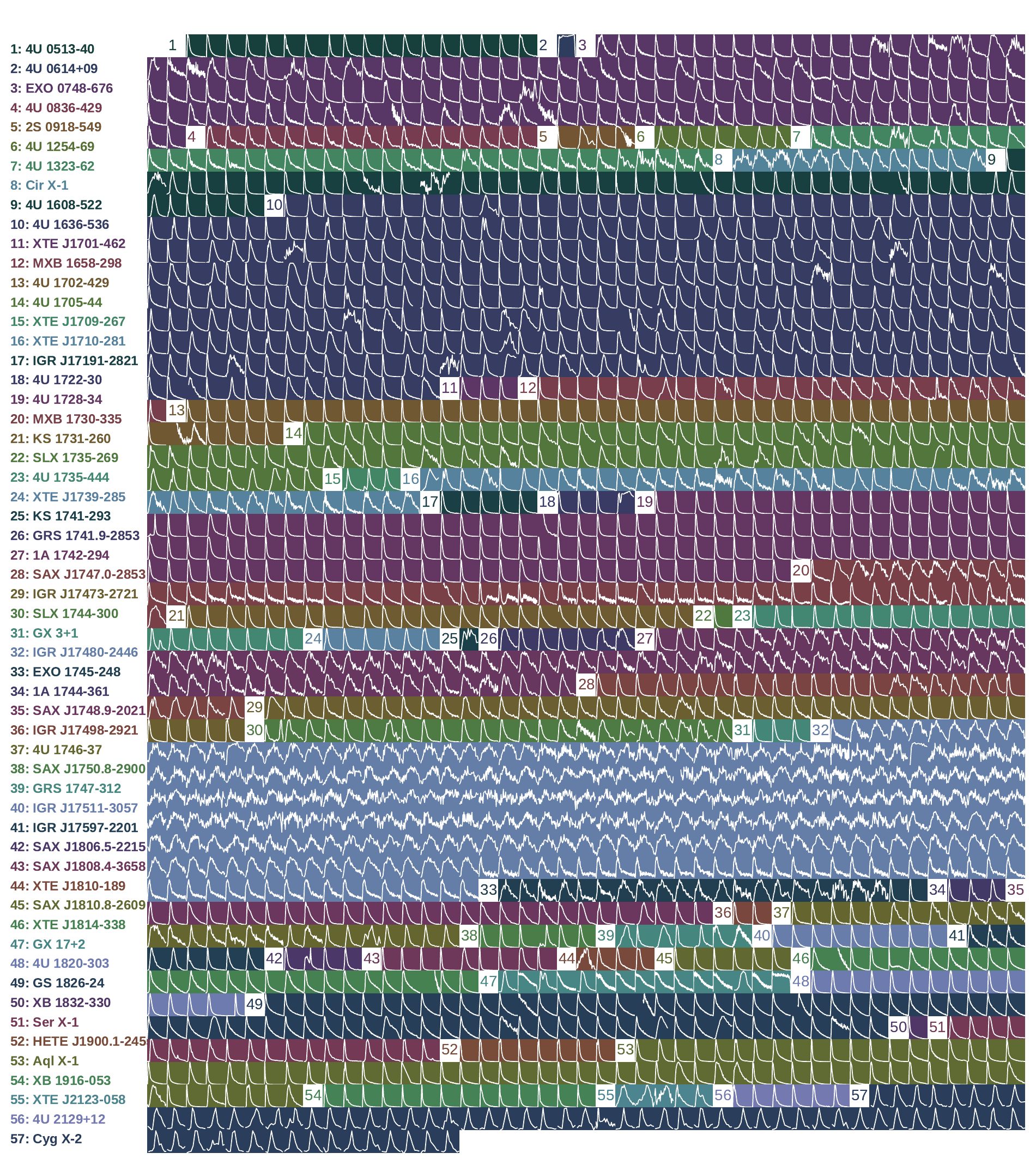}
 \caption{Light curves from Standard-1 data for \nbrst\ bursts from MINBAR catalogue 
 for which high $\tres$ data were available. of bursts from the previous figure. The sources are ordered by their coordinates (right ascension first). 
 Time span matches the adopted  on-burst window for each burst, the y-axis scale is different for each burst.}
 \label{fig:allbursts}
\end{figure*}

\newpage

\startlongtable
\begin{deluxetable*}{lhhlllllllp{0.7cm}p{1.0cm}lllllll}
\setlength{\tabcolsep}{2pt}
\renewcommand*{\arraystretch}{0.8} 
\tabletypesize{\scriptsize}
\tablenum{2}
\tablecaption{Data and $\mathrm{FA_{up}}$ for one of the sources. This table is available in its entirety in machine-readable form.
 \label{table:data}}

\tablehead{
& & & & & & & & & & & & & & & &  & &   \\
1& & &2&3&4&5&6&7&8&9&10&11&12&13&14a&14b&14c&14d\\
 & & & & & & & & & & &  &  &  & & \multicolumn{4}{c}{min $\mathrm{FA_{up}}$ (\%)}   \\
\cline{16-19}
\colhead{\parbox{1.5cm}{Source}} & & &
\colhead{\parbox{0.5cm}{\centering Entry}} & 
\colhead{\parbox{2.0cm}{\centering MINBAR TOA }}&  
\colhead{\parbox{1.0cm}{\centering $T_\mathrm{peak}$ (MET)}}& 
\colhead{\parbox{0.5cm}{\centering Rise (s)}}&
\colhead{\parbox{0.5cm}{\centering Half-max (s)}} & 
\colhead{\parbox{0.5cm}{\centering Tail (s) }} & 
\colhead{\parbox{0.5cm}{\centering Peak S/N}} &  
\colhead{\parbox{2.1cm}{\centering Data mode}} & 
\colhead{\parbox{1.0cm}{\centering Notes }} & 
\colhead{\parbox{0.3cm}{\centering Max $P_\mathrm{m}$ }} & 
\colhead{\parbox{0.3cm}{\centering $\nu$  }} & 
\colhead{\parbox{0.5cm}{\centering $T_\mathrm{win}$ }} & 
\colhead{\parbox{0.4cm}{\centering 0.5 (s) }} & 
\colhead{\parbox{0.4cm}{\centering 1.0 (s) }} & 
\colhead{\parbox{0.4cm}{\centering 2.0 (s) }} & 
\colhead{\parbox{0.4cm}{\centering 4.0 (s) }}  }
\startdata
HETE J1900.1$-$2455 & 1 & 1118 & 3301 & 2005-07-21 23:00:32 & 364604449.9 & 14.5 & 30.5 & 413.5 & 1607.7 & \parbox[t]{2.3cm}{E\_125us\_64M\_0\_1s}& \parbox[t]{1cm}{gaps, bad LC, freq cov}&   34.8 & 383.50 & 4.0 & 4.6 &4.5 &3.3 &2.4\\ 
 & 2 & 1179 & 3362 & 2006-03-20 11:34:06 & 385472056.4 & 7.5 & 9.5 & 118.0 & 510.1 & \parbox[t]{2.3cm}{SB\_125us\_8\_249\_1s}& \parbox[t]{1cm}{}&   29.2 & 228.00 & 4.0 & 4.6 &3.3 &2.4 &1.7\\ 
 & 3 & 1484 & 3667 & 2008-02-10 20:32:51 & 445293182.9 & 9.0 & 5.5 & 86.0 & 435.8 & \parbox[t]{2.3cm}{E\_125us\_64M\_0\_1s}& \parbox[t]{1cm}{}&   35.2 & 12.00 & 0.5 & 7.2 &5.2 &3.8 &2.7\\ 
 & 4 & 1635 & 3818 & 2009-04-02 08:57:54 & 481280277.9 & 2.0 & 5.5 & 76.0 & 729.4 & \parbox[t]{2.3cm}{SB\_125us\_8\_249\_1s}& \parbox[t]{1cm}{}&   43.5 & 376.25 & 4.0 & 6.1 &4.5 &3.3 &2.5\\ 
 & 5 & 1637 & 3820 & 2009-04-04 19:06:51 & 481489619.9 & 6.0 & 3.5 & 97.0 & 777.7 & \parbox[t]{2.3cm}{E\_125us\_64M\_0\_1s}& \parbox[t]{1cm}{}&   28.5 & 1471.00 & 2.0 & 6.6 &4.8 &3.5 &2.6\\ 
 & 6 & 1763 & 3944 & 2010-07-07 21:04:32 & 521154292.9 & 12.0 & 6.5 & 118.5 & 672.8 & \parbox[t]{2.3cm}{E\_125us\_64M\_0\_1s}& \parbox[t]{1cm}{}&   32.6 & 674.50 & 2.0 & 7.1 &5.2 &3.8 &2.7\\ 
 & 7 & 1779 & 3960 & 2010-09-20 05:29:04 & 527578164.4 & 7.5 & 8.5 & 92.0 & 427.1 & \parbox[t]{2.3cm}{SB\_125us\_8\_249\_1s}& \parbox[t]{1cm}{}&   27.6 & 1313.50 & 4.0 & 6.0 &4.3 &3.1 &2.3\\ 
 & 8 & 2170 & 8257 & 2011-09-29 23:43:46 & 559957439.4 & 7.5 & 5.0 & 74.5 & 554.3 & \parbox[t]{2.3cm}{E\_125us\_64M\_0\_1s}& \parbox[t]{1cm}{}&   30.3 & 776.00 & 4.0 & 7.0 &5.1 &3.7 &2.8\\
\enddata
\tablecomments{Table 2 columns: (1) source name; 
(2) burst entry \# in MINBAR database;
(3) burst TOA in MINBAR database;
(4) time of the burst peak;
(5) $dt_\mathrm{rise}$, the size of the rise window;
(6) $dt_\mathrm{halfpeak}$, the time span between peak and the half-peak on the trailing side;
(7) $dt_\mathrm{decay}$, time span between the the peak and burst decay;
(8) peak S/N;
(9) observing modes used for TBO search;
(10) notes, including: manual RBT windows faint bursts or bursts with peculiar shapes, presence of data gaps, partial GTI coverage, 
missing data for at the edges of on-burst window, covariance between Fourier harmonics, local failures in LC modeling;
(11) maximum measured power $P_\mathrm{m}$, low-frequency noise excluded;
(12) frequency of maximum $P_\mathrm{m}$;
(13) $T_\mathrm{win}$ of the maximum  $P_\mathrm{m}$;
(14) best upper limit on FA in four window sizes (14a-d).}
\end{deluxetable*}

\newpage

\startlongtable
\begin{deluxetable*}{lhllchcccccccccc}
\renewcommand*{\arraystretch}{1.0}
\tabletypesize{\scriptsize}
\tablenum{4}
\tablecaption{Fractional amplitudes of candidates per burst}
\tablehead{ 
&  & & & & & & & & & & & & & & \\
1& &2&3&4&&5&6a&6b&6c&6d&6e&7a&7b&7c&7d\\
& & & & & & & \multicolumn{5}{c|}{\centering Indepenent time windows} & \multicolumn{4}{c}{\centering Max FA (\%)} \\
\cline{8-16}
\colhead{\parbox{2.5cm}{ Source}} &  &
\colhead{\parbox{0.5cm}{Entry}} &
\colhead{\parbox{1.5cm}{\centering MINBAR TOA, UT}} &
\colhead{\parbox{0.5cm}{\centering $\nu$ (Hz)}} &  &
\colhead{\parbox{0.7cm}{\centering Where}} &  
\colhead{\parbox{0.3cm}{\centering any}}&
\colhead{\parbox{0.3cm}{\centering 0.5\,s}}&
\colhead{\parbox{0.3cm}{\centering 1.0\,s}}&
\colhead{\parbox{0.3cm}{\centering 2.0\,s}}&
\colhead{\parbox{0.3cm}{\centering 4.0\,s}}&
\colhead{\parbox{0.3cm}{\centering 0.5\,s}}&
\colhead{\parbox{0.3cm}{\centering 1.0\,s}}&
\colhead{\parbox{0.3cm}{\centering 2.0\,s}}&
\colhead{\parbox{0.3cm}{\centering 4.0\,s}}
}
\startdata
4U 0513-40 & 1614 & 3797 & 2009-01-08 12:23:21 & 1378.75 &  & ---T &  1  &  0 &  0 &  0 &  1 &  -- &  -- &  -- &  $30_{25}^{35}$ \\
\hline
EXO 0748-676 & 196 & 2396 & 1998-06-28 15:28:19 & 86.00 &  & ---T &  1  &  1 &  0 &  0 &  0 &  $78_{60}^{98}$ &  -- &  -- &  -- \\
 & 1232 & 3415 & 2006-09-08 21:57:59 & 113.00 &  & ---T &  1  &  0 &  0 &  1 &  0 &  -- &  -- &  $34_{29}^{41}$ &  -- \\
 & 1381 & 3564 & 2007-07-15 23:32:38 & 147.50 &  & ---T &  1  &  0 &  0 &  1 &  0 &  -- &  -- &  $45_{38}^{53}$ &  -- \\
 & 949 & 3132 & 2003-09-21 23:14:21 & 288.00 &  & ---T &  1  &  0 &  0 &  1 &  0 &  -- &  -- &  $19_{16}^{22}$ &  -- \\
 & 35 & 2237 & 1996-08-15 21:47:35 & 327.50 &  & ---T &  1  &  0 &  0 &  1 &  0 &  -- &  -- &  $22_{19}^{25}$ &  -- \\
 & 961 & 3144 & 2003-10-13 00:04:11 & 333.75 &  & ---T &  1  &  0 &  0 &  0 &  1 &  -- &  -- &  -- &  $28_{23}^{33}$ \\
 & 1314 & 3497 & 2007-03-31 07:49:15 & 339.00 &  & ---T &  1  &  0 &  0 &  1 &  0 &  -- &  -- &  $52_{43}^{62}$ &  -- \\
\enddata
\tablecomments{Table 4 columns: (1) source name; 
(2) burst entry \# in MINBAR database;
(3) burst TOA in MINBAR database;
(4) frequency range of candidates from a particular frequency group;
(5) region of occurence (R -- rise, B -- burst, T -- tail);
(6) number of independent window detections, (6a) of any $T_\mathrm{win}$,
(6b-6e) in four window sizes separately;
(7) Maximum FA in time windows of four sizes (7a-7d). }
\end{deluxetable*}


\section{Individual sources}


\subsection{Individual sources with previously detected TBO}
\label{subsec:knownTBO}

\subsubsection{EXO 0748$-$676}
\label{subsec:EXO0748}

\citet{Galloway2010} reported on two strong TBO candidates from the rise of two bursts
out of 157 bursts searched. The candidates (with the powers of 59.68 and 48.26) 
were detected in one independent time window per burst at the frequencies of 552 and 552.5\,Hz.
The estimate of the significance of the pair of candidates separated by $<1$\,Hz was 
based on the conservative number of trials and led to $6.3\sigma$ significance. 

Earlier, \citet{Villarreal2004} had reported a $5.35\sigma$-equivalent 45-Hz oscillation 
in the stacked spectra of 38 bursts. This candidate does not show up in the larger burst 
sample of \citet{Galloway2010}, and its origin is unclear. 

\begin{figure}
 \centering
 \includegraphics[scale=0.72]{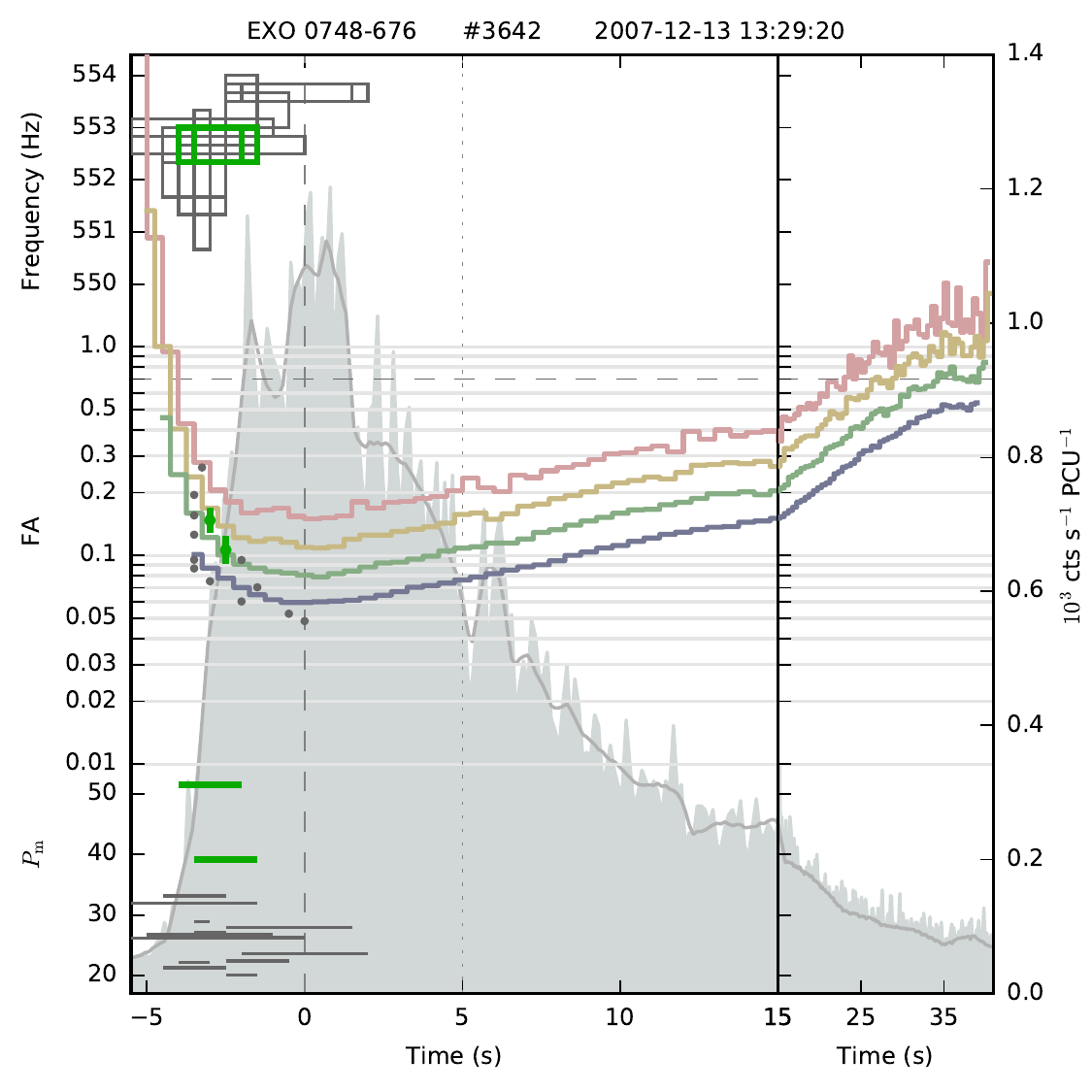}
\caption{Oscillation candidate from EXO 0748$-$676. 
} \label{fig:EXO0748}
\end{figure}

Our sample consists of 159 bursts, the majority with long tails ($\sim 90$\,s). 
We detect candidates at 551.5--552.5\,Hz from the same two bursts as \citet{Galloway2010}.
In both bursts the candidates come from a few time windows, all of them 
dependent (e.g. Fig.~\ref{fig:EXO0748}). The sub-threshold candidates hint to a frequency evolution. The highest 
powers of candidates in 551.5--552.5\,Hz frequency range are 57.5 and 51.4 (56.8 
and 49.2 on non-normalized data, respectively), whereas maximum $P_\mathrm{m}$ 
outside this frequency region is 43.6. None of the simulations had the same 
$P_\mathrm{m}$ as the strongest candidate, however, we did not make a significance 
estimate for the pair of candidates close in frequency.

Our analysis procedure does not find TBO candidates in the 552--554\,Hz frequency 
region from the two fainter bursts mentioned in \citet{Galloway2010} (even 
sub-threshold). 

EXO 0748-676 is remarkable as a prolific source of type II low-frequency. 
Low-frequency candidates come from 2--14.5\,Hz. Sometimes they are
confined to 2-3 Hz, sometimes they chaotically occupy all frequencies up to 13\,Hz, 
and sometimes they occur at distinctly separate frequencies, e.g. 5 or 9\,Hz.  

The source yielded a few dozen (glimmer) candidates with $p=0.01$, none of them close to 
the 45\,Hz of \citet{Villarreal2004}. Some of the candidates at widely separated 
frequencies come from the same burst, sometimes even from the same time windows.

\subsubsection{4U 1608$-$522}

\begin{figure}
 \centering
 \includegraphics[scale=0.72]{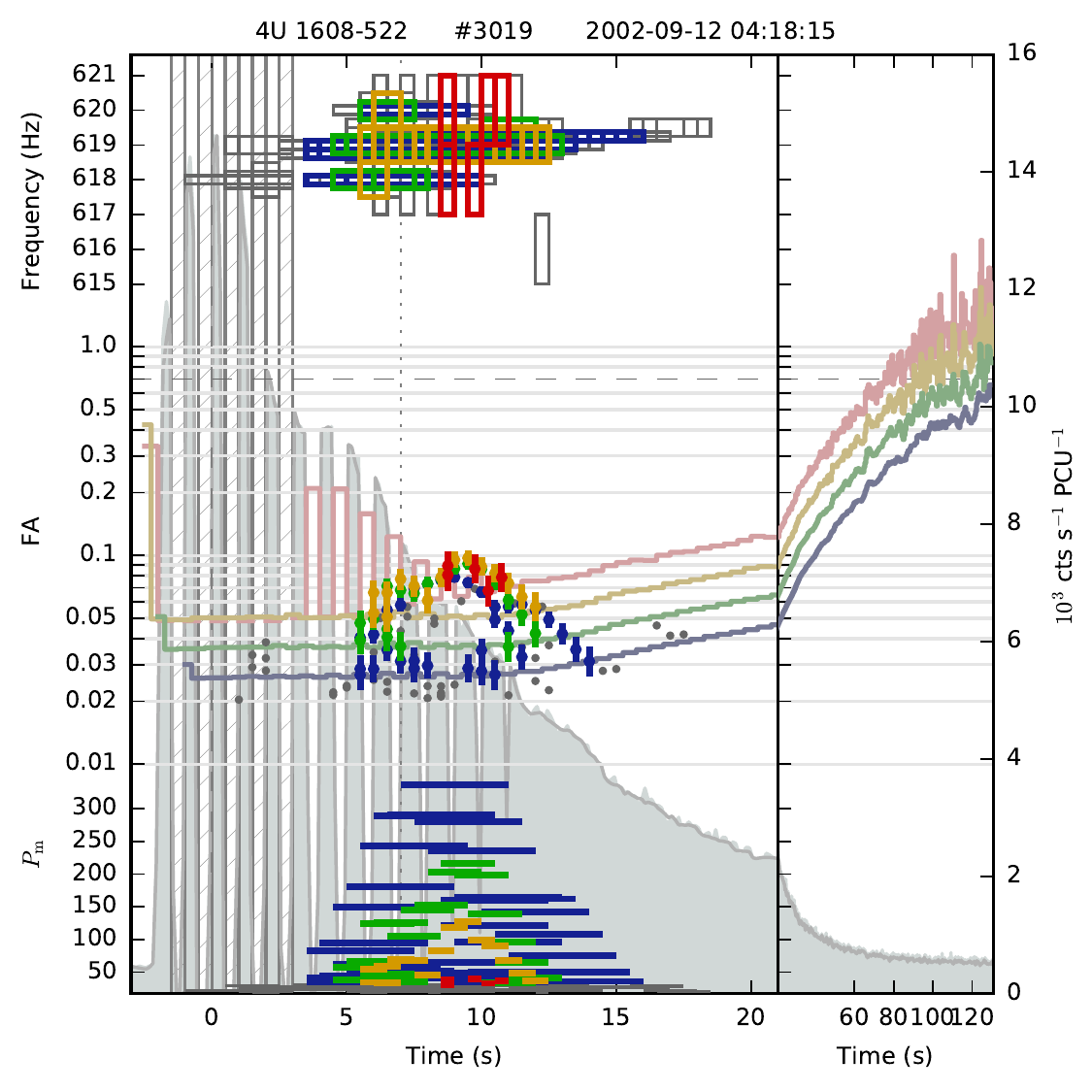}
   \caption{Burst from 4U 1608$-$522, with strong oscillation signal. The data is marred by gaps.   }
 \label{fig:4U1608}
\end{figure}

TBOs from 4U 1608$-$522 have been detected at 619\,Hz in multiple bursts in the rise and 
after the burst peak by \citet{Galloway2008}. The authors report large gradual frequency 
drifts, and FAs of 5--15\%. 

Our sample has 52 bursts, some of them very strong, suffering from data gaps and reduced average
noise power (going down to 1.6). TBOs were detected at 616--620\,Hz in seven bursts (e.g. Fig.~\ref{fig:4U1608}). Two 
bursts had TBO signals in one independent time window (one of them had more sub-threshold 
candidates). The oscillations are mostly detected in the B region; one burst has TBOs 
starting in the rise. The gradual frequency drift throughout the TBO duration and FAs of
3--12\% are consistent with \citet{Galloway2008} and \citet{Ootes2017}.

In addition to TBOs, we record several type I low-frequency 
and about a dozen glimmer candidates ($p=0$).

\subsubsection{4U 1636$-$536}

\begin{figure}
 \centering
 \includegraphics[scale=0.72]{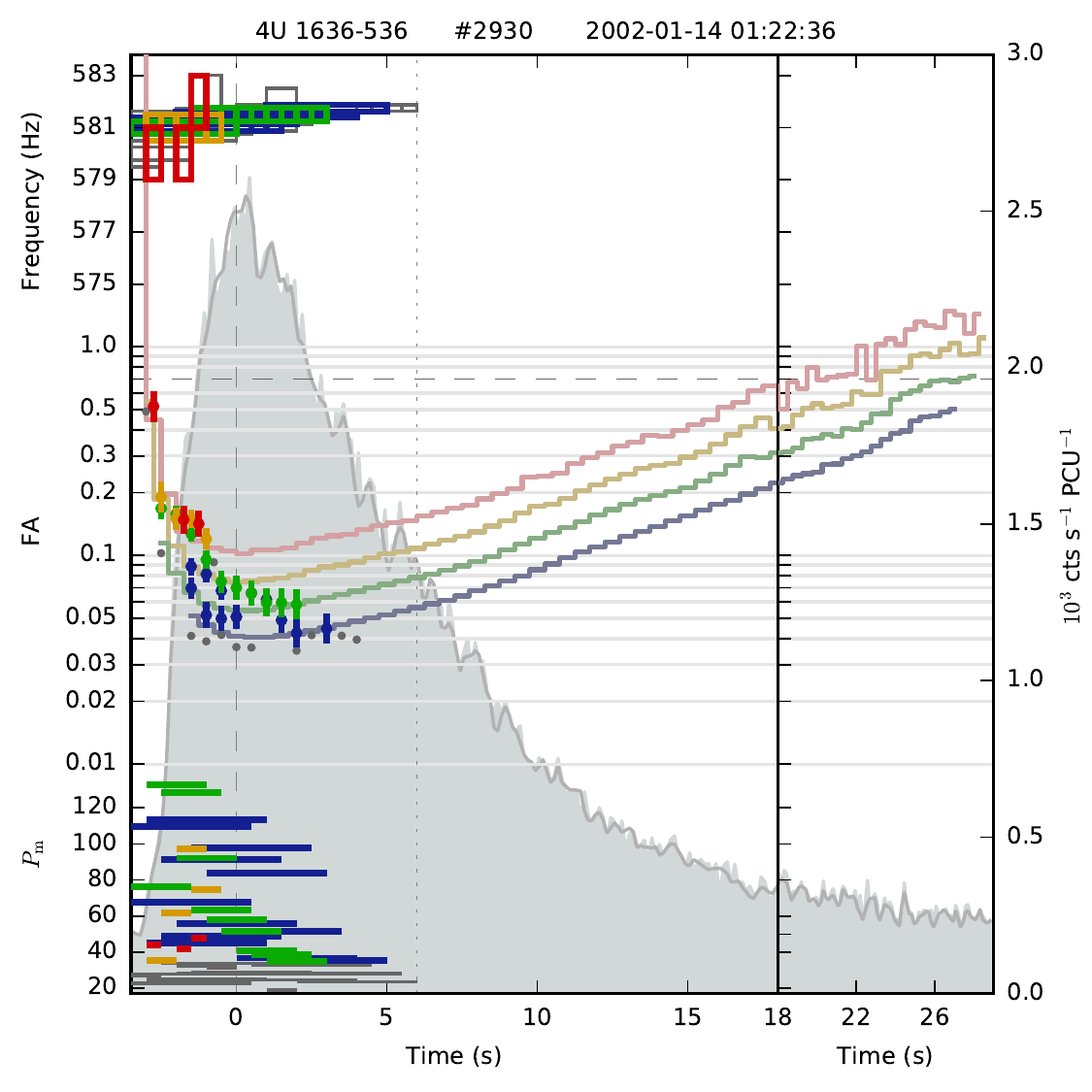}
   \caption{Example of a burst from 4U~1636$-$536 with a high-FA candidate on the burst rise. }
 \label{fig:4U1636_1}
\end{figure}

4U 1636$-$536 is one of the most prolific and best-studied TBO sources. Our sample contains 
368 bursts from 4U 1636$-$536, forming the largest sample among the 57 sources that we have 
in total. Some of the bursts are quite bright, with average noise power dropping as low as 1.7.

TBOs at 576--582\,Hz were detected from 75 bursts, most of them in the RB regions (the 
only detection in the T region is at its left edge). About 30\% of the TBO detections 
are in one independent time window. FAs on the order of 5--15\%, can reach up to 50\% on 
the rise (Fig.~\ref{fig:4U1636_1}). The same large FAs on the rise were previously 
reported by \citet{Strohmayer1998a}. The FAs that we find broadly coincide with the 
values reported in  \citet{Ootes2017}, \citet{Galloway2008}, and \citet{Miller2000}.

In addition to TBO candidates, we detect a few low-frequency candidates and a large, 
but insignificant number of noise candidates ($p=1$). \citet{Miller1999} reported a significant 
signal at 290\,Hz from the sum of 0.75-s intervals on the rise of five bursts. None of our 
noise candidates were close to 290\,Hz.

\subsubsection{MXB 1658$-$298 (X 1658$-$298)}
\label{subsec:MXB1658}

TBOs at 567\,Hz were discovered by \citet{Wijnands2001}, who detected them in six 
bursts out of 14 observed. The TBOs had small (0.5--1\,Hz) frequency drift and FAs 
on the order of 10\%. A larger sample of bursts was later explored by \Gwt.

Our sample yielded 26 bursts, four of them with TBO candidates in the 566.75--567.25\,Hz 
frequency range. The candidates are rather weak, with peak powers of 35--45,
in one independent time window per burst. Some of them occur on the rise, some a few seconds 
after the burst peak. Formally, two TBOs are labeled as coming from the tail region, 
but those bursts had a sharp intensity drop, so that the B region was narrow. 
FAs on the order of 10\% are broadly consistent with the values reported in 
\citet{Wijnands2001} and \Gwt\ for all bursts except for burst \#2519. There, we have FAs 
3 times smaller (consistent with \citet{Wijnands2001}).

Interestingly, there is a discrepancy in burst detections. \Gwt\ does not confirm one 
burst with a detection reported in \citet{Wijnands2001}, but has one more burst with a 
detection from 2001. We do not have any noticeable sub-threshold candidates in three of 
the bursts with detections reported in these two papers. $\FAup$ are similar to or even lower 
than the reported detections.

The standard threshold does not yield any low-frequency candidates. A small number of 
noise candidates is not statistically significant.

\begin{figure}
 \centering
 \includegraphics[scale=0.72]{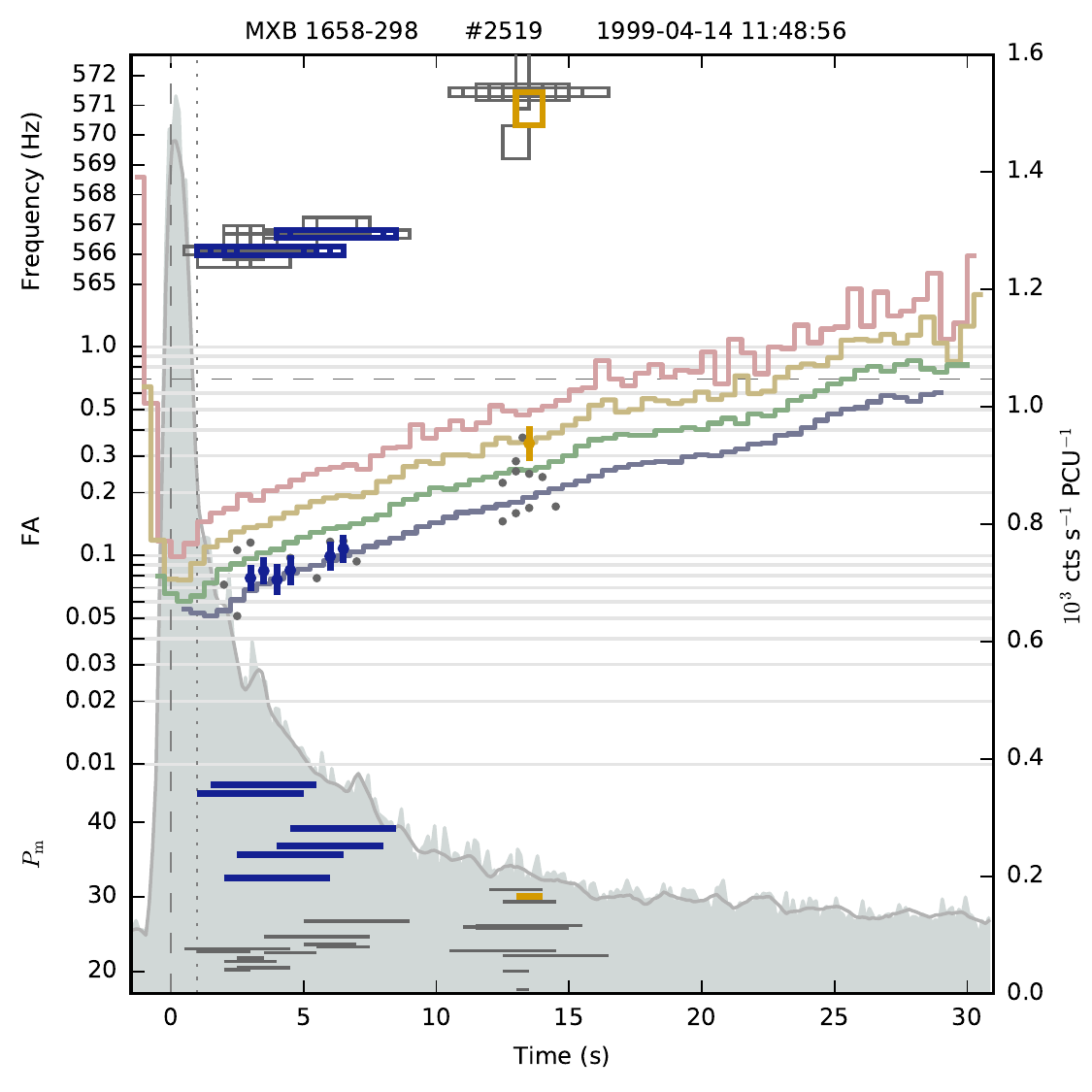}
\caption{The burst with a TBO frequency step reported in \citet{Wijnands2001}. 
On this figure the tail panel was merged with the burst panel to keep
the time scale uniform. The
threshold probability was increased by a factor of 3.7 for this plot; otherwise the 
candidate around 571\,Hz lies below the selection threshold. 
This candidate is detected in one independent 
time window and may be a fortuitous noise spike or glimmer, not connected to TBOs earlier in the burst.}
 \label{fig:MXB1658}
\end{figure}

\citet{Wijnands2001} reported a burst (\#2519) with oscillations reappearing at a frequency 
larger by about 5 Hz (571.5\,Hz), with similar signal strength (maximum $P_\mathrm{m}=32$  
for 2-s sliding windows with 0.25-s  offset using the $Z^2$ statistic). This candidate does 
not exceed our detection threshold 
(corresponding to $P_\mathrm{m}=33.62$ for 2-s windows), however we do detect a 
bunch of sub-threshold candidates in the same region. The candidate with the smallest 
probability has $P_\mathrm{m}=30$ in 1-s window at 571\,Hz (30.2 on non-normalized data).  
It is definitely not as strong as the TBOs earlier in the burst (Fig.~\ref{fig:MXB1658}). 
The discrepancy between our values and those of \citet{Wijnands2001} can be readily 
explained by the different choice of FFT windows and using FFT vs. $Z^2$. 

None of the remaining bursts yielded candidates within the 10\,Hz region of the TBO 
frequency range. It is hard to tell whether the 571-Hz candidate is related to TBOs. 
Two outcomes are possible: a) it is a TBO, as is stated in \citet{Wijnands2001};
b) it is a glimmer candidate. \citet{Wijnands2001} 
estimate its significance taking into account only trials in the 10-Hz frequency 
region around the TBOs. However, it is unclear whether this is a correct choice, since 
this region was picked after the candidate was found on a broader search from 100 to 
1200\,Hz. Our analysis with the probability threshold multiplied by a factor of 3.7 
yields nine more candidates, eight of them below 1000\,Hz and one above. Some of 
these candidates are stronger than the 571-Hz one. The simulated data does not have,
on average this many candidates: only one simulation run had as many as 10 candidates ($p=0.01$). 
Thus, it is possible that MXB 1658$-$298 has glimmer candidates, and the peak at 571-Hz 
is one of them. It is also worth noting that this candidate has a softer spectrum than the 
TBOs earlier in the burst \citet{Wijnands2001}.

The most solid way to prove that step candidate is a TBO would be to detect it once again at the 
same frequency preferably with larger power.

\subsubsection{4U 1702$-$429}

\begin{figure}
 \centering
 \includegraphics[scale=0.72]{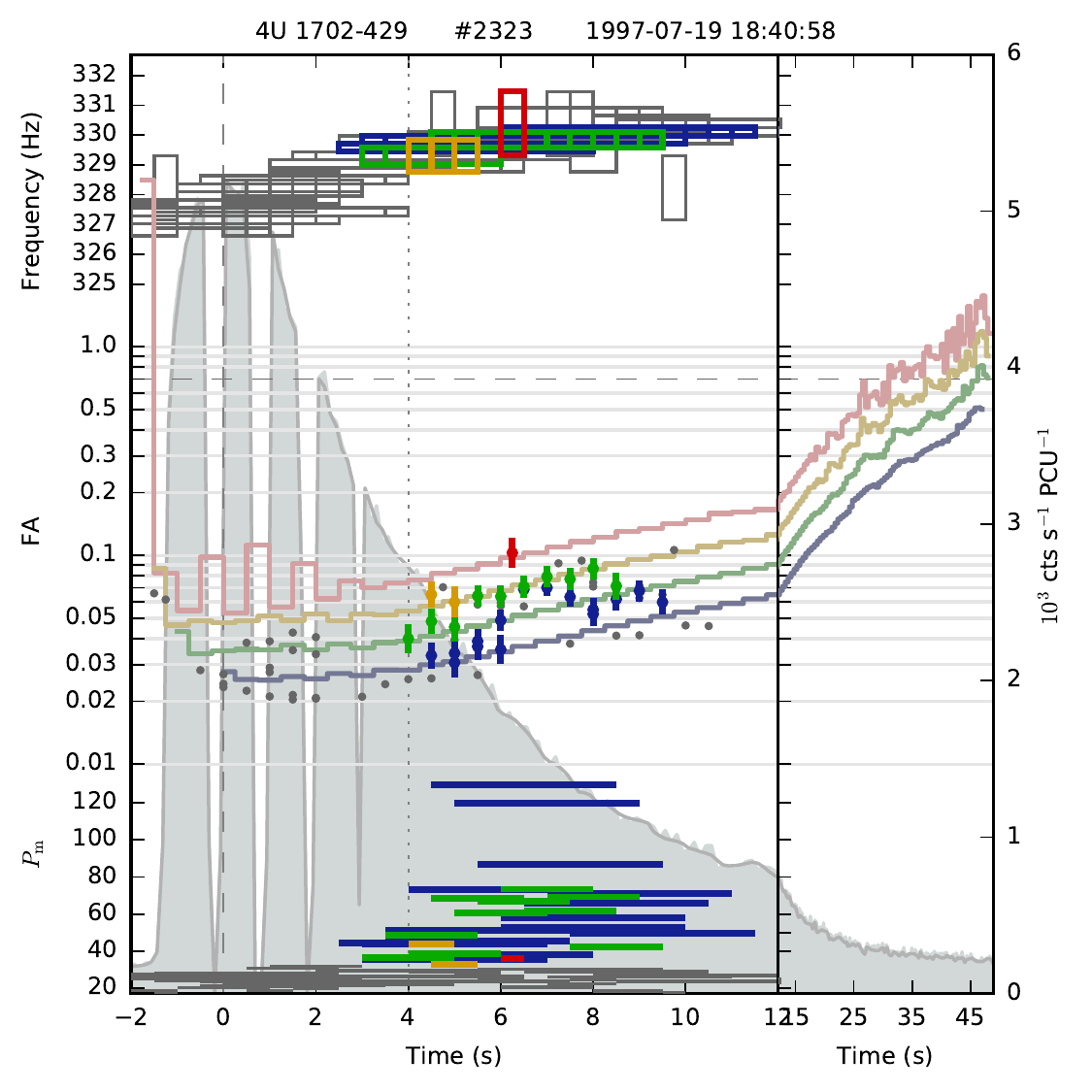}
   \caption{Example of a burst from 4U~1702$-$429 with multiple sub-threshold candidates. }
 \label{fig:4U1702}
\end{figure}

Oscillations around 329\,Hz were discovered by \citet{Markwardt1999} in 5 out of 6 
bursts observed at the time. The TBO frequency was gradually increasing during all 
bursts, and the reported FAs ranged from a few \% up to 18\%, 

Our sample contains 50 bursts from this source, some of them with gaps and noise power 
as low as 1.8. Among these, 32 yielded TBO signals at 326.00--330.5\,Hz in the R and 
B regions.

FA of approximately 3--15\% broadly match the valued reported by \citet{Ootes2017} and 
\citet{Galloway2008}. Some of the bursts have detections in one independent window, with or
without multiple sub-threshold detections in independent windows. There are also many 
sub-threshold detections from bursts with strong TBOs (e.g. Fig.~\ref{fig:4U1702}).
We see large gradual frequency rises; however some of this frequency evolution may be 
biased by the rapid variation of the count rate during the burst rise or gaps.

In addition to TBOs, our burst sample yielded some type I low-frequency candidates at 
2--3\,Hz due to the unmodeled spikes on the rise of several bursts, and statistically
insignificant number of noise candidates.

\subsubsection{IGR J17191$-$2821}

\begin{figure}
 \centering
 \includegraphics[scale=0.72]{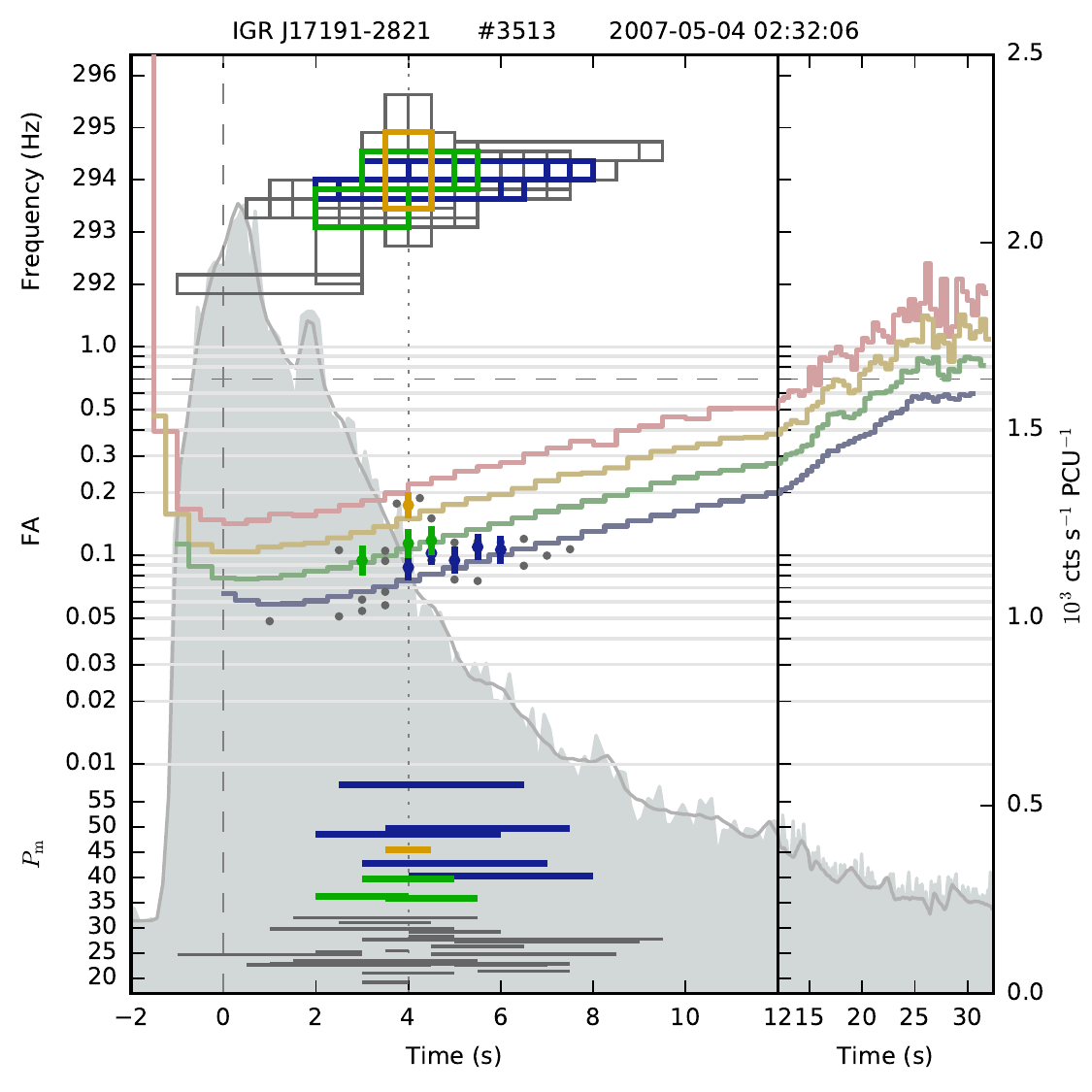}
   \caption{An example of TBOs from IGR J17191$-$282 with their typical large and gradual 
   frequency evolution. }
 \label{fig:IGRJ17191}
\end{figure}

TBOs at 294\,Hz were discovered by \citet{Altamirano2010a} in three bursts out of 
five observed (one of them showed significant oscillations only in part of the energy 
band). Two bursts exhibited a large gradual frequency drift, 2--3\,Hz over about 10\,s. 
The authors reported 5--10\% rms amplitude in the 2--17\,keV energy range. 

Our sample consists of the same bursts as in \citet{Altamirano2010a}. We detect TBOs in two 
bursts at the same frequency in the B region. The burst with the weakest TBOs from 
\citet{Altamirano2010a} had sub-threshold candidates in the  TBO frequency range. The 
FAs of the detections are broadly similar to the ones measured by \citet{Altamirano2010a}, 
despite the differences in time window sizes and energy cuts. For burst \#3513 
(Fig.~\ref{fig:IGRJ17191}), we do not record TBOs closer to the burst rise, having only 
one sub-threshold candidate there.

In addition to TBOs, one type I low-frequency candidate was detected.

\subsubsection{4U 1728$-$34}

\begin{figure}
 \centering
 \includegraphics[scale=0.72]{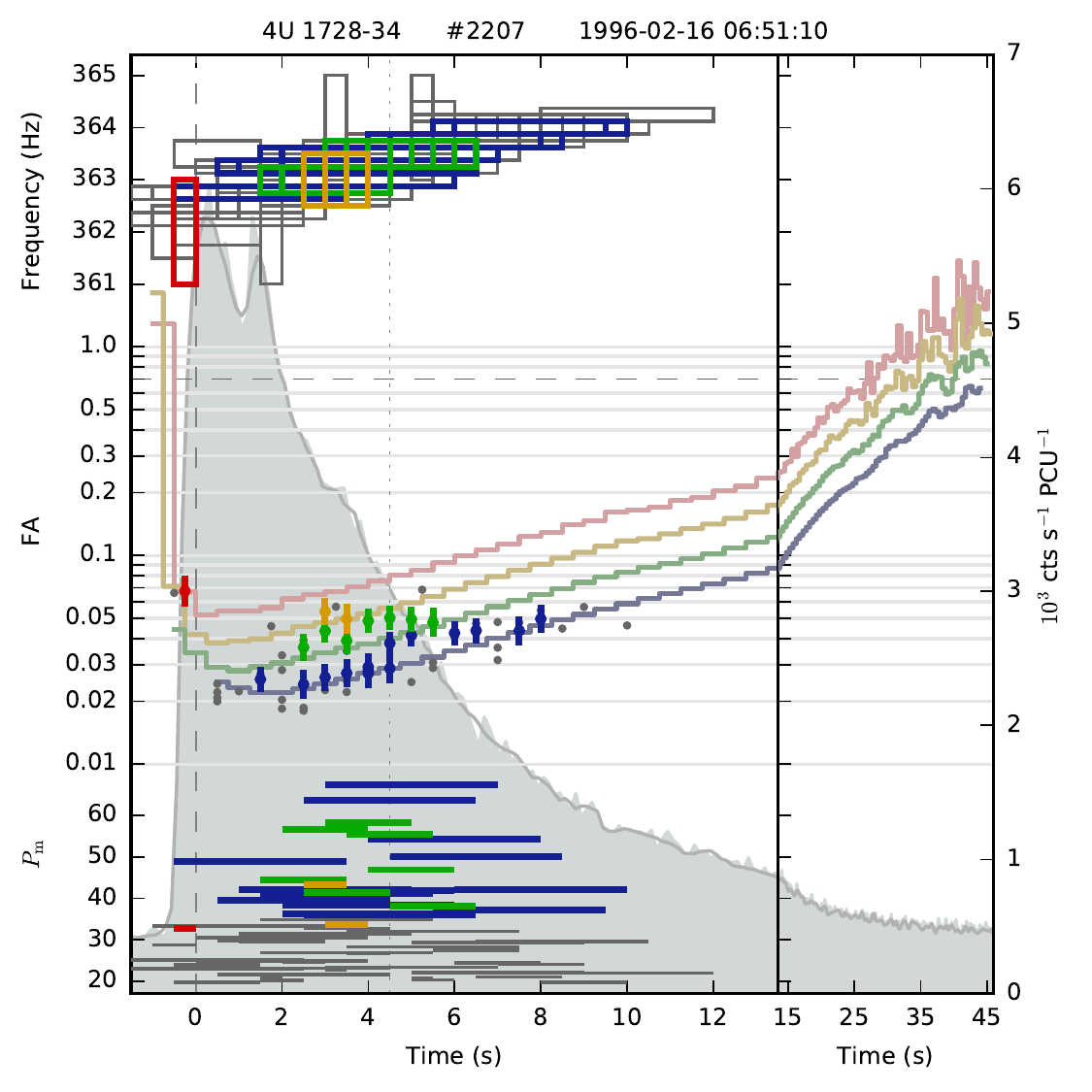}
   \caption{Example of TBOs from 4U~1728$-$34, showing typical gradual frequency drift.}
 \label{fig:4U1728}
\end{figure}

TBOs from 4U 1728$-$34 were discovered by \citet{Strohmayer1996} at 363--364\,Hz. 
The oscillations are characterized by a gradual few-Hz upward frequency drift and FAs 
as as large as 10\%.  A larger sample of bursts were subsequently searched for TBOs 
by \citet{vanStraaten2001}, \citet{Galloway2008}, and \citet{Ootes2017}.

Our sample contains 141 bursts from this source. Some of them have data gaps and average noise 
power as low as 1.8.  Thirty-four bursts yielded TBO candidates in the frequency range of
362--364 Hz. All TBO candidates except one were detected in the R and B regions. The only 
one from the T region is on its left edge. Some of the bursts show detections in multiple 
independent time windows with gradual frequency drift over the course of the TBO train 
(Fig.~\ref{fig:4U1728}). Sometimes the frequency evolution is biased by data gaps. 
Some of the bursts have fainter detections in one independent time window, on the rise or in 
the B region.

FAs of the TBO candidates are broadly comparable to the values reported in \citet{Galloway2008}
and \citet{Ootes2017}, but are consistently smaller than the ones in  \citet{vanStraaten2001}, 
by a factor of about 1.5 although the FA evolution throughout the burst is similar. This is 
explained by the differences in data processing: \citeauthor{vanStraaten2001} added power from 
several frequency bins in 4-Hz windows around maximum power. 

In addition to TBO candidates, our analysis yielded multiple low-frequency candidates, both 
type I and type II, and statistically insignificant number of noise candidates.

\subsubsection{KS 1731$-$260}

\begin{figure}
 \centering
 \includegraphics[scale=0.72]{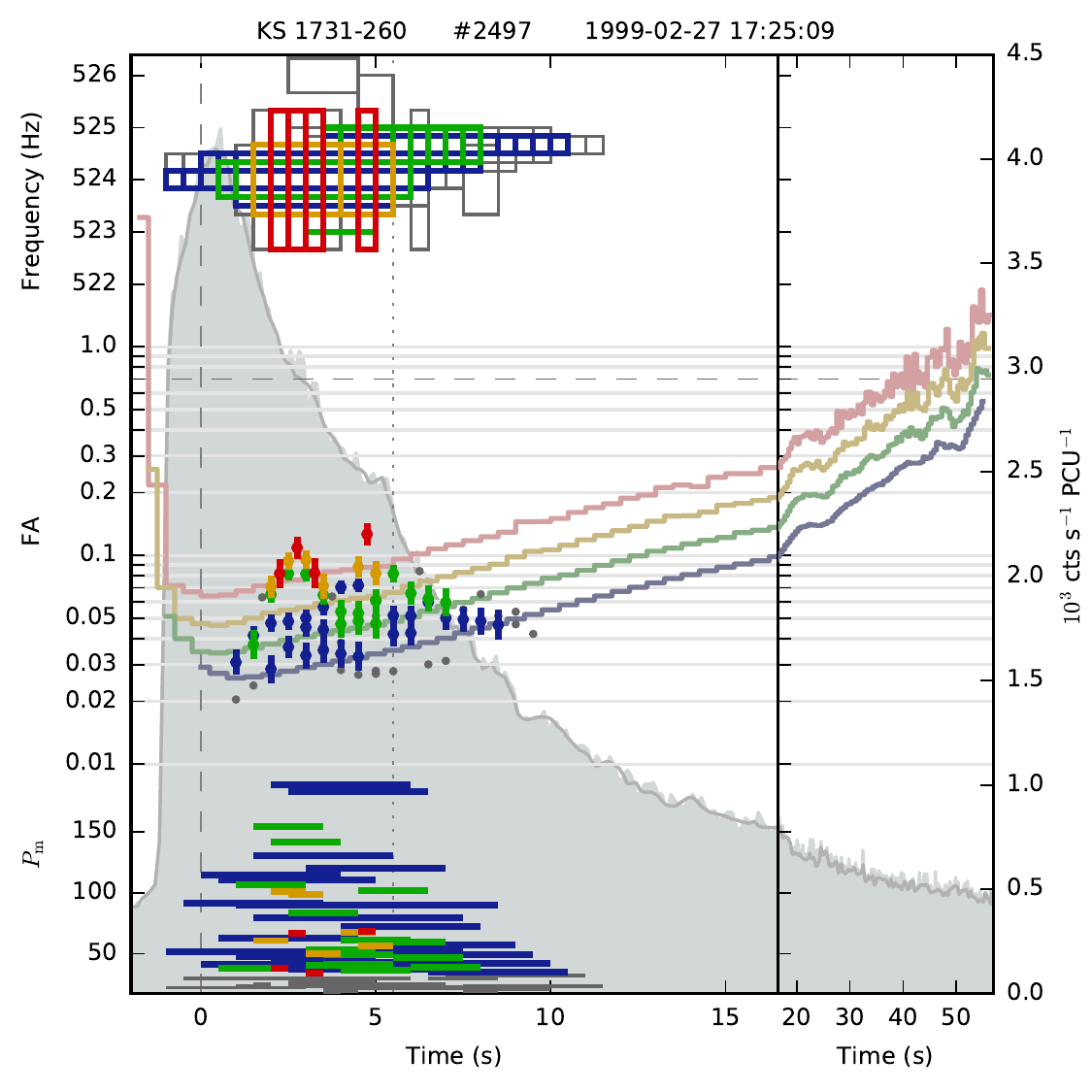}
   \caption{The burst with the strongest TBOs from KS 1731$-$260. }
 \label{fig:KS1731}
\end{figure}

Oscillations at 523.93\,Hz were discovered by \citet{Smith1997} in the single burst observed at 
the time. Later, \Gwt\ searched for TBOs in 26 more bursts and found them in three bursts. 
\citet{Ootes2017}, using different window sizes, also found oscillations in six bursts out of 27.

Because of the GTI requirement, our sample consisted of 26 bursts. TBOs were detected in three 
of them, at frequencies of 523.5--524.25\,Hz, all of them right after the burst peak (e.g. Fig.~\ref{fig:KS1731}). 
Two bursts yielded detections in multiple independent time windows, one in a single time window but 
with more independent sub-threshold candidates on the rise. One more burst had sub-threshold 
candidates only. FAs of 4--14\%  are broadly consistent with the values reported by 
\citet{Galloway2008} and \citet{Ootes2017}. 

The source yielded also a small, statistically insignificant number of noise candidates.

\subsubsection{GRS 1741.9$-$2853}

\begin{figure}
 \centering
 \includegraphics[scale=0.72]{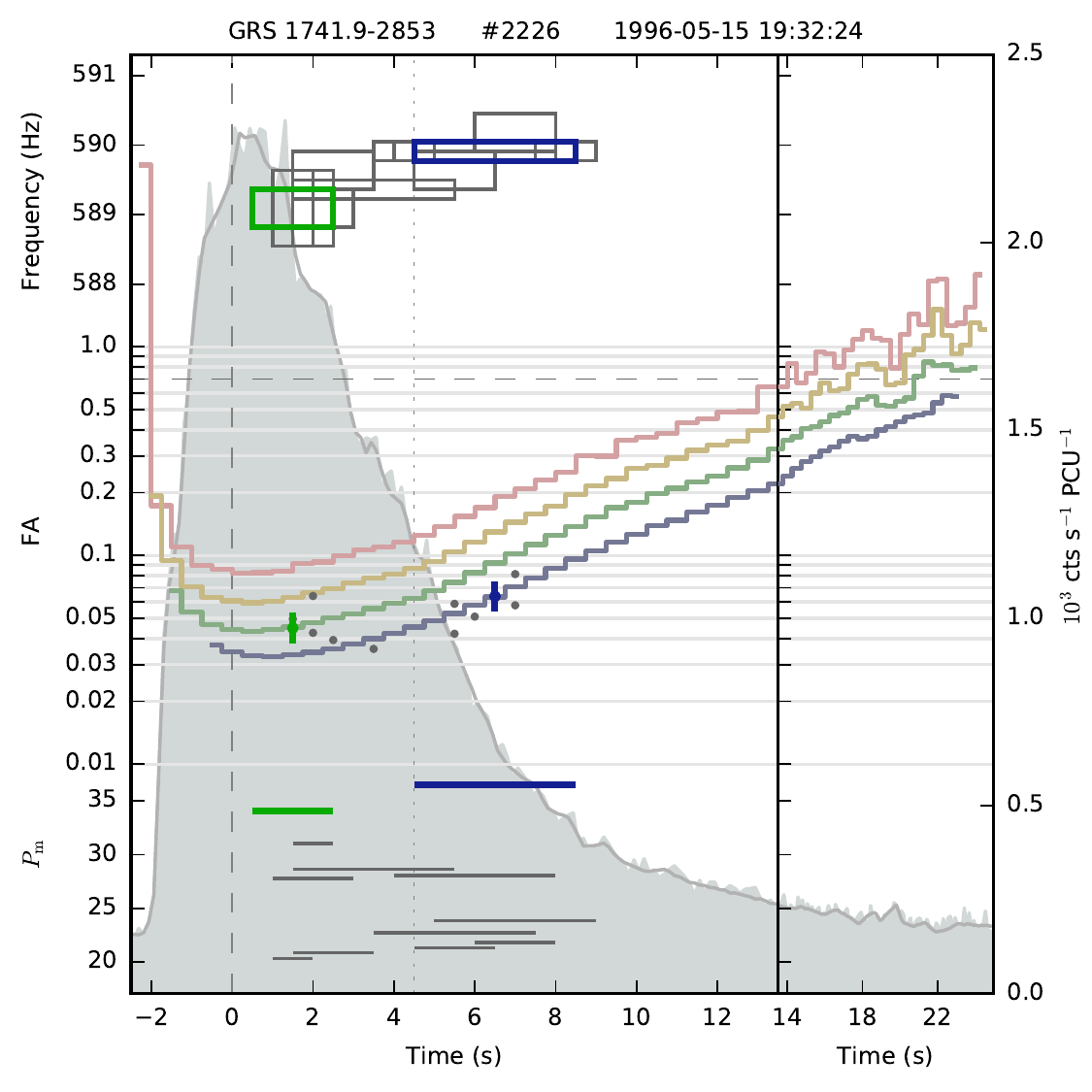}
   \caption{Example of oscillations from a faint TBO source GRS 1741.9-2853. }
 \label{fig:GRS1741}
\end{figure}

\citet{Strohmayer1997a} reported on 589-Hz TBOs in three bursts from GRS 1741.9$-$2853. The FA of 
detections were up to 13\%, but these were for favorable energy cuts and custom time window intervals. 
The source was not in outburst again before the end of the RXTE mission. In 2013, \citet{Barriere2015} 
observed GRS 1741.9$-$2853 with NuSTAR. Unfortunately, the 2.5-ms dead time of NuSTAR hindered 
TBO detection.  No oscillations were found, with the upper limits from simulations of the
injected signals being higher than the detections in \citet{Strohmayer1997a}.

Our sample consists of seven bursts. We detected TBO candidates at 589.00--589.75\,Hz. 
For both bursts, the candidates came from the B region, 
with FA of about 5\%, broadly comparable with the values reported in \citet{Galloway2008}. 
In one burst, candidates were detected in two independent time windows, but of different length and 
not at the same frequency (Fig.~\ref{fig:GRS1741}). The other burst yielded a detection in one 
independent window and another independent sub-threshold candidate.
The third burst with TBOs from \citet{Strohmayer1997a} was not covered by GTI.

The candidates aroung 589.5\,Hz are not strong, with $P_\mathrm{m}<40$. 
All three independent-window candidates come from different frequencies, but the spread is smaller than 1\,Hz. 
For the strongest candidate, 5\% of simulations yielded the same or a larger number of candidates
with equal or larger $P_\mathrm{m}$. However, selecting all candidates above threshold yields
$p=0$. The fact that the candidates are grouped in frequency speaks in favor of their TBO nature,
however a strict estimate of the significance of this grouping is beyond the scope of this paper.

In addition to TBO candidates, GRS 1741.9$-$2853 has some type II low-frequency candidates in 
the 2--4 Hz range, and one noise candidate ($p=0.3$) at 1829.25\,Hz.

\subsubsection{IGR J17480$-$2446}
\begin{figure}
 \centering
 \includegraphics[scale=0.72]{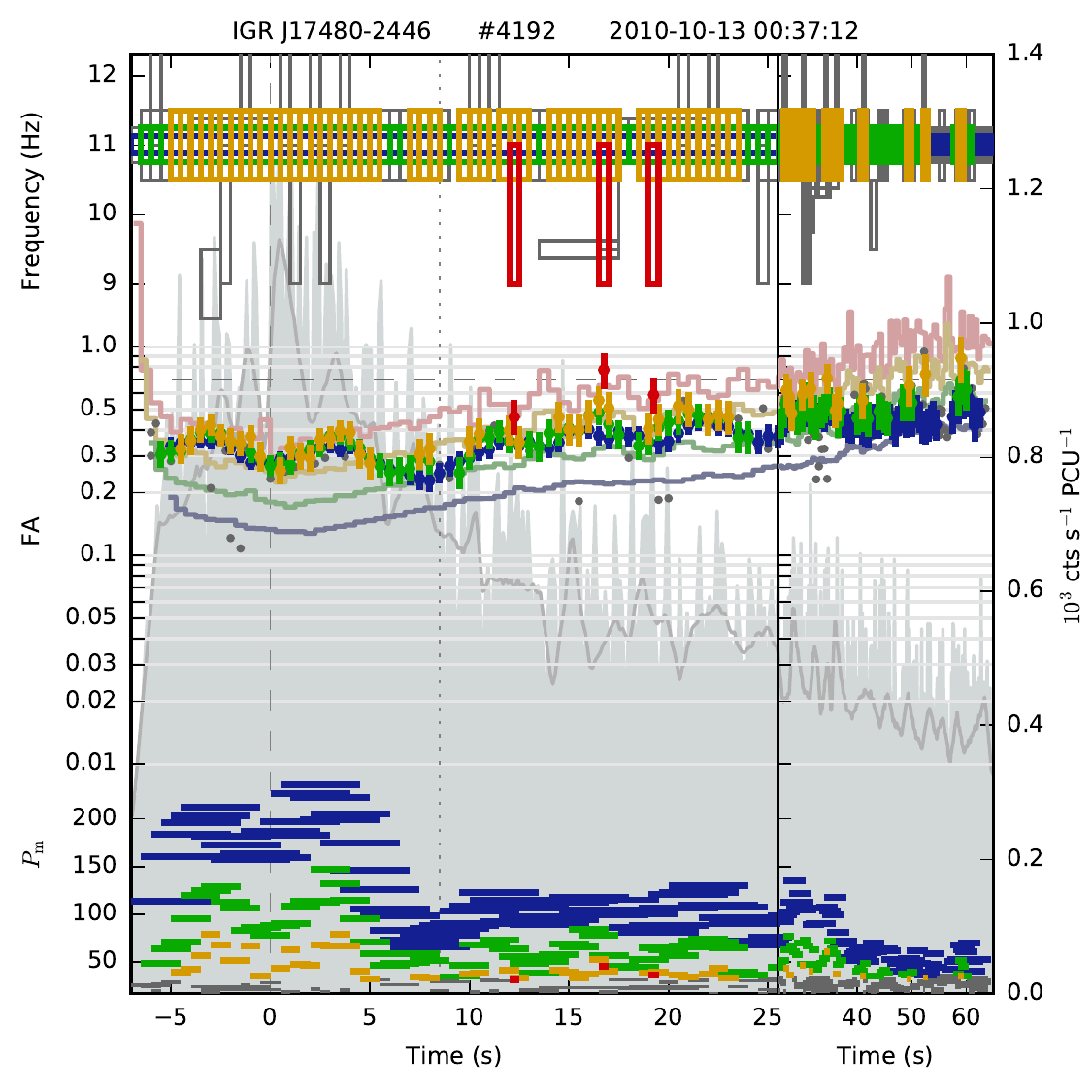}
   \caption{A burst with strong TBOs from IGR J17480$-$2446. The constant frequency and roughly 
   constant FAs make these oscillations similar to the ones from accreting MSPs XTE J1814$-$338 
   and IGR J17511$-$3057, although those pulsars spin an order of magnitude faster.}
 \label{fig:IGRJ17480}
\end{figure}

TBOs from an unusually slowly spinning accreting pulsar IGR J17480$-$2446 were discovered by \citet{Cavecchi2011}. 
Very strong 11-Hz oscillations were detected in one burst, with FAs of 30\% and no frequency drift. The
remaining 230 bursts explored by \citet{Cavecchi2011}  also yielded TBOs in FFT windows from 10 to 300\,s, 
with FAs down to 3\%. The search was conducted on barycentered data using the APP ephemeris. 

Our sample contained 297 bursts, with median peak S/N of only 6.4. Most of the times for the on-burst 
windows were set manually and were short, of about 5\,s. Our FFT windows are shorter and the upper 
limits on FA are consequently much larger than in \citet{Cavecchi2011}; we find characteristic upper 
limits on the FA on the order of 70\%.   Using our analysis procedure, only two bursts had candidates 
in the range of 10--11 Hz. Similarly to XTE J1814$-$338 and IGR J17511$-$3057, the range of TBO frequencies
reflects the coarseness of the Fourier grid, with detections at 10\,Hz coming from 0.5-s windows.

One of the bursts (\#4192) had strong oscillations throughout the entire on-burst window 
(Fig.~\ref{fig:IGRJ17480}), with typical FA of 30\%, maximum up to 90\%. The average FA on 10--20-s 
timescales would have matched the one reported in \citet{Cavecchi2011}. Another yielded a relatively faint 
($P_\mathrm{m}\approx40$) detection in a single independent 4-s window (more if one considers 
sub-threshold candidates),  For this burst, calculated FAs of $\sim 90$\% are most probably affected by an
overestimated pre-burst background level.

Burst \#4192 had several low-frequency (2--3\,Hz) candidates of type II. Overall, the burst sample 
yielded a large but statistically insignificant number of noise candidates.

\subsubsection{IGR J17498-2921}

\begin{figure}
 \centering
 \includegraphics[scale=0.72]{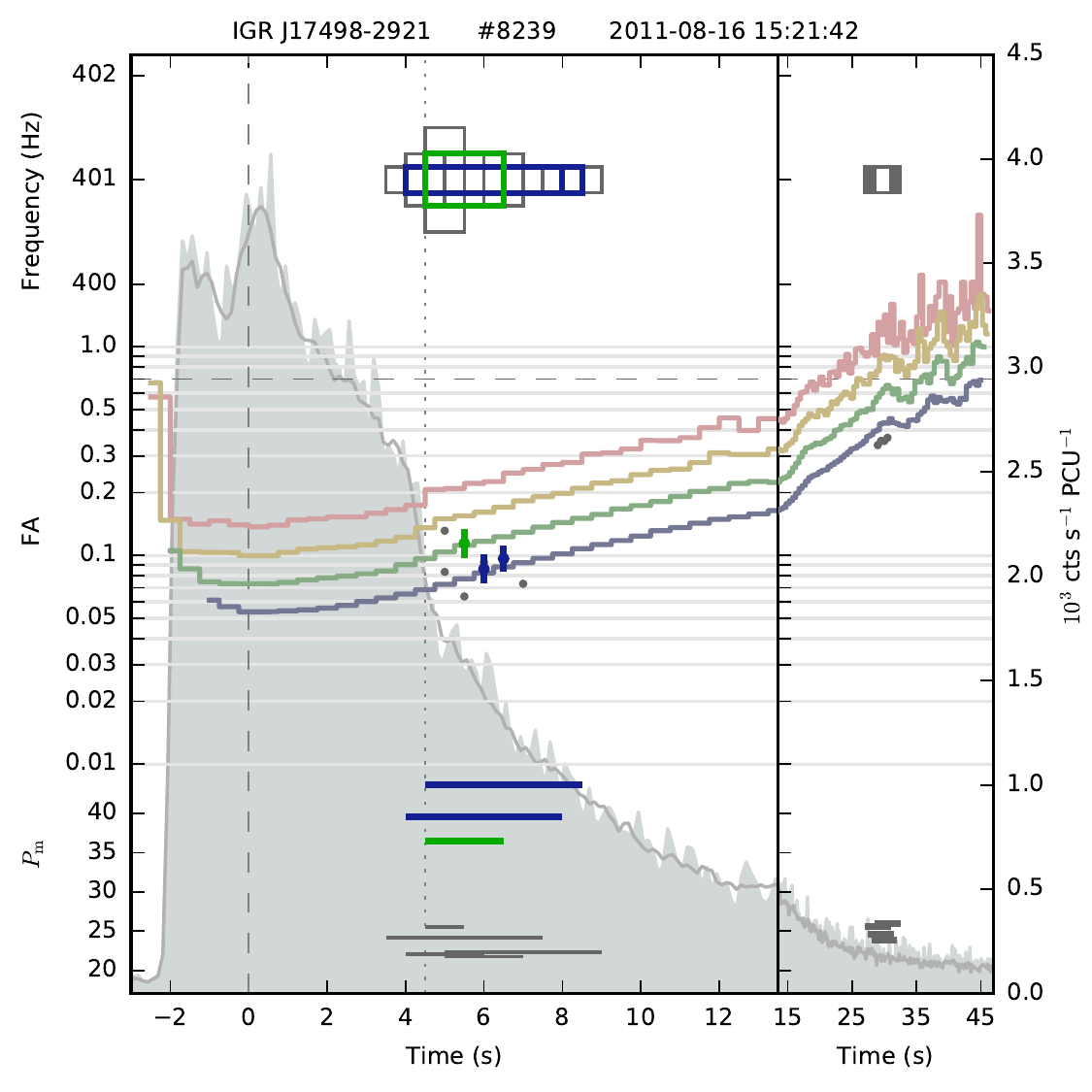}
   \caption{Oscillations from a rather faint TBO source, the accreting millisecond pulsar IGR J17498$-$2921.}
 \label{fig:IGRJ17498}
\end{figure}

IGR J17498$-$2921 is an accreting MSP with TBOs at 401\,Hz discovered by \citet{Linares2011}. 
\citet{Chakraborty2012} analysed 12 bursts from IGR J17498$-$2921 and detected TBOs from two bursts 
in averaged 1-s spectra.  The PCA field of view contains several other bursters and the ten bursts 
without oscillations may be from another source, however the authors argue that this is unlikely. 

The MINBAR catalogue lists only two bursts from IGR J17498$-$2921. From both bursts we detected 
TBOs in the B region, without frequency drift. There were also some sub-threshold candidates 
on the tail.  The FAs of 10\% were consistent with \citet{Chakraborty2012}. 
One burst had two independent-window detections, the other only one independent detection, 
but with sub-threshold candidates at the same frequency in the tail 
(FA of 30\%, Fig.~\ref{fig:IGRJ17498}). In addition to TBOs, only one low-frequency 
candidate was detected. 

\subsubsection{SAX J1750.8$-$2900}

\begin{figure}
 \centering
 \includegraphics[scale=0.72]{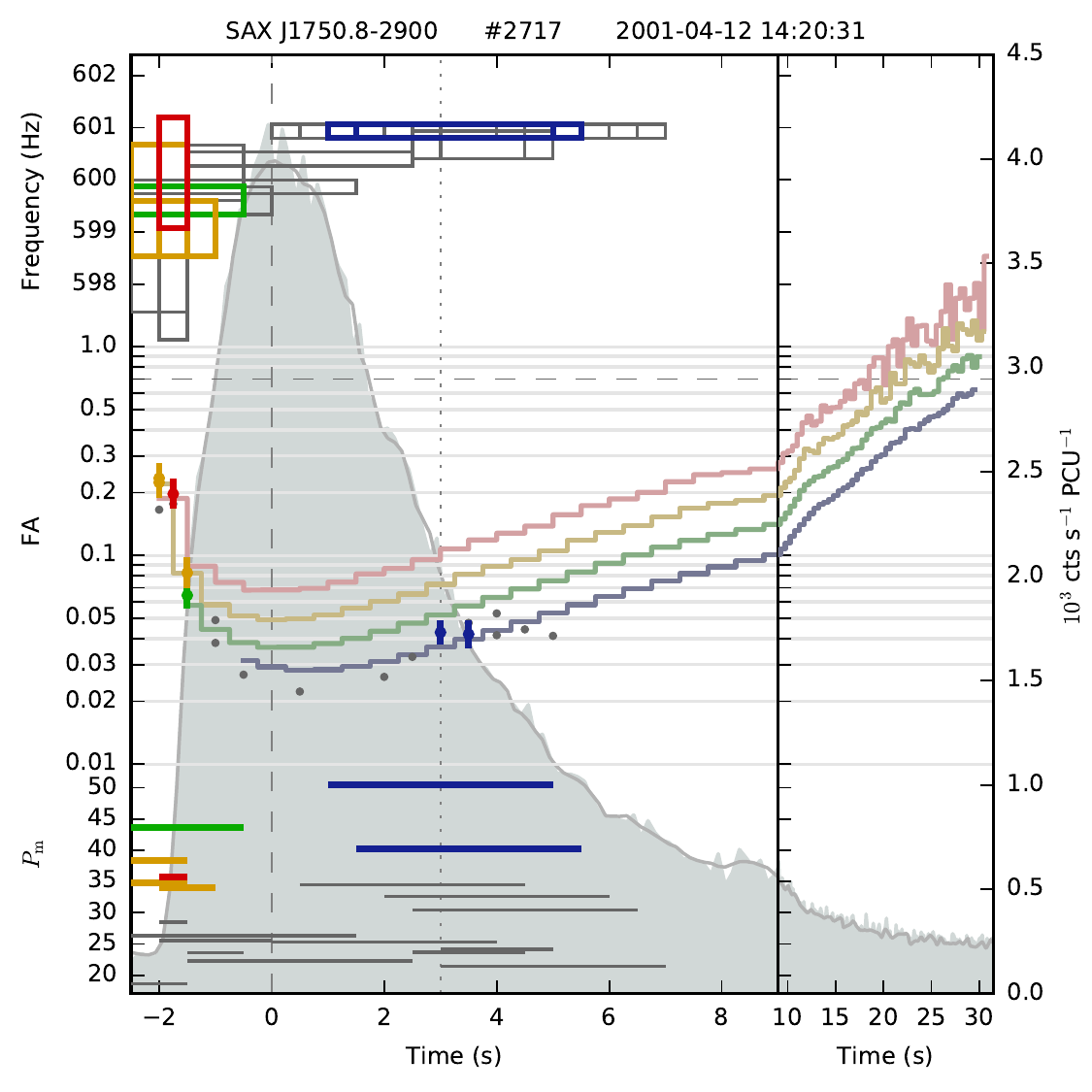}
   \caption{The only burst from SAX J1750.8$-$2900 with TBOs above the detection threshold.}
 \label{fig:SAXJ1750}
\end{figure}

TBOs at 601\,Hz were discovered by \citet{Kaaret2002}  in one of the four bursts studied. 
The authors used merged event lists from the event and burst catcher modes, without any 
energy selection and searched for signal in 4-s windows overlapping by 0.125\,s in the 
frequency range between 200 and 1200\,Hz. 
TBOs were found in both the rise and decay, with a maximum power of 49.3 five seconds 
after the burst rise. No fractional amplitudes were reported. 

We detected TBOs from the same burst in two independent time windows, with similar maximum 
power, five seconds after the burst rise. Similarly to \citet{Kaaret2002},
we record frequency drift and the disappearance of the TBO signal during the burst peak. 
TBOs on the rise occur in shorter windows and have
larger FAs than TBOs right after the burst peak (Fig.~\ref{fig:SAXJ1750}).

\Gwt\ found TBOs on the rise of two more bursts. These bursts have only sub-threshold 
candidates in our analysis, with FA similar to the ones measured by \Gwt. The first burst 
has frequency behavior similar to Burst \#2717, whilst TBOs from 
the second one do not have noticeable frequency drift and also appear at burst peak. 

In 2008, SAX J1750.8$-$2900 went into outburst again, adding two more bursts to the 
MINBAR sample. No TBO candidates were recorded from these bursts, even at the sub-threshold level. 

For the strongest candidate, none of the simulations have candidates of similar strength.
For both candidate groups, before and after the burst peak the peak power was over 40. The 
highest $P_\mathrm{m}$ outside the frequency region around 600\,Hz was smaller than 30. Despite 
the absence of candidates detected in independent time windows at the same frequency, the power 
of the candidates, the close
proximity of their frequencies and the presence of sub-threshold candidates in two more bursts speak in
favour of these candidates being TBOs.

No low-frequency or noise candidates were detected from this source.

\subsubsection{IGR J17511$-$3057}

\begin{figure}
 \centering
 \includegraphics[scale=0.72]{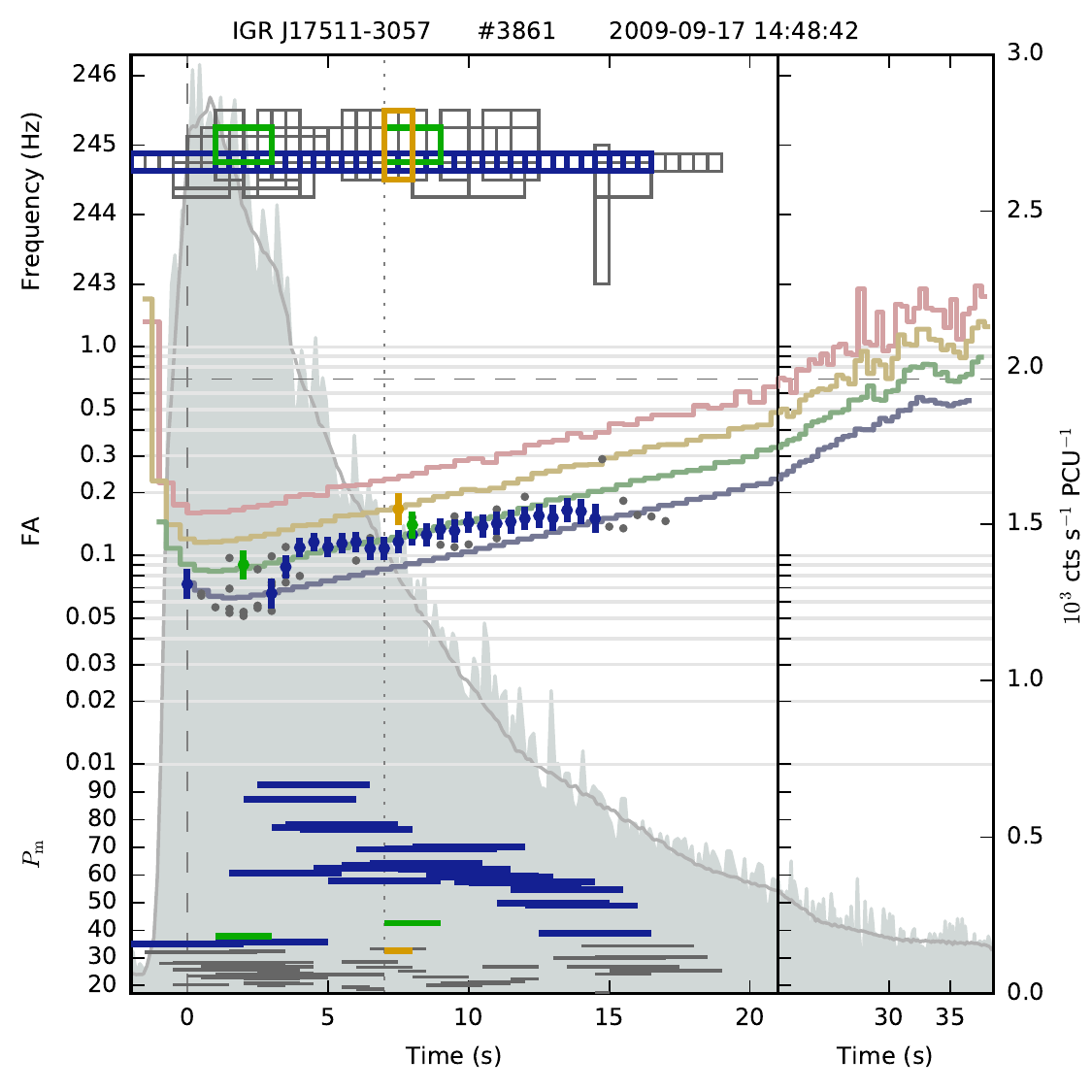}
   \caption{An example of TBOs from IGR J17511$-$3057, showing the main features of its TBOs: constant 
   frequency and dip in FAs during the burst peak. }
 \label{fig:IGRJ17511}
\end{figure}

IGR J17511$-$3057 is an accreting MSP with TBOs at 245\,Hz discovered by \citet{Altamirano2010b}. 
Burst oscillations were seen in all bursts in the sample. For fainter bursts, the oscillations are 
detected earlier in the burst. For brighter bursts, TBOs often disappear at the burst peak. 
The authors note small (0.1\,Hz) frequency drift on the rise and report FAs of 5--12\%, with 
FAs on the tail larger than on the rise and peak. 

The MINBAR database lists 9 bursts for this source, all of them with TBOs in the same regions 
as in \citet{Altamirano2010b}. One of the bursts has TBOs in one independent time window only. 
We do not observe any frequency drift on the rise, although our Fourier frequency resolution 
is rather coarse. Similarly to XTE J1814$-$338, the frequency range of the detections reflects 
the coarseness of the Fourier frequency grid.

Like \citet{Altamirano2010b}, we note a dip in FAs during burst peaks (Fig.~\ref{fig:IGRJ17511}), 
although in a smaller number of bursts than they do.  Our FAs are also consistent with 
\citet{Altamirano2010b}.

No low-frequency or noise candidates were detected from this source.

\subsubsection{SAX J1808.4$-$3658}
\begin{figure}
 \centering
 \includegraphics[scale=0.72]{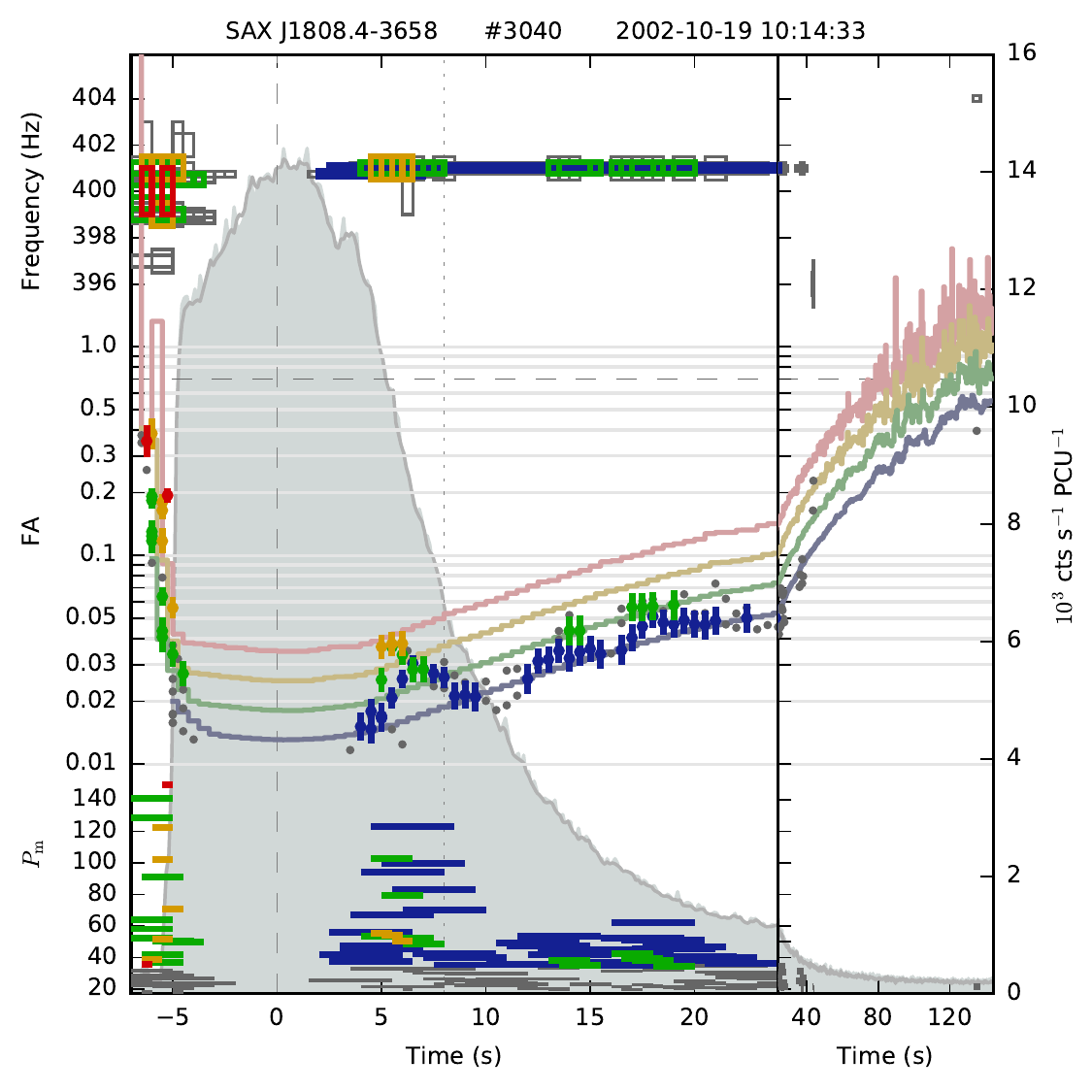}
 \includegraphics[scale=0.72]{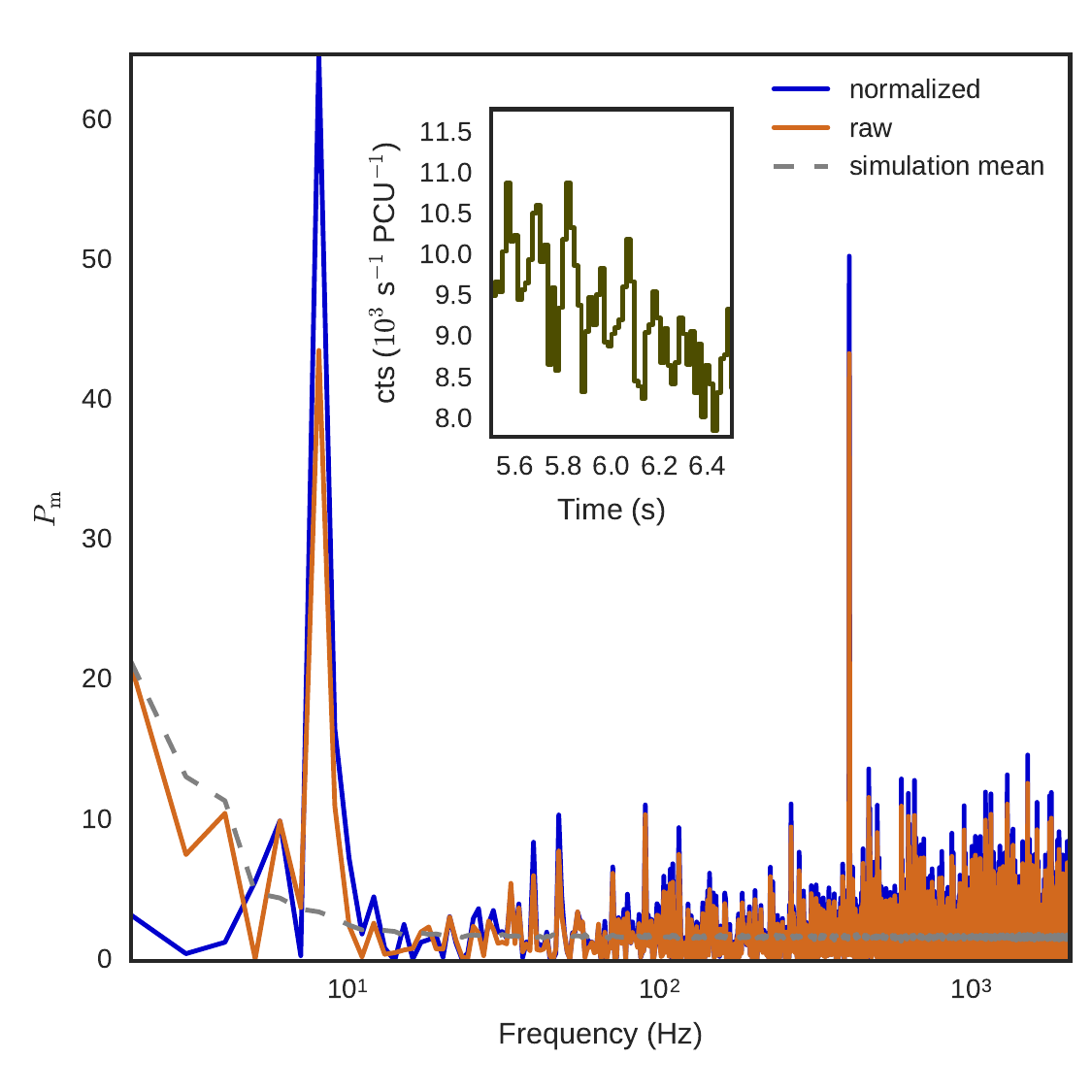}
   \caption{\textit{Top:} TBOs from J1808.4$-$3658, showing rapid frequency drift at burst 
   onset, disappearance at burst peak and stable frequency at the burst tail. \textit{Bottom:} 
   1-D power spectrum in 1-s windows starting $\sim5.5$\,s after burst peak
   in the same burst. In addition to the TBO, a strong glimmer candidate is evident at 8\,Hz. 
   The inset shows the LC in the same 1-s window. The oscillation with 8 periods per second is 
   visible. The consistent offset between the normalized and raw PS is due to the large 
   influence of dead time.}
 \label{fig:SAXJ808}
\end{figure}

SAX J1808.4$-$3658 is an accreting MSP with APPs and TBOs at 401\,Hz  
\citep{Wijnands1998,Chakrabarty2003}.

\MB\ lists 9 Type I bursts, three of which have not been analysed for TBO 
behaviour before. We find TBOs in seven bursts. The source is very bright and some 
observing sessions suffer from data gaps. Typical behavior is as follows: TBOs start 
at the burst onset and rapidly drift in frequency up or down by a few Hz within a 
single FFT window of 0.5--1\,s (the amount of perceived drift may be biased by 
frequency covariance). The FAs on the rise are on order of 10--40\%. Oscillations 
disappear during the burst peak, even accounting for the dead time which
lowers the noise power to 1.6. Then oscillations reappear at frequencies slightly 
higher or lower and are fairly stable in frequency with wave-like variations of FAs, 
which at the same time increase slightly on the tail. The FAs after burst rise are 
on the order of few \%.  One burst did not show oscillations on the rise, another 
had only sub-threshold candidates on the rise. Such TBO properties are consistent 
with ones reported previously \citep{Chakrabarty2003,Bhattacharyya2006c,Bhattacharyya2007a,Galloway2008}

One of the bursts has type II low-frequency noise, a strong TBO, and a peculiar 
low-frequency glimmer candidate at 8\,Hz, with peak power exceeding 60 
(Fig.~\ref{fig:SAXJ808}). Note that the standard detection threshold yields $p=0.22$, 
however the power of glimmer candidate is much larger than any power in simulated datasets.
The oscillation profile folded with 8-Hz frequency 
has a sinusoidal shape.

\subsubsection{XTE J1814$-$338}

\begin{figure}
 \centering
 \includegraphics[scale=0.72]{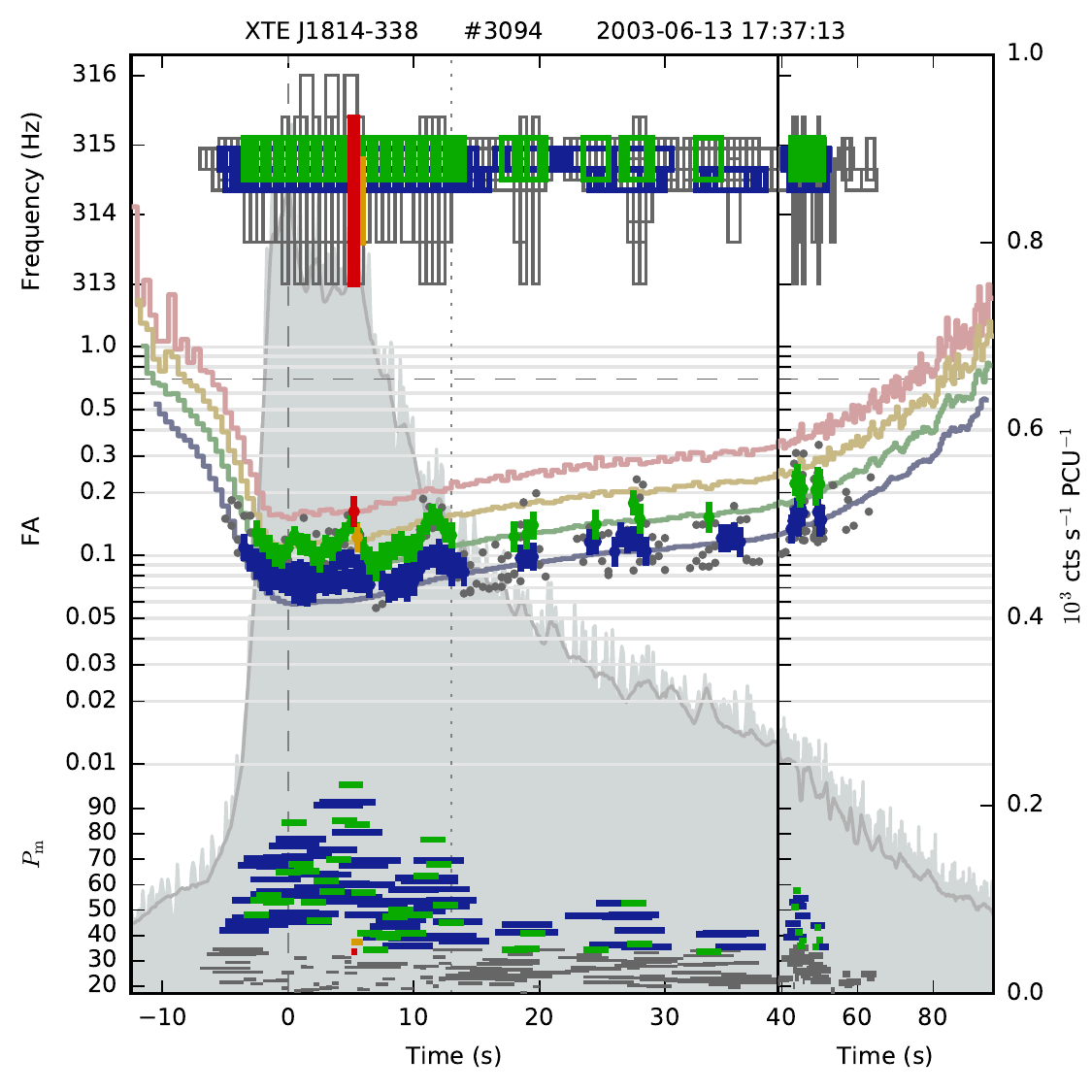}
   \caption{An example of TBOs from XTE J1814$-$338, showing the common features of its TBOs: constant 
   frequency, presence throughout the tail, large-scale FA constancy and small-scale wave-like 
   variations in FAs.}
 \label{fig:XTEJ1814}
\end{figure}

The accreting millisecond pulsar XTE J1814$-$338 is one of the best studied TBO sources 
\citep{Strohmayer2003,Watts2005,Watts2006,Watts2008}.
Our sample consisted of 28 bursts, all of which have been studied before. We detect TBOs with 
power above the threshold in 26 bursts. The two remaining bursts did not have 
GTI coverage during the burst rise and peak, but had weak sub-threshold candidates during the 
tail. 

The oscillation frequency (314\,Hz) does not change by more than the Fourier frequency 
resolution during the burst; the apparent frequency range in Table~\ref{table:det} 
reflects the coarseness of the Fourier frequency grid.

The oscillations do not disappear during the burst peak and are often present in the 
burst tail. There are also many sub-threshold candidates. FAs tend to be roughly constant 
throughout the duration of the oscillation, until it disappears under 
the rising $\FAup$ (Fig.~\ref{fig:XTEJ1814}). On top of the constant level, there are e
vident wave-like variations of FAs. 
In general, our measurements of FAs of $\sim10$\% are broadly consistent with 
\citet{Galloway2008,Watts2008}.

\subsubsection{HETE J1900.1$-$2455}

\begin{figure}
 \centering
 \includegraphics[scale=0.72]{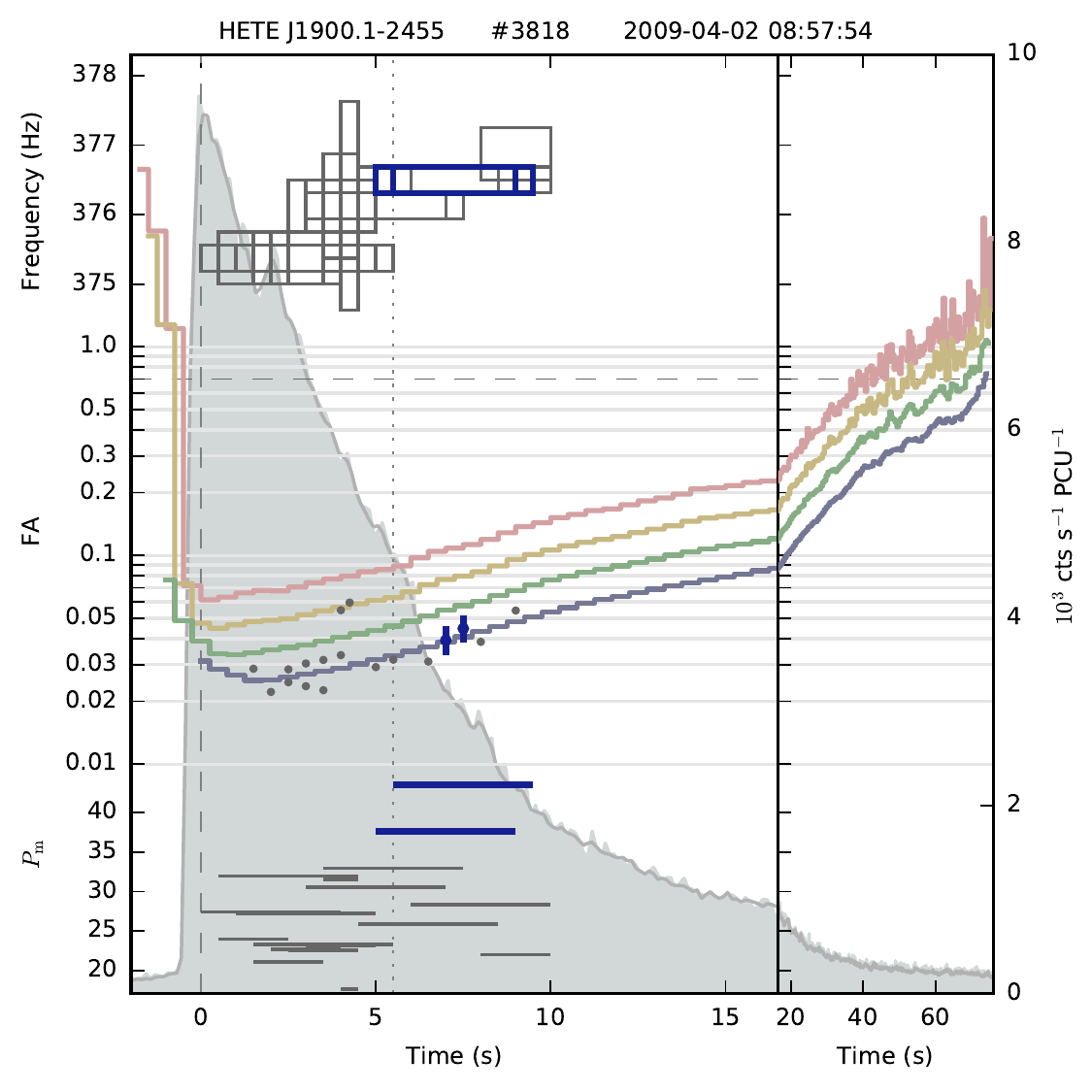}
   \caption{The only known burst with oscillations from the accreting MSP HETE J1900.1$-$2455. This candidate would not
    be significant by our blind (broad frequency range) search.}
 \label{fig:HETE1900}
\end{figure}

HETE J1900.1$-$2455 is an accreting MSP with spin frequency of $\sim377$\,Hz \citep{Patruno2012}. 
TBOs in a single burst were discovered by  by \citet{Watts2009}. The authors detected significant 
signal in four consecutive independent 2-s windows for 2--30 keV photons.
The reported FAs were 3.5\% in the same energy range. 

Our sample consisted of 8 bursts, one of them with data gaps. The bursts are quite bright 
and during the peak the mean noise power drops to 1.7. Despite accounting for the influence of dead 
time, our search did not result in any new TBO detections -- relatively
faint TBO candidates were detected at 376.25\,Hz only in the same burst as in \citet{Watts2009}. 
Only one independent time window yielded $P_\mathrm{m}$ above the detection 
threshold (Fig.~\ref{fig:HETE1900}), although sub-threshold candidates extend longer both in time 
and the frequency maps the same frequency evolution as in \citet{Watts2009}.  The FA was 4.5\%, 
comparable to the value in \citet{Watts2009}, noting that we have made different choices of 
windows and energy ranges.

HETE J1900.1$-$2455 is a good illustration of the advantages of using external information to 
narrow down the frequency range searched. 
TBOs at 376.25\,Hz would have been deemed insignificant by our broad frequency range analysis 
-- 3\% of our simulation runs had at least one candidate with probability equal to or smaller 
than the probability of the TBO candidate.

The source also has some low-frequency noise and an insignificant number of noise candidates.

\subsubsection{Aql X-1}

\begin{figure}
 \centering
 \includegraphics[scale=0.72]{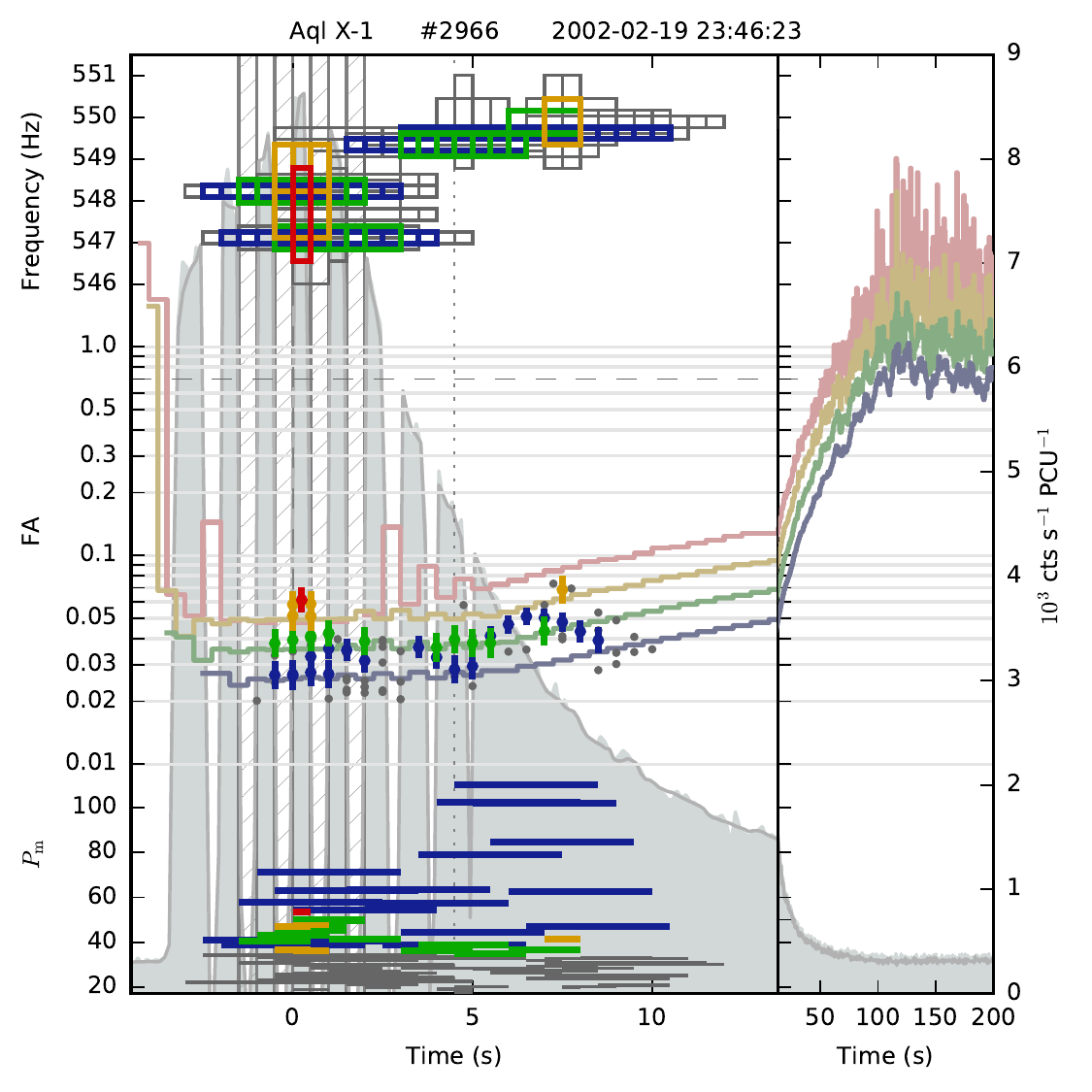}
   \caption{Example of a burst from Aql X-1, with data gaps distorting the observed TBO frequency evolution. }
 \label{fig:AqlX1}
\end{figure}

TBOs at 549\,Hz were discovered by \citet{Zhang1998} in RXTE Burst Catcher data. Later, 
\citet{Casella2008} reported on strong ($P_\mathrm{m}=120$) APPs at 550.27\,Hz in a single 
150-s time window which was not close to any burst. 

Our sample consisted of 73 bursts, about half of them suffering from data gaps (e.g. Fig.~\ref{fig:AqlX1}). Noise power
at the burst peak often drops to as low as 1.7. TBOs were detected in 8 bursts in  the R and 
B regions in the frequency range 547.4--550\,Hz. 

For several bursts, TBOs were detected in one independent window only. For brighter TBOs, there 
is a hint of a gradual frequency drift ($\sim1$\,Hz over few seconds), although gaps in the 
data have a large adverse effect on the observed frequencies. FAs of TBOs are on the order 
of 4--7\%,  broadly consistent with the values reported in \citet{Zhang1998}, \citet{Galloway2008} 
and \citet{Ootes2017}.

In addition to TBO candidates, we detected multiple low-frequency candidates at 2--6\,Hz and 
an insignificant number of noise candidates.

\subsection{Individual sources with tentative TBO detections reported by previous papers}
\label{subsec:tentTBO}

In this Section we discuss sources for which TBO detections have been claimed, or tentatively claimed, by previous 
works and which were classified as tentative in the review article by \citet{Watts2012}.  There are 8 sources in this 
category:  4U 0614+09, 4U 1254$-$69 (XB 1254$-$690), MXB 1730$-$335 (Rapid Burster), XTE J1739$-$285, 1A 1744$-$361, 
SAX J1748.9$-$2021, GS 1826$-$24 and XB 1916$-$053 (X 1916$-$53, 4U 1916$-$053).  

\subsubsection{4U 0614$+$09}

\citet{Strohmayer2008} found a 415-Hz signal in a 10-s window in the tail of one of the two bursts in 
their sample. The bursts were detected in 2006 and 2007 with Swift Burst Alert Telescope (BAT). 
The oscillations had FAs of 12.3\% and occurred in the 13--20\,keV energy range, in one of the 
10-s windows. The signal has $4\sigma$ significance assuming a conservative number of trials. 

The RXTE sample consists of only one burst, different to those observed by \citet{Strohmayer2008}. 
The burst was extremely bright, resulting in the telemetry rate being heavily saturated, which
caused large data gaps. The burst started with a very bright sub-second spike, followed by a gap,
which is an indication of PRE. This spike was not included in the on-burst window although it 
is part of the real burst rise. Our LC modeling did not reproduce a short gap at about 4\,s from 
the burst start, thus all candidates from the time windows covering that moment were discarded.   
We detected no candidates above the specified threshold. Our count rates imply much stronger 
upper limits on FA, around 2\%.

Our non-detection does not challenge the TBO claim of \citet{Strohmayer2008}, since even sources 
with strong TBO records have bursts that are apparently devoid of oscillations. However, it remains the 
case that TBOs were detected essentially only in one independent time window, from one burst.  
Detecting the oscillations at similar frequencies from more bursts would strengthen the conclusion.

\subsubsection{4U 1254$-$69 (XB 1254$-$690)}

\citet{Bhattacharyya2007b} reported on a tentative 95\,Hz candidate from the rising phase of 
one of the five bursts recorded. The $P_\mathrm{m}=24.3$ TBO candidate was found in 
the first 1-s interval after the burst start and had FA of $0.31\pm 0.07$. The signal was confined 
to a 1\,s time window and no significant frequency evolution was found, according to the authors. 
The significance was estimated to be 95\%, considering the number of harmonics and windows searched.

We have two more bursts compared to \citet{Bhattacharyya2007b}. Using our formal detection criterion, 
our analysis did not yield any candidates in the same data. However we did confirm similar powers 
($P_\mathrm{m}$ of about 25.5) at the same frequency of 95\,Hz.  The sub-threshold candidates at 
this frequency are  strongest in 1-s windows;  there are less significant sub-threshold candidates 
at other window sizes, but not in independent windows. The FA of the maximum-power signal 
is similar to the one in \citet{Bhattacharyya2007b}. 

Lowering the detection threshold to 25.43 on 1\,s window and correspondingly at other windows 
(multiplying the threshold probability by a factor of 30), we get five additional noise candidates ($p=0.3$), 
both below and above 1000\,Hz, all in the bursts analysed by \citet{Bhattacharyya2007b}. 
\citet{Bhattacharyya2007b} did not find any significant oscillations in other bursts up to 2048 Hz, 
however our data suggest otherwise. This may be explained by differences in the choice of windows 
and oversampling factor.

Some of our sub-threshold candidates also occur in 1-s time windows and are stronger than the power 
reported by \citet{Bhattacharyya2007b}. We conclude there is not enough evidence to classify the 95-Hz 
oscillation candidate as TBO and not as a noise candidate.

\subsubsection{MXB 1730$-$335 (Rapid Burster)}

\citet{Fox2001} described a tentative 306.5\,Hz TBO candidate in the sum of spectra from the rising 
part of 31 bursts, with 1.8\% of it being a chance detections according to their simulations. 
According to the authors, various tweaks to the data selection parameters affected the detection 
significance in different ways. The candidate was not detected in single bursts, and was not confirmed in 
two subsequent outbursts.

Our sample consists of 57 bursts, some of them with data gaps. We did not detect any candidates at 
frequencies close to 306.5\,Hz. The source yielded some type II low-frequency candidates within 
2--6\,Hz, and a large, but not significant number of noise candidates ($p=0.16$).
One of the candidates (at 18.25\,Hz) is quite strong, with $P_\mathrm{m}=46.5$ (Fig.~\ref{fig:glimmer}). 
This candidate has a power of 44.52 on non-normalized data. Only 2\% of simulation runs have one or 
more candidates of the same or larger significance.

\subsubsection{XTE J1739$-$285}
A tentative sub-ms oscillation from XTE J1739$-$285 was found by \citet{Kaaret2007}.
The authors reported a 1122-Hz candidate in one of the six bursts examined. The authors used a
different energy cut for PCU0 compared to the other PCUs in order to to minimize background, 
since PCU0 had recently lost its propane layer. \citeauthor{Kaaret2007} used 4-s FFTs with a 
0.125-s step. The maximum candidate power was  42. The significance of the candidate (equivalent 
to 3.97$\sigma$ of normal distribution) was estimated with simulations based on LC modeling, taking
into account dead time. The candidate was not confirmed by \citet{Galloway2008}, who used 
non-overlapping 4-s windows and potentially different energy cuts.

Our sample had the same bursts as in \citet{Kaaret2007} and \citet{Galloway2008}. We do not find 
any candidates from this source, having, on average 0.29 candidates from our simulation runs. 

The maximum power from the burst in \citet{Kaaret2007} 
is 29.5 in a 4-s window at the same frequency in a similar place during the burst tail. 
Increasing the detection threshold probability by a factor of 16.5 to match $P_\mathrm{m}=29.5$ 
(29.9 on non-normalized data) in 4-s windows yielded 5 additional candidates ($p=0.6$), some of which 
were more significant than the 1122-Hz one. 
Thus, in our analysis there is not enough evidence to classify the 1122-Hz candidate
as a TBO and not a noise candidate.

The tentative detection of \citet{Kaaret2007} presents an interesting case 
since it is rather strong, its significance was established using simulations but it was not
confirmed using different energy cuts and window overlap. It may be possible that the custom  
energy cuts \citet{Kaaret2007} were more sensitive to potential oscillations. However, 
it also may be the case that their noise model was not entirely correct (no analysis of model 
applicability was given) or that the detection is not related to TBOs (e.g. a glimmer candidate). 

Detection of the 1122-Hz signal in more bursts in the future would serve as the 
strongest corroboration of its TBO nature. It is also worth re-examining the existing RXTE
data, investigating the influence of energy cuts on the 1122-Hz candidate's power and the 
distribution of $P_\mathrm{m}$ in general.

\subsubsection{1A 1744$-$361}

\begin{figure}
 \centering
 \includegraphics[scale=0.72]{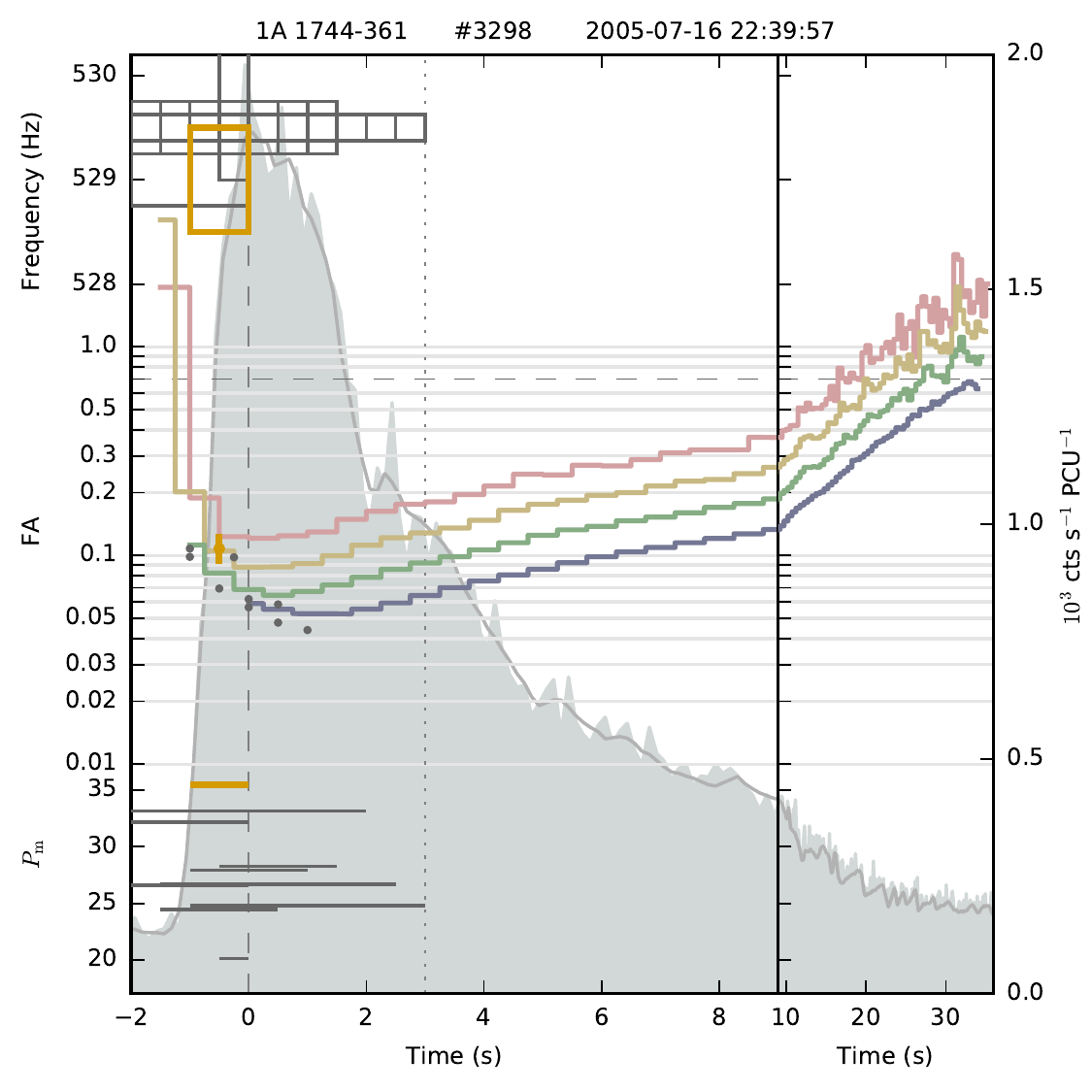}
   \caption{A TBO candidate from 1A~1744$-$361, previously reported by \citet{Bhattacharyya2006}. 
   The short duration, moderate power, but at the same time potential frequency drift make it hard 
   to classify this candidate either as noise or a TBO.}
 \label{fig:1A1744}
\end{figure}

A burst oscillation candidate was found by \citet{Bhattacharyya2006} in the single burst observed 
by RXTE. The signal appeared at $\sim 529$\,Hz in the rise of burst \# 3298. Splitting the peak 
into 4-s power spectra, and using a $Z^2$ spectrum, indicated a small ($<0.5$\,Hz) step in frequency. 
The highest rms FAs were 10.3\% in the $>3$\,keV band and 15\% for $>8$\,keV.
The candidate was also found by \citet{Galloway2008}, who reported FA of $11.3\pm1.8$\%.

Our analysis on a sample of three bursts yielded exactly one candidate, matching the one from 
\citet{Bhattacharyya2006}. The candidate was at 529\,Hz, with $P_\mathrm{d}=35.4$ (34.24 on 
non-normalized data) in one 1-s window (Fig.~\ref{fig:1A1744}). There are also sub-threshold candidates with different FFT 
windows.  The sub-threshold detections are somewhat later and higher in frequency, but none of 
them occurs in an independent time window. The associated FA is similar to that reported previously: 
$11\pm2$\%. Our sample contains two more bursts compared to the previous analysis. The additional 
bursts are a factor of a few fainter, and the upper limits on FA are about three times larger 
than the FA of detection. 

For the simulation runs, 10\% of them have one or more candidates above the detection threshold. 
For the higher threshold corresponding to $P=35$ in 1-s windows, only 1\% of simulations had one or 
more candidate. 

Based on the power alone, the 529-Hz candidate is not strong enough to classify as a TBO in our analysis. 
However, the presence of sub-threshold candidates hinting at frequency drift makes this candidate interesting. 
A definitive answer requires the detection of candidates at similar frequencies from future bursts.

\begin{figure*}
 \centering
 \includegraphics[scale=0.72]{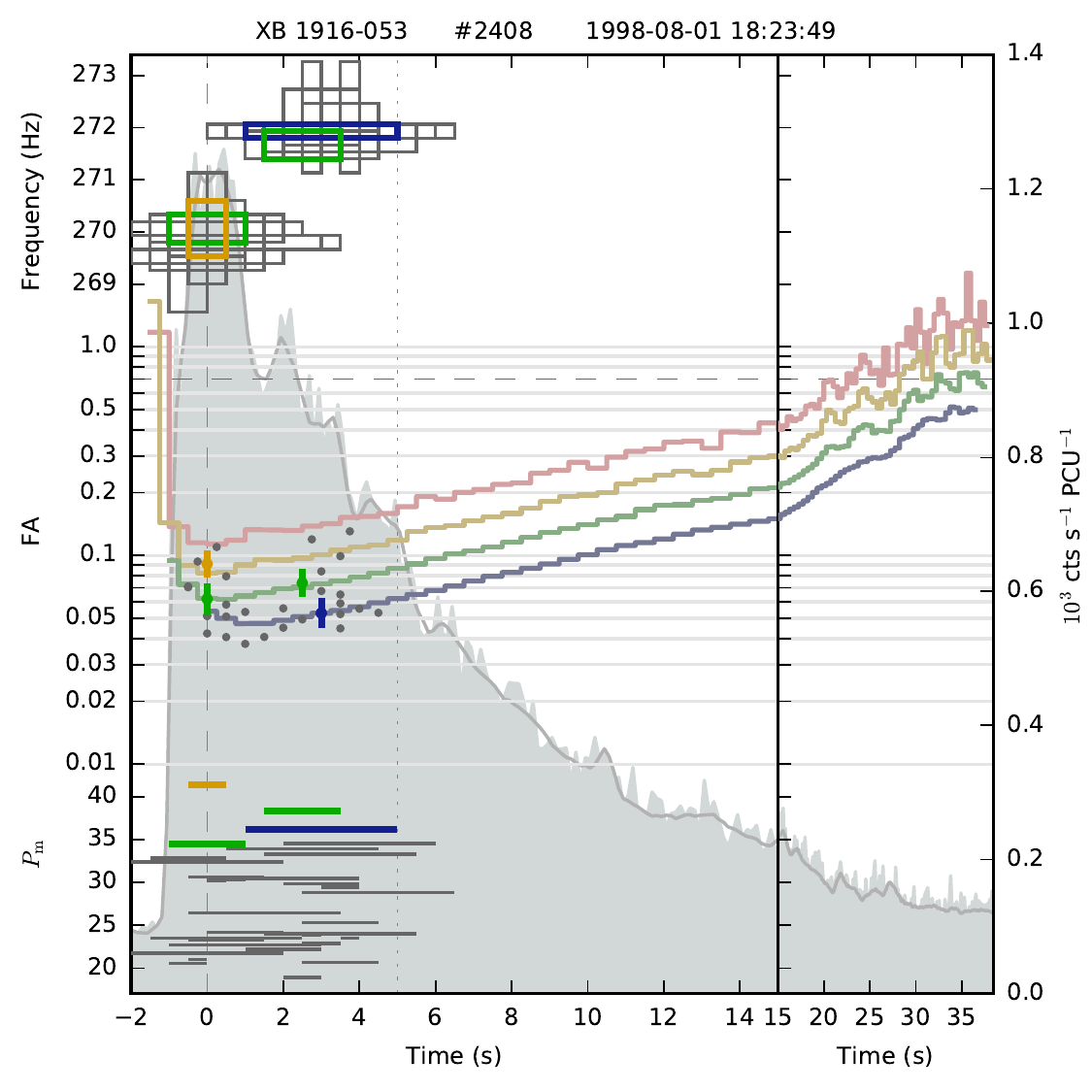}\includegraphics[scale=0.72]{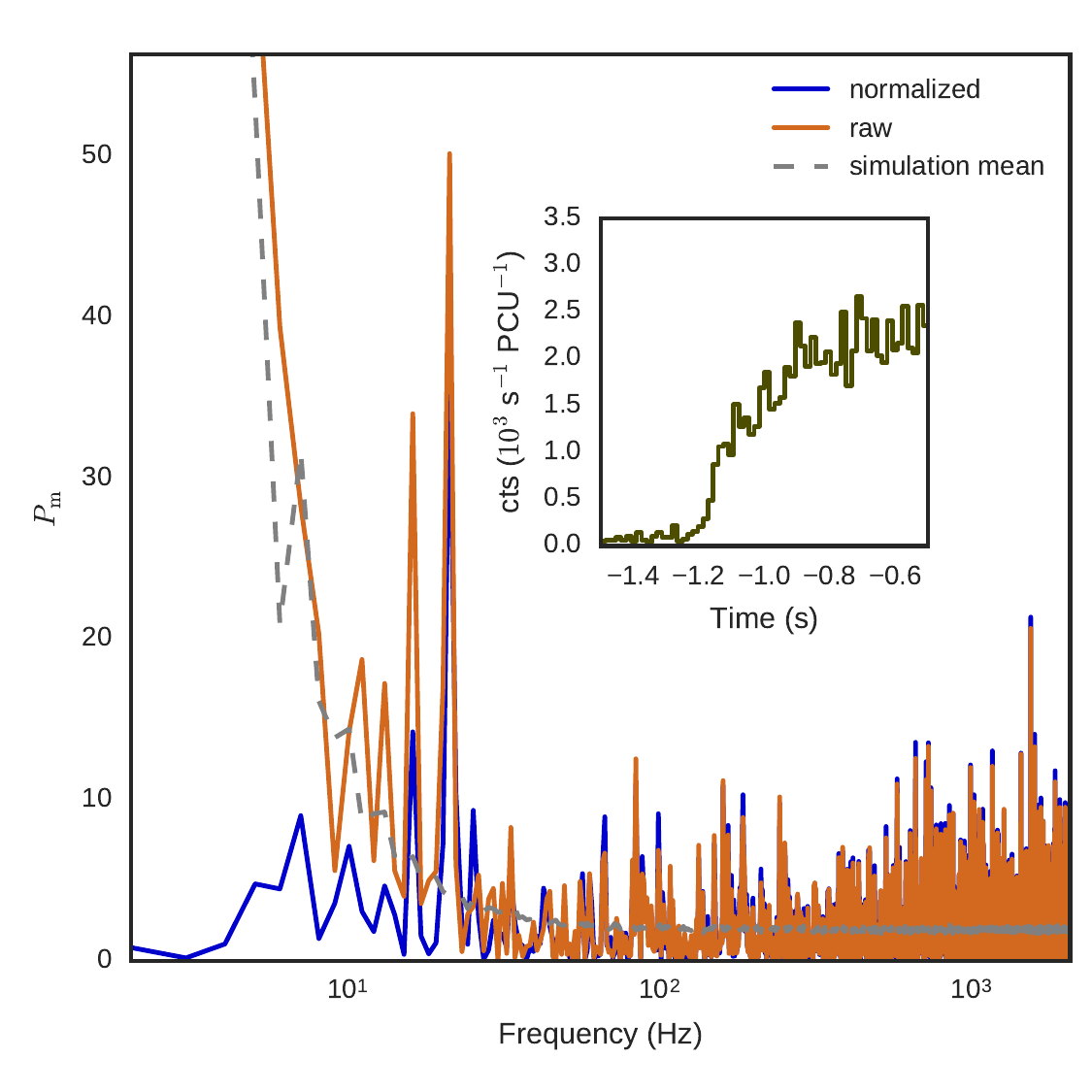}
   \caption{\textit{Left:} TBO candidate from XB 1916$-$053, previously reported by \citet{Galloway2001}. 
   \textit{Right:} 1D PS spectrum featuring a strong 21-Hz candidate during the rise of another burst. 
   After renormalization,    the signal power drops from 50.19 to 37.28. } 
 \label{fig:XB1916}
\end{figure*}

\subsubsection{SAX J1748.9$-$2021}

\citet{Kaaret2003} reported on a $P_\mathrm{m}=38.7$ TBO candidate at 409.7 Hz in one of the 15 bursts 
observed. The authors used merged photon TOA lists from event and burst catcher modes and computed FFTs in 
3-s successive time windows with 0.25-s steps. The oscillations lasted for about 4\,s and did not show any 
obvious frequency evolution. The significance of the detection was estimated to be equivalent to $4.4\sigma$ 
of the normal distribution, however the number of time windows searched was not taken into account.
Later, \citet{Altamirano2008a} found intermittent APPs at 442.36\,Hz in the persistent emission, with 
Leahy-normalized power as large as 100 with favorable data selection. The authors repeated the analysis 
of \citet{Kaaret2003}, but without window overlap. Taking into account the number of time windows searched, 
the significance of the 409.7-Hz candidate dropped to $\leq2.5\sigma$.

Our sample consisted of 29 bursts from the 2001 and 2010 outbursts. One low-frequency and two noise 
candidates were detected, none of them close to 409 or 442\,Hz. For the same burst as \citet{Kaaret2003}, 
we detected at 409.75\,Hz a maximum power of 33.2 in a 4-s window.  

Multiplying the detection probability by a factor of 2.5 to match $P_\mathrm{m}=33.2$ in 4-s windows 
yielded two more candidates, some of which were more significant than the 409.75-Hz one. About half 
of the simulation runs have the same or a larger number of candidates.

The weakness of 409.7-Hz candidate, and, more importantly the detection of a strong pulsation signal 
at a distinct frequency by \citet{Altamirano2008a} leads us to conclude that the candidate reported 
by \citet{Kaaret2003} was a spurious detection.

\subsubsection{GS 1826$-$24}

\citet{Thompson2005} reported on a 4.7-sigma detection of 611-Hz oscillations in the summed spectra of 
three burst tails during their simultaneous observations with Chandra. The authors searched for oscillations 
in FFT windows of several sizes and used different energy cuts. The signal was detected in part of the
RXTE energy band, in 0.25 time windows. 

Our sample contained 77 bursts which are rather long (2 min on average). Some of bursts were not covered by 
GTI or suffered from data gaps. Our analysis yielded several type II low-frequency candidates during the burst 
tails and a large, but not significant ($p=0.63$) number of noise candidates, none of them around 611\,Hz.

\subsubsection{XB 1916$-$053 (X 1916$-$53, 4U 1916$-$053)}
\label{subsec:XB1916}

\citet{Galloway2001} searched for TBOs in six bursts from XB 1916$-$053. The search was carried out in 
power spectra made from 0.5-s sliding windows overlapping by 0.25\,s. Power spectra were oversampled in 
frequency by a factor of 8. The authors reported TBOs at 269.4\,Hz with a single-trial significance equivalent
to $4.6\sigma$ of normal distribution; the significance accounting for the number of time windows searched and 
frequency oversampling was not given. After its onset, the signal becomes weaker and drifts in frequency by 
about 1.5\,Hz over the next second, then disappears and reappears $\sim 1$\,s later about 0.5\,Hz higher, 
with the first 0.25-s window yielding two peaks 3 Hz apart. \citet{Galloway2008} did not find any additional 
TBO candidates from the same frequency range on a larger sample of bursts.

The MINBAR catalogue lists 14 bursts from XB 1916$-$053 which were suitable for our analysis. Like Galloway et al.,
we detect candidates at $\sim 271$\,Hz in burst \#2408. The candidates appear in two independent time windows, of different size and 
separated by 1.75\,Hz (see Fig.~\ref{fig:XB1916}, left). Overall, we record 11 candidates above the adopted threshold, at 
frequencies between 6.5 and 1588\,Hz. Six of them come from the independent time windows. For the strongest candidate,
the one in 2-s window at 270\,Hz, 2\% of the simulation runs had the same or larger number of candidates of at least the 
same power. Adjusting the power threshold to match the power of second (third and so on) candidate results in $p=0$. 
Five out of 11 candidates have frequencies not higher than 21\,Hz and their $P_\mathrm{m}$ changed substantially after renormalization 
due to large mean  values of Fourier coefficients (see Fig.~\ref{fig:XB1916}, right). One of these five candidates is stronger  
than the one at 271.5\,Hz.

The highest-power candidate alone is marginally significant on our analysis. The presence of another candidate close 
in frequency and time makes its TBO nature more likely, however the quantitative estimate of it is beyond the scope of this work.

\subsection{Sources without previously detected TBOs}
\label{subsec:other}

\subsubsection{Unremarkable sources}

The following twelve sources yielded no TBO candidates in our analysis:
4U~2129+12, Cir~X-1, GRS~1747$-$312, KS~1741$-$293, SLX~1735$-$269, SAX~J1747.0$-$2853,
XB~1832$-$330, XTE~J1709$-$267, XTE~J1739$-$285, XTE~J1810$-$189, GX~3+1, and SAX~J1806.5$-$2215.
All of these sources had $<30$ bursts with total burst duration of $<15$\,min per source. The median peak S/N of the bursts differed by
two orders of magnitude, from 8 to 580, and the median upper limits on FAs in 1-s time windows at the burst peaks (hereafter, 
\textit{characteristic} $\FAup$) 
were anywhere from 5\% to 81\%. The average number of candidates from simulated bursts ranged from 0.02 to 1.1. 
Of these sources, GX~3+1 and SAX~J1806.5$-$2215 each had type I low-frequency candidates 
in one burst. 

2S~0918$-$549, 4U~0513$-$40, IGR~J17597$-$2201, and Ser X-1 also had a relatively small number of bursts ($<20$), with 
parameters comparable to the previous groups and a similar $\sim 1$ average number of simulated candidates per source. These
sources yielded some type I low-frequency candidates and one or two noise candidates. A significant fraction of simulation 
runs ($p=0.16$--$0.5$) had the same or a larger number of candidates. All noise candidates from real data had powers relatively close 
to the selection thresholds. 

The following group of sources yielded type II low-frequency candidates at frequencies of 2--4\,Hz: 
4U~0836$-$429,  4U~1323$-$62, and 4U~1705$-$44. Those sources had longer total observing bursting durations (18--70\,min), moderate 
median peak S/N ratios ($\sim 50$) and characteristic $\FAup$ of about 10\%. Several noise candidates recorded had a large chance of being 
due to random noise fluctuations, with $p$ ranging from 0.4 to 0.8.

1A~1742$-$294, XTE~J1701$-$462, 4U~1735$-$444, 4U~1746$-$37, and XTE~J2123$-$058 yielded a small number of noise candidates each 
and no low-frequency candidates. The number of bursts, total duration, median peak S/N and characteristic FA varied by a factor of few within this group, however for all
sources the large fraction of simulated bursts ($p$ of 0.13--0.83) had the same or a larger number of noise candidates.
A negative result from TBO searches from 4U~1746$-$37 was reported previously in \citet{Ootes2017}. 

GX 17+1 had unusually long bursts (average duration 4.5 min). Six candidates were detected with $p=0.7$. 
About two thirds of our sample was previously examined by \citet{Kuulkers2002}: who did not find any signal beyond $P_\mathrm{m}=42.8$ in 
time windows of 0.25 and 2\,s, For all bursts in the sample, the maximum power above 10 Hz did not exceed 37.5.

\subsubsection{Sources with somewhat larger number of noise or low-frequency candidates}

The relatively faint sources XTE J1710$-$281 and SLX~1744$-$300 had a marginally significant  number 
of noise candidates per source ($p$ of 0.02 and 0.01, respectively). 
None of the candidates were particularly strong, and their frequencies were scattered between 20 and 1980\,MHz. No low-frequency 
candidates were recorded.

A stronger source, 4U~1722$-$30, yielded a rather strong single candidate. The candidate was 
recorded at 22\,Hz at burst peak in a 0.5-s time window covering a small dip in the LC. This dip was well modeled
and had no low-frequency type I candidates assosiated with it. 
The candidate had a power of 36.49, with an unnormalized power of 33.59.
Only 1 out of 100 simulations yielded 1 or more candidates with the same or larger power. 
A similarly strong candidate coming from a 1-s time window with a data gap was recorded from 4U~1820$-$303 at 280\,MHz. 
The candidate had $P_\mathrm{m}=39.7$ (34.69 in unnormalized data) and $p=0.02$.
It is worth mentioning that NICER has observed PRE bursts from this source recently, and no oscillations have been 
detected \citep{Keek2018}. Overall, although these candidates are rather strong, their power depends greatly
on proper LC modeling.

Two other faint sources, EXO 1745$-$248 and Cyg~X-2 had remarkably numerous and strong low-frequency candidates (type II).
For EXO 1745$-$248, these candidates were recorded within most of the on-burst windows at multiple frequencies 
between  2 and 12\,Hz. Sometimes the candidates were grouped in time, similarly to IGR J17473$-$2721. Only one noise candidate ($p=0.4$) was detected from this 
source, at 14.5\,Hz, with multiple low-frequency candidates from the same time windows at 2--8\,Hz. 
 
The low-frequency candidates from Cyg X-2 exhibited various behavior: during some bursts they were confined to a single frequency,
during others they were spread chaotically within 2--12\,Hz or occurred at two separate frequencies (e.g. 3 and 9\,Hz).
Few glimmer candidates were recorded, with $p=0.01$.

\subsubsection{IGR J17473$-$2721: an interesting pair of candidates}
 \label{subsec:IGRJ17473}

\begin{figure}
 \centering
 \includegraphics[scale=0.72]{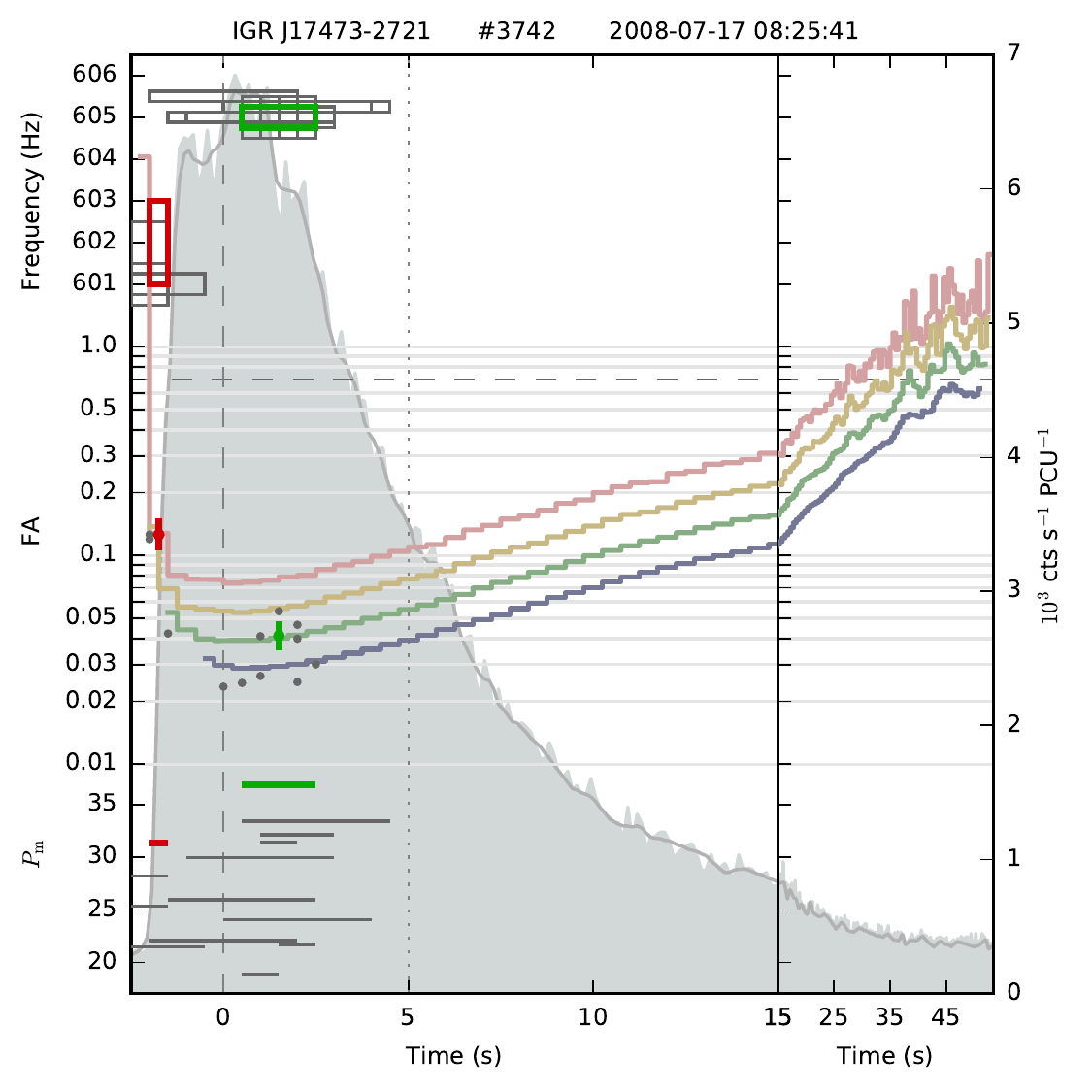}
   \caption{An interesting pair of oscillation candidates from IGR J17473$-$2721. The candidates are rather faint and not significant 
   based on their power alone. However, the occurence within the burst (marked as PRE in the MINBAR catalogue) and 
   frequency separation resemble typical TBO behavior.}
 \label{fig:IGRJ1747}
\end{figure}

IGR J17473$-$2721 generated an interesting pair of candidates at 602 and 605\,Hz occurring few seconds apart in 
the same burst (Fig.~\ref{fig:IGRJ1747}). The first candidate came from a 0.5-s window on the burst rise, had 
$P_\mathrm{m}=31.3$ (29.3 on non-normalized data), and FA of 13\%.   The second candidate came from a 1-s window 
 right after the burst peak and had $P_\mathrm{m}$ of 36.8 (33.8 on non-normalized data) and  FA of 4\%.  
The other 43 bursts did not have any candidates at similar frequencies.  

The moderate power (more than half of simulations had the same or larger number of candidates of at least the same 
significance) and the lack of detections in multiple time windows or bursts do not allow us to classify these candidates as 
TBOs, however the other properties (candidates framing burst peak for the PRE burst, second candidate being few-Hz higher 
than the first one) are similar to confirmed TBOs (cf. e.g. SAX J1750.8$-$2900). 

IGR J17473$-$2721 also had many low-frequency candidates at frequencies of 2--6.25\,Hz. 
Sometimes low-frequency candidates were spread uniformly throughout an on-burst window, sometimes they formed distinct groups.

\subsubsection{SAX J1810.8$-$269: new TBO source}
\label{subsec:SAXJ1810}

\begin{figure}
 \centering
 \includegraphics[scale=0.72]{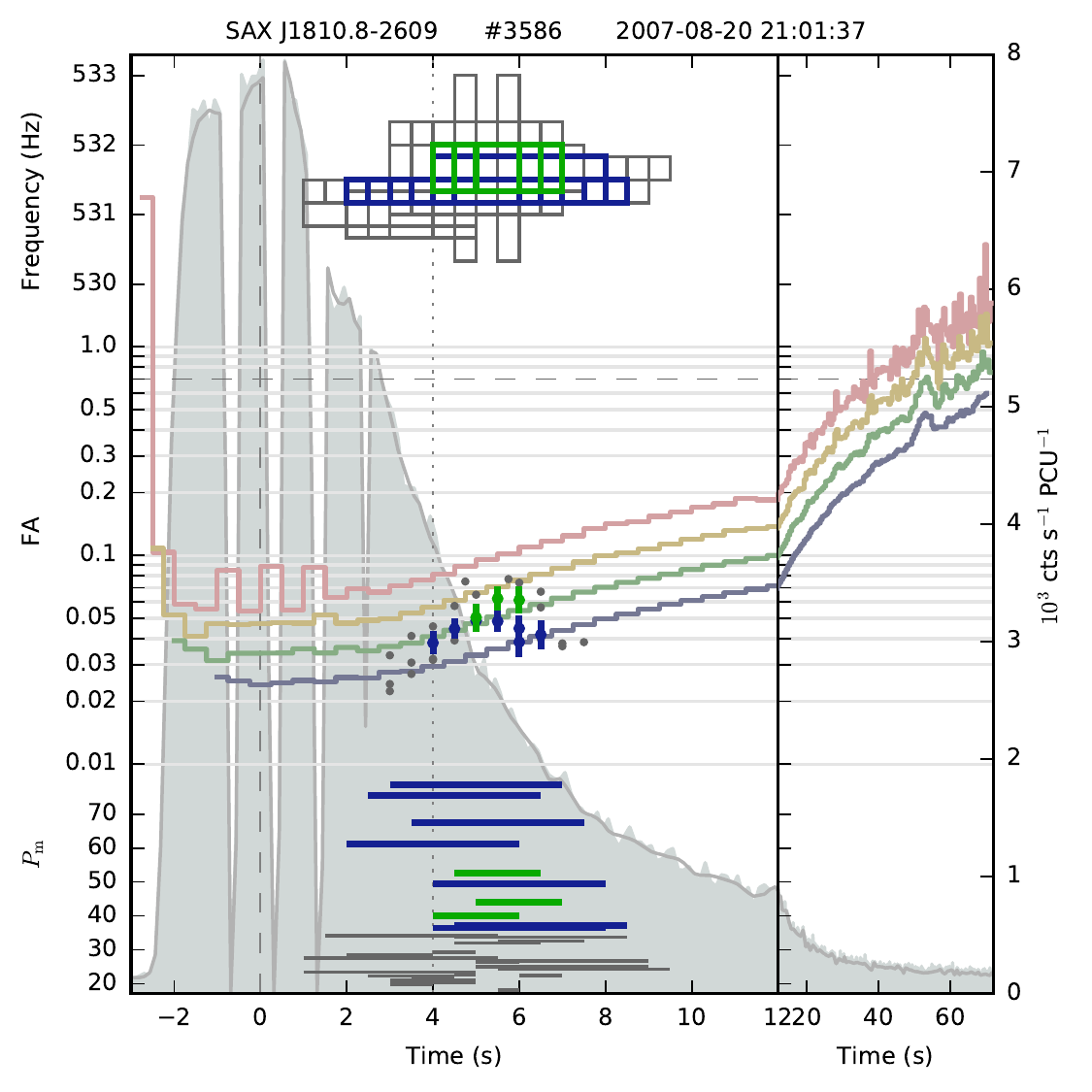}
   \caption{TBOs from a new oscillation source SAX~J1810.8$-$269.}
 \label{fig:SAXJ1810}
\end{figure}

The MINBAR catalogue lists six bursts from SAX J1810.8$-$ 269. The bursts are relatively bright and 
some of them have data gaps. Strong oscillations ($P_\mathrm{m}=78.58$, 74.28 on the non-normalized data) 
were discovered at $\sim 531$\,Hz in the B region of one of the bursts (Fig.~\ref{fig:SAXJ1810}). For the adopted detection
threshold, the signal is present in one independent 4-s time window. None of the simulations had a signal with similar power. 
Subthreshold candidates with $P_\mathrm{m}\gtrsim24$ are detected in two consecutive 4-s windows, 
for the significance estimates of that see \citet{Bilous2018}, which reports this discovery in more detail. 

No low-frequency or noise candidates were detected from this source. 

\bibliographystyle{apj} 
\bibliography{lit}

\end{document}